\documentclass[onecolumn,12pt, draftclsnofoot]{IEEEtran}

\ifCLASSINFOpdf
\else
   \usepackage[dvips]{graphicx}
\fi
\usepackage{url}
\usepackage{hhline}
\usepackage{graphicx}
\usepackage{multicol}
\usepackage{graphicx}
\usepackage{subfigure}
\usepackage{amssymb}
\usepackage{amsmath}

\usepackage{calc}
\usepackage{array}
\usepackage{optidef}
\usepackage{epstopdf}
\usepackage{mathtools}
\usepackage{mathrsfs}
\usepackage{bm}
\usepackage{cite}
\usepackage{boldline}
\usepackage{amsthm}
\usepackage{algorithm}
\usepackage{algorithmicx, algpseudocode}
\usepackage{amsthm}

\newtheorem{remark}{Remark}

\usepackage{amsmath}
\usepackage{footmisc}
\DeclareMathOperator*{\maximize}{maximize}

\usepackage{booktabs,makecell,tabularx}

\newcolumntype{C}[1]{>{\centering\arraybackslash}p{#1}}
\newcolumntype{L}{>{\raggedright\arraybackslash}X}

\usepackage{siunitx}
\usepackage{etoolbox}
\newrobustcmd{\B}{\bfseries}

\usepackage{comment}
\usepackage{algorithm}% http://ctan.org/pkg/algorithms
\usepackage{algpseudocode}% http://ctan.org/pkg/algorithmicx
\usepackage{enumitem} % for '\newlist' and '\setlist' macros

\usepackage{colortbl}     % preamble
\usepackage{subfigure}
\usepackage{color, soul}
\usepackage{stfloats}
\usepackage{float}
\usepackage{multirow}
\usepackage{wrapfig}
\usepackage{booktabs}
\usepackage{slashbox}
\usepackage{optidef}
\definecolor{LightBlue}{rgb}{0.75,0.936,1.00}
\sethlcolor{LightBlue}
\definecolor{LightCyan}{rgb}{0.88,1,1}

\usepackage{xcolor}

\usepackage[colorlinks=true,linkcolor=blue,citecolor=blue,urlcolor=blue]{hyperref}

\begin{document}

\bstctlcite{IEEEexample:BSTcontrol}

\title{Robust Rate-Splitting Design for Mixed Dual-Polarized Integrated Satellite-Terrestrial Networks Under Polarization Mismatch}
%\title{Robust RSMA Design for Quad-Polarized Integrated Satellite-Terrestrial Networks}

\author{Jaehyup Seong, Juhwan Lee, Jungwoo Lee, Sean Kwon, and Wonjae Shin %\vspace{-9mm}

    %\thanks{Manuscript received Month XX, 2022. The associate editor coordinating the review of this letter and approving it for publication was XXXX 2022. This work was supported by the Institute of Information \& Communications Technology Planning \& Evaluation (No. 2021-0-00260) grant, and the Basic Science Research Programs under the National Research Foundation of Korea (NRF) funded by the Ministry of Science and ICT (No. 2019R1C1C1006806, No. 2021R1A4A1030775).}
    \thanks{Jaehyup Seong and Wonjae Shin are with the School of Electrical Engineering,
    Korea University, Seoul 02841, South Korea 
    (email: jaehyup@korea.ac.kr; wjshin@korea.ac.kr).
    
    Juhwan Lee and Jungwoo Lee are with the Department of Electrical and Computer Engineering, 
    Seoul National University, Seoul 08826, South Korea (e-mail: sgsyk649@snu.ac.kr; junglee@snu.ac.kr).

    Sean Kwon is with the Department of Electrical Engineering,
California State University Long Beach, Long Beach, CA 90840, USA
(e-mail: sean.kwon@csulb.edu).
}
\thanks{A part of this work was presented in
part at the IEEE GLOBECOM, Taipei, Taiwan, Dec. 2025 \cite{11432517}.}
}

%\markboth{Submitted to IEEE Transactions on Wireless Communications}
\maketitle

\begin{abstract}
Dual-polarized transmission offers a promising approach to improve spectral efficiency in multi-antenna networks by reusing frequency and time resources across orthogonal polarization domains. 
Building upon this advantage, this paper investigates interference management in mixed dual-polarized integrated satellite-terrestrial networks (MDP-ISTN), comprising a circularly polarized (CP) satellite sub-network and a linearly polarized (LP) terrestrial sub-network. 
To this end, we employ rate-splitting multiple access (RSMA), which enables flexible non-orthogonal transmission through partial interference decoding and partial interference treating-as-noise.
Specifically, to jointly mitigate both inter-network interference between the CP low Earth orbit (LEO) satellite and LP terrestrial sub-networks as well as intra-network interference within each sub-network, we propose an {\emph{MDP-RSMA}} framework that incorporates inter-network rate-splitting (RS) with a super-common message together with intra-network RS. 
Moreover, we account for practical challenges in MDP-ISTN, including Faraday rotation, polarization mismatch, channel depolarization, antenna depolarization, and imperfect channel state information at the transmitter.
To maximize the minimum user rate among all satellite and terrestrial users, we formulate a robust precoder optimization problem and develop a weighted minimum mean square error (WMMSE)-based algorithm tailored to the proposed MDP-RSMA.
Numerical results demonstrate that the proposed scheme significantly improves the minimum user rate over several baseline schemes across diverse MDP-ISTN scenarios.
\end{abstract}
%\vspace{-1.5mm}

%\vspace{-3.5mm}
%\vspace{-3mm}
\section{Introduction}
%\vspace{-1.5mm}
Advances in satellite technology, together with the increasing demand for high data rates in wireless communications, have stimulated substantial interest in integrated satellite-terrestrial networks (ISTN) \cite{6829945, 10559954,  jamshed2025tutorial}. % 줄인거
% , thanks to their capability to provide reliable communication services over wide coverage areas 
The integration of satellite and terrestrial networks can substantially enhance connectivity in remote or rural regions where terrestrial infrastructure is limited. 
To accommodate the growing traffic demand under spectrum scarcity, sharing radio resources such as frequency and time between the satellite and terrestrial networks is emerging as a promising approach. 
Furthermore, by exploiting orthogonal polarization dimensions, dual-polarized transmission can provide an additional degree of freedom and has the potential to double the achievable capacity in ISTN \cite{1603707, 5962379, MIMOSat:surv:11, castellanos2023linear,lee2026integrated}. %줄인거

%Furthermore, dual-polarized transmission, which exploits orthogonal polarization dimensions, can provide an additional degree of freedom and has the potential to effectively double the achievable capacity in ISTN \cite{1603707, 5962379, MIMOSat:surv:11}.

%Dual-polarized transmission exploits two orthogonal polarizations, such as vertical and horizontal polarizations (VP/HP) for linearly polarized (LP) transmission \cite{ElecWaveAntenna:02, Antenna5G:20, Sena:twc:19}, or right- and left-hand circular polarizations (RHCP/LHCP) for circularly polarized (CP) transmission \cite{EduCP:TEdu:03, Zhang:TVT:15}.
Dual-polarized transmission exploits two orthogonal polarization states. For LP transmission, vertical and horizontal polarizations (VP/HP) can serve as an orthogonal pair \cite{ElecWaveAntenna:02, Antenna5G:20, Park:twc:15, Sena:twc:19, de2021irs}. For CP transmission, right- and left-hand circular polarizations (RHCP/LHCP) form an orthogonal pair \cite{EduCP:TEdu:03, Zhang:TVT:15}. %줄인거
When LP transmission, which is widely utilized in terrestrial networks, is applied to satellite communications (SATCOM), the transmitted signal is affected by Faraday rotation while traversing the ionosphere. This phenomenon rotates the polarization plane, leading to a misalignment with the receive antenna orientation and consequent power loss \cite{Linda:icc:15, ITU:23}. Furthermore, LEO satellites move at high velocities, reaching approximately \num{7.56} km/s at an altitude of \num{600} km. The rapid satellite motion causes dynamic antenna orientation differences between the satellite and users, resulting in time-varying polarization mismatch \cite{DongHyun:tcom:22}. % 줄인거 
This mismatch deteriorates polarization orthogonality, thereby increasing cross-polar coupling and degrading the system performance. 
Conversely, CP transmission provides robustness to ionospheric Faraday rotation. Moreover, CP transmission is significantly less sensitive to antenna orientation mismatch than LP transmission, making it particularly suitable for LEO SATCOM  \cite{Sella:tvt:06, Liolis:tcom:10, Milcom:15, Zhu:GC:18, Qian:tccn:21}. %줄인거

%Therefore, in the ISTN, it is advisable to employ CP for the satellite sub-network and LP for the terrestrial sub-network, thereby making a mixed dual-polarized ISTN (MDP-ISTN) with dual-CP and dual-LP. 
%However, aggressive frequency reuse can lead to significant interference both between and within polarized sub-networks. 
%Within each sub-network, the practical limitations of polarized antennas, such as limited cross-polarization discrimination (XPD) \cite{Coldrey:vtc:08, }, hinder their ability to fully discriminate the desired polarization. As a result, cross-polar interference between orthogonal polarizations, as well as co-polar interference between identical polarizations, is induced. Moreover, interference between the CP-based satellite sub-network and the LP-based terrestrial sub-network exacerbates the degradation of user data rates. 

%Accordingly, CP transmission is typically employed for the LEO satellite networks, while LP transmission is commonly adopted in the terrestrial networks, thereby forming a mixed dual-polarized ISTN (MDP-ISTN) with CP and LP transmission. 
Accordingly, CP transmission is typically employed in LEO satellite networks, while LP transmission is commonly adopted in terrestrial networks. The combined use of CP and LP transmission in ISTN forms a mixed dual-polarized ISTN (MDP-ISTN). % 줄인거
However, aggressive reuse of radio resources leads to significant signal interference both between and within polarized sub-networks. 
Within each sub-network, depolarization can arise from non-ideal antennas and the radio propagation environment. Ideally, a polarized antenna transmits or receives only the intended co-polarized component. In practice, however, transmit and receive antennas exhibit nonzero cross-polar responses, resulting in antenna depolarization. In addition, signal reflections, diffractions, and scattering in the propagation environment cause channel depolarization. 
The severity of antenna and channel depolarization is commonly characterized by their respective cross-polarization discrimination (XPD) values. The XPD is defined as the ratio of the average co-polarized signal power to the average cross-polarized signal power \cite{5962379, Coldrey:vtc:08}.
Consequently, antenna and channel depolarization generate cross-polar interference, while resource sharing among users gives rise to co-polar interference within each sub-network.
Moreover, the dual-CP LEO satellite sub-network and dual-LP terrestrial sub-network employ heterogeneous pairs of orthogonal polarization states. Across the sub-networks, polarization orthogonality is not preserved between the CP and LP states, resulting in severe inter-network interference that further degrades system performance. For instance, transmissions using either RHCP or LHCP interfere with receptions using both VP and HP, and vice versa.
To overcome such interference issues in MDP-ISTN, accurate channel state information (CSI) at the transmitter (CSIT) is required. However, acquiring accurate CSIT is particularly challenging for the satellite networks due to the high velocity of LEO satellites and significant propagation delay \cite{you2020massive, li2021downlink, you2022beam, 10787138, Jaehyup:jsac:24}. % 줄인거

% Mixed 좀 더 갖오?
%Furthermore, the inter-network interference between CP-based satellite sub-network and LP-based terrestrial sub-network can also deteriorate user rates.  
% CSI error in SATCOM
% Furthermore, due to the long round-trip delay and the satellite mobility especially for the low-earth orbit (LEO) satellite, accurate satellite channel state information (CSI) is challenging to acquire at the gateway (GW). 

% RSMA를 통해 간섭문제 & CSI-robustness 해결한다
%Recently, rate-splitting multiple access (RSMA) has arisen as a promising transmission scheme, owing to its flexible interference management capabilities and robustness to CSI uncertainty \cite{clerckx2016rate, Mao:eurasp:18, Bruno:wcl:19, Mao:tut:22, Shin:Net:24}.
%Following the one-layer RSMA principle, all user messages are divided into common and private parts at the transmitter. 
%The common messages are combined and encoded into a common stream that is decodable by all users, whereas the private messages are independently encoded into distinct private streams. 
%The users decode and remove the common stream using successive interference cancellation (SIC), then decode their desired private streams while treating other signals as noise. 
%As such, RSMA is capable of adaptively mitigating interference by partially decoding it and partially treating it as noise. 
%Resorting to this flexibility, RSMA has been widely utilized in SATCOM and ISTN to manage interference while ensuring robustness to CSI errors 

Recently, rate-splitting multiple access (RSMA) has arisen as a promising multiple access (MA) technique owing to its flexible interference management capability and robustness to CSI uncertainty. 
Following the one-layer RSMA principle, each user message is divided into common and private parts at the transmitter. The common parts of all users are combined and encoded into a common stream decodable by all users. In contrast, the private parts are independently encoded into user-specific private streams. % 줄인거
At the receiver side, each user first decodes and removes the common stream using successive interference cancellation (SIC). Each user then decodes its intended private stream while treating the remaining interference as noise. %줄인거 
By flexibly adjusting the portion between the common and private parts, RSMA has been shown to provide high spectral efficiency across diverse network scenarios while remaining robust against CSI uncertainty \cite{clerckx2016rate, Mao:eurawc:18, Shin:Net:24}.

% RSMA를 LP에서 사용한 연구 있었다.
%\vspace{-3mm}
\subsection{Related Works}
%\vspace{-1mm}

Inspired by the aforementioned advantages of RSMA, its application to dual-LP terrestrial networks has been investigated in \cite{Sena:gc:22,Sena:wcl:22,Sena:twc:23,sun2025dual}. These studies have demonstrated the advantages of RSMA in managing co-polar interference arising from resource sharing among users and cross-polar interference caused by channel depolarization, compared with conventional MA schemes, including spatial division MA (SDMA), non-orthogonal MA (NOMA), and orthogonal MA (OMA). Specifically, the authors of \cite{Sena:gc:22,Sena:wcl:22} have proposed SIC free dual-polarized RSMA approaches that independently transmit the common and private streams over orthogonal polarization domains. These approaches enable users to recover the two streams independently without SIC, thereby reducing receiver complexity by eliminating SIC operations and avoiding performance degradation caused by residual SIC errors. Also, \cite{Sena:twc:23} has investigated dual-polarized RSMA, where superimposed common and private streams are transmitted over both polarization domains. The proposed approach has been shown to achieve a higher sum rate and a lower outage probability than conventional single-polarized RSMA. In \cite{sun2025dual}, a stacked intelligent metasurface-assisted dual-polarized RSMA transceiver architecture has been proposed. It has been shown that the proposed architecture significantly improves both sum rate and physical-layer security over the considered baseline schemes.

Moreover, robust precoding for jointly managing intra- and inter-network interference has been investigated in ISTN \cite{Longfei:twc:23,jh:tvt:24,10908461}, as well as geostationary orbit (GEO) and LEO satellites coexisting networks \cite{xu2024distributed} and space-air-ground integrated networks (SAGIN) \cite{zhang2024interference,zhang2025metars}.
Specifically, \cite{Longfei:twc:23} investigated both coordinated and cooperative approaches for ISTN. The coordinated approach shares CSI between the networks, whereas both CSI and user data are shared in the cooperative approach. For both approaches, precoding schemes incorporating single-layer rate-splitting (RS) have been proposed to mitigate intra- and inter-network interference. The trade-off between the signaling overhead incurred by user data sharing and the resulting communication performance has been investigated. In \cite{jh:tvt:24}, the coordinated approach presented in \cite{Longfei:twc:23} has been extended by incorporating a multilayer RS strategy for LEO satellite and terrestrial networks. Moreover, an RSMA-based distributed precoding framework for an ISTN scenario, where neither CSI nor user data sharing is available, has been introduced in \cite{10908461}. 

Furthermore, in \cite{xu2024distributed,zhang2024interference,zhang2025metars}, multi-layer RS strategies have been investigated to jointly manage intra- and inter-network interference in GEO-LEO coexisting satellite networks and SAGIN. Specifically, the authors in \cite{xu2024distributed} have introduced distributed RSMA (D-RSMA), where the GEO satellite provides multicast services while the LEO satellites provide unicast services. By sharing user data across the GEO and LEO satellite networks, portions of the multicast GEO message and unicast LEO messages are combined and jointly encoded into a super-common stream at the GEO satellite. The encoded super-common stream allows users in both networks to partially decode and cancel inter-network interference. Meanwhile, common streams transmitted by the LEO satellites are employed to manage intra-network interference. Building on the D-RSMA principle, fully distributed RSMA (FD-RSMA) has been developed for a SAGIN comprising LEO satellite and aerial networks, both supporting unicast transmission \cite{zhang2024interference}. In FD-RSMA, inter-network user data sharing enables the construction of super-common streams for inter-network interference management, while network specific common streams are employed to manage intra-network interference. As an extension of FD-RSMA, MetaRS has been proposed to jointly optimize user access-point associations, precoders, and rate allocations for the super-common and network specific common streams in SAGIN \cite{zhang2025metars}.

\subsection{Motivations and Contributions}
%\vspace{-1mm}

Despite recent advances in dual-polarized RSMA precoder design and robust precoding for intra- and inter-network interference management, existing studies have limitations that hinder their application to practical MDP-ISTN. In practical MDP-ISTN, Faraday rotation, polarization mismatch, antenna depolarization, and channel depolarization reduce the desired co-polar signal power while generating cross-polar interference within each sub-network. 
Meanwhile, multiple users sharing the same polarization within each sub-network generate co-polar interference. Reusing radio resources across the dual-CP satellite and dual-LP terrestrial networks additionally introduces inter-network interference. Therefore, jointly managing co-polar, cross-polar, and inter-network interference within a unified framework is essential for ensuring reliable communication services in practical MDP-ISTN. 

However, existing studies that manage intra- and inter-network interference \cite{Longfei:twc:23,jh:tvt:24,10908461,xu2024distributed,zhang2024interference,zhang2025metars} have not explicitly accounted for practical polarization-related effects. Consequently, cross-polar interference within each sub-network and interference between sub-networks employing distinct pairs of orthogonal polarization states have not been addressed.
Existing work on dual-polarized RSMA precoder design under channel depolarization focuses primarily on co-polar and cross-polar interference management in dual-LP terrestrial networks \cite{Sena:gc:22,Sena:wcl:22,Sena:twc:23,sun2025dual}. However, inter-network interference arising in MDP-ISTN remains unaddressed.
Moreover, the high mobility of LEO satellites induces time-varying polarization mismatch due to dynamic antenna orientation differences between the satellite and users, while ionospheric propagation induces Faraday rotation \cite{DongHyun:tcom:22, ITU:23}. These specific effects of the dual-CP LEO satellite network, which have not been addressed in dual-LP terrestrial network studies \cite{Sena:gc:22, Sena:wcl:22,Sena:twc:23,sun2025dual}, should also be carefully incorporated into the precoder design for MDP-ISTN.

To tackle these challenges, we propose the MDP-RSMA framework to jointly manage co-polar and cross-polar interference within the dual-CP LEO satellite and dual-LP terrestrial networks, as well as inter-network interference between them. We establish a practical MDP-ISTN model that incorporates Faraday rotation, polarization mismatch, antenna depolarization, and channel depolarization. We further account for imperfect CSIT at the LEO satellite due to its high mobility and the absence of inter-network user data sharing due to the associated signaling, backhaul, and processing overhead. Under these practical constraints, we develop a robust MDP-RSMA precoder based on an inter-network RS strategy that requires no user data sharing between networks, combined with intra-network RS within each sub-network. The main contributions of this paper are summarized as follows:
\begin{itemize}
\item 
We put forth a new MDP-RSMA framework that effectively mitigates both inter- and intra-network interference in MDP-ISTN without user data sharing between the LEO satellite and the terrestrial base station (BS). Specifically, each SU message is split into super-common, common, and private parts. The super-common parts of the SU messages are combined and encoded into a super-common stream at the LEO satellite, with its power adaptively allocated according to the satellite interference experienced by the CUs. With the super-common stream first decoded and removed via SIC by all users, inter-network interference can be flexibly managed in the absence of inter-network user data sharing. Alongside this super-common stream design, intra-network RS with common and private streams is employed to manage co- and cross-polar interference within each sub-network.

\item We model the practical MDP-ISTN by accounting for antenna and channel depolarization in both satellite and terrestrial links, together with Faraday rotation and polarization mismatch in the satellite links. To guarantee fairness among all users, we formulate a max-min fairness problem based on the proposed MDP-RSMA framework. %To address the non-convexity of the formulated problem, we develop a weighted minimum mean square error (WMMSE)-based algorithm tailored to MDP-RSMA that jointly optimizes the beamforming vector and power allocation for super-common, common, and private streams.
To address the non-convexity of the formulated problem, we first establish a problem-specific weighted minimum mean
square error (WMMSE) reformulation that reflects the proposed message-splitting, decoding, and interference management structure. Leveraging this reformulation, we develop an alternating optimization (AO)-based algorithm. The resulting AO procedure alternates between updating the stream-specific equalizers and weights and jointly optimizing the precoders and the rate portions of the super-common and common streams.

\item We numerically demonstrate the effectiveness of the proposed
MDP-RSMA in maximizing the minimum user spectral efficiency under practical constraints, including imperfect CSIT, Faraday rotation, polarization mismatch, channel and antenna depolarization, and the absence of inter-network user data sharing. Extensive numerical results across diverse MDP-ISTN scenarios confirm its consistent performance advantages over competitive interference management strategies based on RSMA and other state-of-the-art MA techniques. % 줄인거

\end{itemize}

%\vspace{-2mm}
%\textit{Notations:} %Throughout the paper, scalars, vectors, and matrices are denoted by standard letters, lowercase boldface letters, and uppercase boldface letters, respectively. 
%\subsection{Notations}
%The operators $\odot$, $\otimes$, $(\cdot)^{\sf T}$, $(\cdot)^{\sf H}$, $\sf{vec}(\cdot)$, $\sf{tr}(\cdot)$, $\mathbb{E}[\cdot]$, and $\exp(\cdot)$ denote the Hadamard product, Kronecker product, transpose, conjugate transpose, column-wise vectorization, trace, expectation, and exponential function, respectively. The matrix $\mathbf{I}_N$ denotes the $N\times N$ identity matrix, and $\mathbf{1}_{N\times1}$ and $\mathbf{0}_{1\times N}$ denote the all-ones and all-zeros vectors, respectively.

%\vspace{-1mm}
\textit{Notations:} %Throughout the paper, scalars, vectors, and matrices are denoted by standard letters, lowercase boldface letters, and uppercase boldface letters, respectively. 
The operators $\odot$, $\otimes$, $(\cdot)^{\sf T}$, $(\cdot)^{\sf H}$, $\sf{vec}(\cdot)$, $\sf{tr}(\cdot)$, $\mathbb{E}[\cdot]$, and $\exp(\cdot)$ denote the Hadamard product, Kronecker product, transpose, conjugate transpose, column-wise vectorization, trace, expectation, and exponential function, respectively. The matrix $\mathbf{I}_N$ denotes the $N\times N$ identity matrix, and $\mathbf{1}_{N\times1}$ and $\mathbf{0}_{1\times N}$ denote the all-ones and all-zeros vectors, respectively.

%\vspace{-3mm}
%\vspace{-3.5mm}
\section{System Model and Problem Formulation}
\begin{figure}[!t]
\centering
 \includegraphics[width=1\linewidth]{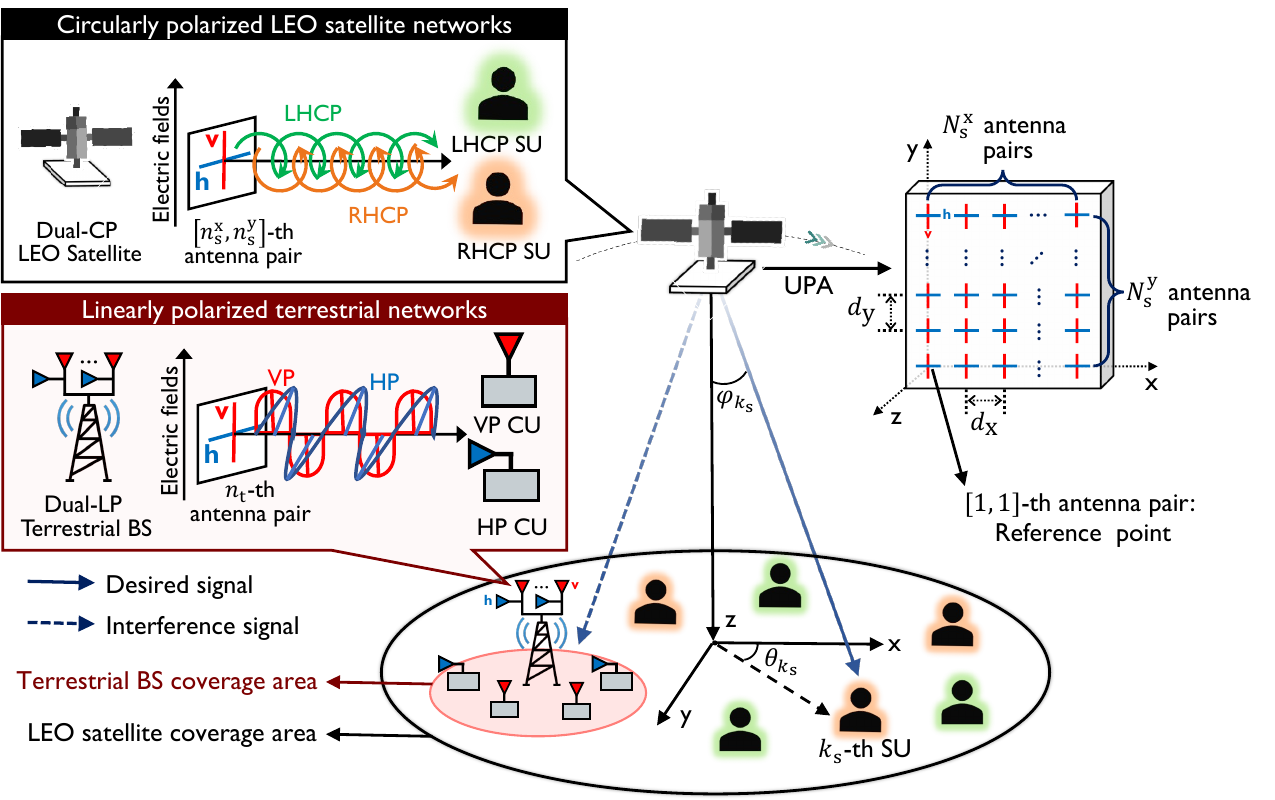}
             %\vspace{-3mm}
 		\caption{System model of the mixed dual-polarized ISTN systems.}
    	\label{Fig_1}%\vspace{-3mm}
\end{figure}

%\vspace{-1mm}
We consider MDP-ISTN, consisting of CP LEO satellite networks and LP terrestrial networks, where frequency and time resources are fully reused, as shown in Fig.~\ref{Fig_1}. The LEO satellite is equipped with a uniform planar array (UPA) comprising $N_{\sf s} \triangleq N_{\sf s}^{\sf x} \times N_{\sf s}^{\sf y}$ pairs of co-located vertical and horizontal antenna elements, where $N_{\sf s}^{\sf x}$ and $N_{\sf s}^{\sf y}$ denote the numbers of antenna pairs along the ${\sf x}$- and ${\sf y}$-axes, respectively. 
Moreover, the terrestrial BS is equipped with $N_{\sf t}$ pairs of co-located vertical and horizontal antenna elements.
At the LEO satellite, each co-located pair of vertical and horizontal antenna elements is jointly excited with equal amplitudes and a $\pm \pi/2$ phase offset to generate right- and left-hand circularly polarized signals, thereby providing robustness against ionospheric Faraday rotation and polarization mismatch induced by rapid LEO satellite motion \cite{Zhang:TVT:15, Milcom:15}. Accordingly, each antenna pair at the LEO satellite generates two orthogonal CP signals, resulting in dual-CP transmission. At the BS, each co-located pair of vertical and horizontal antenna elements is independently excited without phase coupling as in \cite{Sena:gc:22, Sena:wcl:22, Sena:twc:23, Park:twc:15}. As such, the BS directly generates vertically and horizontally polarized signals, thereby enabling dual-LP transmission.

% 1. 캡션 추가

The LEO satellite serves $K_{\sf s}$ SUs, each equipped with a single pair of co-located vertical and horizontal antennas, indexed by 
${\mathcal{K}_{\sf s}} \triangleq \{1,\cdots,K_{\sf s}\}$.
To alleviate signal interference among SUs by exploiting the orthogonality between right- and left-hand circular polarized signals, we consider that half of the SUs, indexed by ${\mathcal{K}_{{\sf s}, {\sf r}}} \triangleq \{1,\cdots,\frac{K_{\sf s}}{2}\}$, employ RHCP by delaying the received signal at the horizontal antenna by $-\pi/2$ relative to the vertical antenna port. The remaining SUs, indexed by ${\mathcal{K}_{{\sf s}, {\sf l}}} \triangleq \{\frac{K_{\sf s}}{2} + 1,\cdots,K_{\sf s}\}$, employ LHCP by delaying the received signal at the horizontal antenna by $\pi/2$ relative to the vertical antenna port. The BS serves $K_{\sf t}$ CUs, each of which employs either vertical or horizontal polarization, indexed by $\mathcal{K}_{\sf t} \triangleq \{1,\cdots,K_{\sf t}\}$, as in \cite{Park:twc:15}. We assume half of the CUs, indexed by ${\mathcal{K}_{{\sf t}, {\sf v}}} \triangleq \{1,\cdots,\frac{K_{\sf t}}{2}\}$, employ vertical polarization, while the remaining CUs, indexed by ${\mathcal{K}_{{\sf t}, {\sf h}}} \triangleq \{\frac{K_{\sf t}}{2} + 1,\cdots,K_{\sf t}\}$, employ horizontal polarization.
We consider a scenario in which the LEO satellite serves SUs lacking reliable terrestrial access, particularly in sparsely populated or remote areas, as in \cite{Longfei:twc:23,jh:tvt:24,10908461}.\footnote{The SUs with reliable terrestrial access can be associated with the terrestrial BS and served as CUs \cite{10908461}.} Consequently, the SUs are assumed to be located outside the effective service region of the terrestrial BS and thus experience negligible interference from the terrestrial BS.
In contrast, each CU experiences both inter-network interference from the LEO satellite and intra-network interference. Under this scenario, we consider a coordinated ISTN architecture in which the CSI of all direct and interfering links is shared between the satellite and the terrestrial BS, whereas user data is not exchanged.

% 2. active mode를 사용하낟 3. RF chain  갯수 확인해보기 4. 점 곱하기 (같은 의미인데, 다른 표현)

%\vspace{-4mm}

% width=81mm
%\begin{figure}[t]
%    \centering
    %\vspace{-11mm}
%    \vspace{-5mm}
%    \includegraphics[height=70mm]{Figures/System Architecture_rev.png}
%    \captionsetup{width=\columnwidth}
%    \caption{System model of MDP-ISTN.}
%    \vspace{-4mm}
%    \label{fig:System Architecture}
    %\vspace{-4mm}
    %\vspace{-5mm}
%\end{figure}
%\vspace{-3mm}
%\vspace{-3mm}
\subsection{Channel Model for MDP-ISTN}\label{AA}
\subsubsection{Satellite Channel Model}
Since LEO satellite links typically exhibit a dominant line-of-sight (LOS) component, we model the LEO satellite-to-user channels according to a Rician fading model. Accordingly, the $2\times2$ channel matrix between the $(n_{\sf s}^{\sf x},n_{\sf s}^{\sf y})$-th antenna pair of the LEO satellite and the $k_{\sf s}$-th SU, for all $k_{\sf s}\in \mathcal{K}_{\sf s}$, at instant $t$ is given by \cite{3gpp_channel,3GPP_ntn_channel,4138008,1033685}
\begin{align}
\label{SU_ch_2by2_mat}
 \mathbf{F}_{k_{\sf{s}}}^{[n_{\sf s}^{\sf x},n_{\sf s}^{\sf y}]}(t) & = \underbrace{\sqrt{\frac{\kappa_{k_{\sf{s}}}}{\kappa_{k_{\sf{s}}}+1}} \cdot \mathbf{F}_{k_{\sf{s}},0}^{[n_{\sf s}^{\sf x},n_{\sf s}^{\sf y}]}(t)}_{\sf{LOS \,\, component}} \nonumber \\
& +  \underbrace{\sqrt{\frac{1}{\kappa_{k_{\sf{s}}}+1}}\cdot\bigg( \sum_{l=1}^{L_{k_{\sf{s}}}}\mathbf{F}_{k_{\sf{s}},l}^{[n_{\sf s}^{\sf x},n_{\sf s}^{\sf y}]}(t)\bigg)}_{\sf{NLOS \,\, components}} \in \mathbb{C}^{2\times2}. 
\end{align}
The path with index $l=0$ corresponds to the LOS component, while the remaining $L_{k_{\sf s}}$ paths correspond to non-line-of-sight (NLOS) components. 
The Rician $K$-factor, denoted by $\kappa_{k_{\sf s}}$, is defined as the power ratio between the LOS component and the aggregate NLOS components. Accordingly, the LOS component and the $l$-th NLOS component of the satellite channel between the $(n_{\sf s}^{\sf x},n_{\sf s}^{\sf y})$-th antenna pair of the satellite and the $k_{\sf s}$-th SU are given by \cite{3gpp_channel, 3GPP_ntn_channel, 4138008,1033685}
\begin{align}
    \label{SU_ch_2by2_mat_los}
\mathbf{F}_{k_{\sf{s}},0}^{[n_{\sf s}^{\sf x},n_{\sf s}^{\sf y}]}(t) &=  \beta_{{k_{\sf{s}}},0} \cdot \mathbf{R}(\zeta_{k_{\sf{s}}})^{\sf T} \begin{bmatrix} 
        \sqrt{1-\chi_{k_{\sf{s}},0}} \!\!&\!\!  \sqrt{\chi_{k_{\sf{s}},0}} \\
\sqrt{\chi_{k_{\sf{s}},0}} \!\!&\!\!  \sqrt{1-\chi_{k_{\sf{s}},0}}
    \end{bmatrix} \mathbf{R}(\zeta_{\sf F}) \nonumber \\
    & \cdot \exp{(-j2\pi f_{\sf{c}} \tau_{k_{\sf{s}},0}^{\sf ref})}
    \!\cdot\! \exp{(j2\pi f_{{k_{\sf{s}}},0}^{\sf{D}}t)} \cdot \exp{(-j2\pi f_{\sf{c}} \Delta \tau^{[n_{\sf s}^{\sf x},n_{\sf s}^{\sf y}]}(\theta_{{k_{\sf{s}}}, 0}, \varphi_{{k_{\sf{s}}}, 0}))}
\end{align}
and
\begin{align}
    \label{SU_ch_2by2_mat_nlos}
\mathbf{F}_{k_{\sf{s}},l}^{[n_{\sf s}^{\sf x},n_{\sf s}^{\sf y}]}(t) &=  \beta_{{k_{\sf{s}}},l} \cdot \mathbf{R}(\zeta_{k_{\sf{s}}})^{\sf T} \bigg(\begin{bmatrix} 
        \sqrt{1-\chi_{k_{\sf{s}},l}} \!\!&\!\!  \sqrt{\chi_{k_{\sf{s}},l}} \\
\sqrt{\chi_{k_{\sf{s}},l}} \!\!&\!\!  \sqrt{1-\chi_{k_{\sf{s}},l}}
    \end{bmatrix} \odot\begin{bmatrix} 
\exp{(j\Phi_{{k_{\sf{s}}},l}^{\sf{v},\sf{v}})}&  \exp{(j\Phi_{{k_{\sf{s}}},l}^{\sf{v},\sf{h}})} \\
\exp{(j\Phi_{{k_{\sf{s}}},l}^{\sf{h},\sf{v}})} &  \exp{(j\Phi_{{k_{\sf{s}}},l}^{\sf{h},\sf{h}})}
    \end{bmatrix}\bigg) \mathbf{R}(\zeta_{\sf F}) \nonumber \\
    & \cdot \exp{(j2\pi f_{{k_{\sf{s}}},l}^{\sf{D}}t)} \!\cdot\! \exp{(-j2\pi f_{\sf{c}} \Delta \tau^{[n_{\sf s}^{\sf x},n_{\sf s}^{\sf y}]}(\theta_{{k_{\sf{s}}}, l}, \varphi_{{k_{\sf{s}}}, l}))},
\end{align}
where $\sf v$ and $\sf h$ denote vertical and horizontal, respectively.

\begin{figure*}[!b]
%\vspace{-3mm}
\noindent\rule{\textwidth}{.5pt}%\vskip3pt
\small \begin{align}
\label{SU_ch_2by2_mat_2}
&\mathbf{F}_{k_{\sf{s}}}^{[n_{\sf s}^{\sf x},n_{\sf s}^{\sf y}]}(t) =  \mathbf{R}(\zeta_{k_{\sf{s}}})^{\sf T}  \Bigg\{ \sqrt{\frac{\kappa_{k_{\sf{s}}}}{\kappa_{k_{\sf{s}}}+1}} \cdot \bigg( {\beta_{{k_{\sf{s}}},0}} \cdot \exp{(-j2\pi\{f_{\sf{c}} \tau_{k_{\sf{s}},0}^{\sf ref} - (f_{k_{\sf{s}},0}^{\sf{D}})^{\sf{su}}t\})} \cdot   \begin{bmatrix} 
        \sqrt{1-\chi_{k_{\sf{s}},0}} & \!\!\! \sqrt{\chi_{k_{\sf{s}},0}} \\
\sqrt{\chi_{k_{\sf{s}},0}} & \!\!\! \sqrt{1-\chi_{k_{\sf{s}},0}}
    \end{bmatrix} \bigg) \\
    & + \sqrt{\frac{1}{\kappa_{k_{\sf{s}}}+1}}
    \cdot\bigg(\sum_{l=1}^{L_{k_{\sf{s}}}}{\beta_{{k_{\sf{s}}},l}} \cdot \exp{(j2\pi(f_{k_{\sf{s}},l}^{\sf{D}})^{\sf{su}}t)} \cdot \begin{bmatrix} 
        \sqrt{1- \chi_{k_{\sf{s}}, l}}\cdot\exp{(j\Phi_{{k_{\sf{s}}},l}^{\sf{v},\sf{v}})} & \!\!\!\!\!\! \sqrt{\chi_{k_{\sf{s}},l}}\cdot\exp{(j\Phi_{{k_{\sf{s}}},l}^{\sf{v},\sf{h}})}  \\
\sqrt{\chi_{k_{\sf{s}},l}}\cdot\exp{(j\Phi_{{k_{\sf{s}}},l}^{\sf{h},\sf{v}})} & \!\!\!\!\!\! \sqrt{1-\chi_{k_{\sf{s}},l}}\cdot\exp{(j\Phi_{{k_{\sf{s}}},l}^{\sf{h},\sf{h}})} 
    \end{bmatrix} \bigg) \Bigg\} \mathbf{R}(\zeta_{\sf F}) \cdot a^{[n_{\sf s}^{\sf x},n_{\sf s}^{\sf y}]}(\theta_{k_{\sf{s}}}, \varphi_{k_{\sf{s}}}). \nonumber
\end{align}
\end{figure*}
\begin{figure*}[!b]
\vspace{-5mm}
\noindent\rule{\textwidth}{.5pt}%\vskip3pt
\small \begin{align}
\label{SU_ch_2by2_mat_3}
\mathbf{F}_{k_{\sf{s}}}^{[n_{\sf s}^{\sf x},n_{\sf s}^{\sf y}]}(t) & =  \mathbf{R}(\zeta_{k_{\sf{s}}})^{\sf T} \Bigg\{ \sqrt{\frac{\beta_{{k_{\sf{s}}}}}{\kappa_{k_{\sf{s}}}+1}} \cdot \bigg( \sqrt{\kappa_{k_{\sf{s}}}} \cdot \exp{j\phi_{k_{\sf{s}}}(t)} \cdot \begin{bmatrix} 
        \sqrt{1-\chi_{k_{\sf{s}},0}} & \!\!\! \sqrt{\chi_{k_{\sf{s}},0}} \\
\sqrt{\chi_{k_{\sf{s}},0}} & \!\!\! \sqrt{1-\chi_{k_{\sf{s}},0}}
    \end{bmatrix}
      \\
    & + \begin{bmatrix} 
        \sqrt{1-\tilde{\chi}_{k_{\sf{s}}}}\cdot \tilde{g}_{k_{\sf{s}}}^{\sf{v},\sf{v}}(t) & \!\!\!\!\!\! \sqrt{\tilde{\chi}_{k_{\sf{s}}}}\cdot \tilde{g}_{k_{\sf{s}}}^{\sf{v},\sf{h}}(t)  \\
\sqrt{\tilde{\chi}_{k_{\sf{s}}}}\cdot \tilde{g}_{k_{\sf{s}}}^{\sf{h},\sf{v}}(t) & \!\!\!\!\!\! \sqrt{1-\tilde{\chi}_{k_{\sf{s}}}}\cdot \tilde{g}_{k_{\sf{s}}}^{\sf{h},\sf{h}}(t) 
    \end{bmatrix} \bigg) \Bigg\} \mathbf{R}(\zeta_{\sf F}) \cdot a^{[n_{x}, n_{y}]}(\theta_{k_{\sf{s}}}, \varphi_{k_{\sf{s}}}) = \mathbf{G}_{k_{\sf{s}}}(t)\cdot a^{[n_{\sf s}^{\sf x},n_{\sf s}^{\sf y}]}(\theta_{k_{\sf{s}}}, \varphi_{k_{\sf{s}}}). \nonumber
\end{align}
%\noindent\rule{\textwidth}{.5pt}%\vskip3pt
%\vspace{-4.5mm}
\end{figure*}

Herein, $f_{\sf c}$ denotes the carrier frequency, and 
$\beta_{k_{\sf s},l}$ and $f^{\sf D}_{k_{\sf s},l}$ represent the complex gain and Doppler shift of the $l$-th propagation path, respectively, for $l \in \mathcal{L}_{k_{\sf s}} \triangleq \{0,1,\cdots,L_{k_{\sf s}}\}$. 
The matrix $\mathbf{R}(\zeta_{k_{\sf s}}) =
\begin{bmatrix}
\cos{\zeta_{k_{\sf s}}} & -\sin{\zeta_{k_{\sf s}}} \\
\sin{\zeta_{k_{\sf s}}} & \cos{\zeta_{k_{\sf s}}}
\end{bmatrix}$ denotes the rotation matrix accounting for the antenna orientation difference $\zeta_{k_{\sf s}}$ between the LEO satellite and the $k_{\sf s}$-th SU due to the rapid movement of the LEO satellite. Moreover, $\mathbf{R}(\zeta_{\sf F}) =
\begin{bmatrix}
\cos{\zeta_{\sf F}} & -\sin{\zeta_{\sf F}} \\
\sin{\zeta_{\sf F}} & \cos{\zeta_{\sf F}}
\end{bmatrix}$
denotes the Faraday rotation matrix in which the Faraday rotation angle in degrees is given by $\zeta_{\sf F} = \frac{\num{108}}{(f_{\sf c}/10^9)^2}$ \cite{3GPP_ntn_channel}. The parameter $\chi_{k_{\sf s}, l}$ models the effect of channel depolarization, satisfying $0 \le \chi_{k_{\sf s}, l} \le 1$, and quantifies the power ratio between co-polarized and cross-polarized signal components \cite{Milcom:15, Coldrey:vtc:08}. Moreover, $\tau^{\sf ref}_{k_{\sf s},0}$ represents the propagation delay of the LOS path between the $k_{\sf s}$-th SU and the reference antenna pair at the satellite, i.e., the $(1,1)$-th antenna pair. The term $\Delta \tau^{[n_{\sf s}^{\sf x},n_{\sf s}^{\sf y}]}(\theta_{k_{\sf{s}},0}, \varphi_{k_{\sf{s}},0})$ represents the time difference of arrival (TDoA) of the LOS path between the reference antenna pair and the $(n_{\sf s}^{\sf x},n_{\sf s}^{\sf y})$-th antenna pair.
For the NLOS components, $\Phi^{i,j}_{k_{\sf{s}},l}$, $\forall i,j \in \{\sf{v},\sf{h}\}$, represents the random phase for the $l$-th path between the $k_{\sf s}$-th SU and the reference antenna pair at the satellite, associated with the link from the $j$-polarized antenna to the $i$-polarized antenna. 
Herein, $\Phi^{i,j}_{k_{\sf{s}},l}$ follows an independent uniform distribution over $[0,2\pi]$ \cite{3gpp_channel, 4138008}.
$\Delta \tau^{[n_{\sf s}^{\sf x},n_{\sf s}^{\sf y}]}(\theta_{k_{\sf{s}},l}, \varphi_{k_{\sf{s}},l})$ denotes the 
TDoA of the $l$-th path between the reference antenna pair and the $(n_{\sf s}^{\sf x},n_{\sf s}^{\sf y})$-th antenna pair. 
The angle pair $(\theta_{k_{\sf{s}},l}, \varphi_{k_{\sf{s}},l})$ denotes the angle-of-departure (AoD) of the $l$-th path toward the $k_{\sf{s}}$-th SU, for all $l \in \mathcal{L}_{k_{\sf s}}$, where $\theta_{k_{\sf{s}},l}$ and $\varphi_{k_{\sf{s}},l}$ correspond to the azimuth and off-nadir angles, respectively.

Since the LEO satellite operates at a high altitude ($\num{200}-\num{2000}$ km), the AoDs of different propagation paths for a given SU can be regarded as identical, i.e., 
$ \Delta \tau^{[n_{\sf s}^{\sf x},n_{\sf s}^{\sf y}]}(\theta_{k_{\sf{s}},l}, \varphi_{k_{\sf{s}},l}) \approx \Delta \tau^{[n_{\sf s}^{\sf x},n_{\sf s}^{\sf y}]}(\theta_{k_{\sf{s}}}, \varphi_{k_{\sf{s}}}), \forall l \in \mathcal{L}_{k_{\sf s}}$ \cite{you2020massive, li2021downlink, you2022beam, 10787138, Jaehyup:jsac:24}. Moreover, owing to the relative motion of the LEO satellite and SUs, the Doppler shift $f^{\sf{D}}_{k_{\sf{s}},l}$ consists of two independent components due to the mobility of the LEO satellite and the $k_{\sf s}$-th SU, denoted by $(f^{\sf{D}}_{k_{\sf{s}},l})^{\sf{sat}}$ and $(f^{\sf{D}}_{k_{\sf{s}},l})^{\sf{su}}$, respectively. Accordingly, the overall Doppler shift can be expressed as 
$f^{\sf{D}}_{k_{\sf{s}},l} = (f^{\sf{D}}_{k_{\sf{s}},l})^{\sf{sat}} + (f^{\sf{D}}_{k_{\sf{s}},l})^{\sf{su}}$ \cite{you2020massive, li2021downlink, you2022beam, 10787138, Jaehyup:jsac:24}.
Due to the high altitude of the LEO satellite, the satellite-induced Doppler shift can be reasonably assumed to be identical across different propagation paths, i.e., 
$(f^{\sf{D}}_{k_{\sf{s}},l})^{\sf{sat}} = (f^{\sf{D}}_{k_{\sf{s}}})^{\sf{sat}}, \forall l \in \mathcal{L}_{k_{\sf s}}$ \cite{you2020massive, li2021downlink, you2022beam, 10787138, Jaehyup:jsac:24}. 
Furthermore, $(f^{\sf{D}}_{k_{\sf{s}}})^{\sf{sat}}$ can be effectively compensated by a proper frequency synchronization technique at the LEO satellite \cite{you2020massive, li2021downlink, you2022beam, 10787138, Jaehyup:jsac:24}. Thus, the channel matrix (\ref{SU_ch_2by2_mat}) can be rewritten as (\ref{SU_ch_2by2_mat_2}) at the bottom of the previous page, where $a^{[n_{\sf s}^{\sf x},n_{\sf s}^{\sf y}]}(\theta_{k_{\sf{s}}}, \varphi_{k_{\sf{s}}}) \triangleq \exp\big(-j2\pi f_{\sf{c}} \Delta \tau^{[n_{\sf s}^{\sf x},n_{\sf s}^{\sf y}]}(\theta_{k_{\sf{s}}}, \varphi_{k_{\sf{s}}})\big)$.\footnote{Owing to the high altitude of the LEO satellite, the considered satellite-to-user links lie in the far-field regime, such that the antenna orientation difference between the satellite and the $k_{\sf s}$-th SU is identical across all satellite antenna pairs, i.e., $\zeta_{k_{\sf s}}=\zeta_{k_{\sf s},1}=\cdots=\zeta_{k_{\sf s},N_{\sf s}}$. For near-field systems, antenna pair-dependent orientation differences should instead be considered, as in \cite{naeem2025polarization}.}

For notational simplicity, from the NLOS components in (\ref{SU_ch_2by2_mat_2}), we define $g_{k_{\sf{s}}}^{i,i}(t)$ for all $ i\in\{\sf{v},\sf{h}\}$ as
$g_{k_{\sf{s}}}^{i,i}(t)
\triangleq 
\sum_{l=1}^{L_{k_{\sf{s}}}}
\beta_{k_{\sf{s}},l}
\exp\!\big(j2\pi (f_{k_{\sf{s}},l}^{\sf{D}})^{\sf{su}} t\big)\sqrt{1\!-\!\chi_{k_{\sf{s}},l}}
\exp\!\big(j(\Phi_{k_{\sf{s}},l})^{i,i}\big)$. We also define $g_{k_{\sf{s}}}^{i,j}(t)$ for all $i\neq j$ and $i,j\in\{\sf{v},\sf{h}\}$ as
$g_{k_{\sf{s}}}^{i,j}(t)
\triangleq 
\sum_{l=1}^{L_{k_{\sf{s}}}}
\beta_{k_{\sf{s}},l} 
\exp\!\big(j2\pi (f_{k_{\sf{s}},l}^{\sf{D}})^{\sf{su}} t\big)\sqrt{\chi_{k_{\sf{s}},l}}
\exp\!\big(j(\Phi_{k_{\sf{s}},l})^{i,j}\big)$.
Since the statistical properties of the NLOS components of the LEO satellite channel are mainly determined by the local propagation environment surrounding the user terminal, it can be modeled as complex Gaussian random variables \cite{you2020massive, li2021downlink, you2022beam, 10787138, Jaehyup:jsac:24}. Accordingly, $g_{k_{\sf{s}}}^{i,i}(t)$ and $g_{k_{\sf{s}}}^{i,j}(t)$ can be modeled as independent and identically distributed (i.i.d.) complex Gaussian random variables following $ \mathcal{CN}\!\big(0,(1-\tilde{\chi}_{k_{\sf{s}}})\beta_{k_{\sf{s}}}\big)$ and $ \mathcal{CN}\!\big(0,\tilde{\chi}_{k_{\sf{s}}}\beta_{k_{\sf{s}}}\big)$, respectively. 
Herein, $\beta_{k_{\sf{s}}}$ denotes the average power of the channel gain and is set as
$\beta_{k_{\sf{s}}} = \frac{G_{{\sf sat}}^{\sf Tx} G_{k_{\sf{s}}}^{{\sf Rx}}}{k_{\sf B} T_{\sf sys} B}\big(\frac{c}{4\pi f_{\sf c} d_{k_{\sf{s}}}^{\sf sat}}\big)^{2}$, where $G_{\sf sat}^{{\sf Tx}}$, $G_{k_{\sf{s}}}^{{\sf Rx}}$, $c$, $d_{k_{\sf{s}}}^{\sf sat}$, $k_{\sf B}$, $T_{\sf sys}$, and $B$ represent the LEO satellite transmit antenna gain, the $k_{\sf{s}}$-th SU receive antenna gain, the speed of light,  the satellite-to-$k_{\sf{s}}$-th SU distance, the Boltzmann constant, the system noise temperature, and the bandwidth, respectively. Moreover, $\tilde{\chi}_{k_{\sf{s}}}$ denotes the expected value of the channel depolarization effect for the NLOS paths, i.e., $ \tilde{\chi}_{k_{\sf{s}}} = \mathbb{E}[\chi_{k_{\sf{s}},l}]$, $\forall l \in \{1,\cdots,L_{k_{\sf{s}}}\}$.
The relationship between the channel XPD over the NLOS paths and $\tilde{\chi}_{k_{\sf{s}}}$ is given by ${\sf{XPD}}_{k_{\sf{s}}}^{\sf sat} = \frac{1-\tilde{\chi}_{k_{\sf{s}}}}{\tilde{\chi}_{k_{\sf{s}}}}$, while the channel XPD over the LOS path is given by ${\sf XPD}_{k_{\sf{s}},0}^{\sf sat} = \frac{1-\chi_{k_{\sf{s}},0}}{\chi_{k_{\sf{s}},0}}$ \cite{Milcom:15, Coldrey:vtc:08}. 

Accordingly, (\ref{SU_ch_2by2_mat_2}) can be rewritten as (\ref{SU_ch_2by2_mat_3}) at the top of this page, where $\phi_{k_{\sf{s}}}(t) = -2\pi\{f_{\sf c}\tau_{k_{\sf{s}},0}^{\sf{ref}} - (f_{k_{\sf{s}},0}^{\sf D})^{\sf su} t\}$, and $\tilde{g}^{i,j}_{k_{\sf{s}}}(t)$, $\forall i,j\in\{\sf v,\sf h\}$, are modeled as i.i.d. complex Gaussian random variables such that $\tilde{g}^{i,j}_{k_{\sf{s}}}(t)\sim\mathcal{CN}(0,1)$.
In (\ref{SU_ch_2by2_mat_3}), $a^{[n_{\sf s}^{\sf x}, n_{\sf s}^{\sf y}]}(\theta_{k_{\sf{s}}}, \varphi_{k_{\sf{s}}})$ is obtained by calculating the TDoA between the reference antenna pair and the $(n_{\sf s}^{\sf x}, n_{\sf s}^{\sf y})$-th antenna pair as follows \cite{Jaehyup:jsac:24}:
\begin{align}
     a^{[n_{\sf s}^{\sf x}, n_{\sf s}^{\sf y}]}(\theta_{k_{\sf{s}}}, \varphi_{k_{\sf{s}}}) &\triangleq e^{-j2\pi f_{\sf{c}} \Delta \tau^{[n_{\sf s}^{\sf x}, n_{\sf s}^{\sf y}]}(\theta_{k_{\sf{s}}}, \varphi_{k_{\sf{s}}})}
     = e^{-j \pi \sin\varphi_{k_{\sf{s}}}[(n_{\sf s}^{\sf x} \! - \! 1) \sin\theta_{k_{\sf{s}}} + (n_{\sf s}^{\sf y} \! - \! 1) \cos\theta_{k_{\sf{s}}}]}. 
\end{align} 
The array response vector of the $k_{\sf{s}}$-th SU, $\mathbf{a}_{k_{\sf{s}}}$, formed by stacking $a^{[n_{\sf s}^{\sf x},n_{\sf s}^{\sf y}]}(\theta_{k_{\sf{s}}},\varphi_{k_{\sf{s}}})$, can be represented as follows: %using the Kronecker product as follows:
\begin{align}
\label{steering vector}
    \mathbf{a}_{k_{\sf{s}}} = \mathbf{a}_{k_{\sf{s}}}^{\sf x} \otimes \mathbf{a}_{k_{\sf{s}}}^{\sf y} \in \mathbb{C}^{(N_{\sf{s}}^{\sf{x}} N_{\sf{s}}^{\sf{y}})\times1}.
\end{align}
In (\ref{steering vector}), $\mathbf{a}_{k_{\sf{s}}}^{\sf x}\in\mathbb{C}^{N_{\sf{s}}^{\sf x}\times1}$ and $\mathbf{a}_{k_{\sf{s}}}^{\sf y}\in\mathbb{C}^{N_{\sf{s}}^{\sf y}\times1}$ respectively indicate
\begin{align}
&\mathbf{a}_{{k_{\sf{s}}}}^{\sf x}
\triangleq 
\left[1,e^{-j\pi\sin\varphi_{k_{\sf{s}}}\cos\theta_{k_{\sf{s}}}},
\cdots ,  e^{-j\pi\left(N_{\sf{s}}^{\sf x}-1\right)\sin\varphi_{k_{\sf{s}}}\cos\theta_{k_{\sf{s}}}}\right]^{\sf T},
\\
&\mathbf{a}_{k_{\sf{s}}}^{\sf y}
\triangleq  
\left[1,e^{-j\pi\sin\varphi_{k_{\sf{s}}}\sin\theta_{k_{\sf{s}}}},
 \cdots ,  e^{-j\pi\left(N_{\sf{s}}^{\sf y}-1\right)\sin\varphi_{k_{\sf{s}}}\sin\theta_{k_{\sf{s}}}}\right]^{\sf T}.
\end{align}
Therefore, the overall channel matrix between the LEO satellite equipped with $N_{\sf{s}}$ antenna pairs and the $k_{\sf{s}}$-th SU equipped with a single antenna pair is given by
\begin{align}
\label{SU_channel}
    \mathbf{F}_{k_{\sf{s}}}(t) = \mathbf{a}_{k_{\sf{s}}} \otimes \mathbf{G}_{k_{\sf{s}}}(t)\in \mathbb{C}^{2N_{\sf{s}}\times2}.
\end{align}
For notational simplicity, we rewrite (\ref{SU_channel}) as $\mathbf{F}_{k_{\sf{s}}} = \mathbf{a}_{k_{\sf{s}}} \otimes \mathbf{G}_{k_{\sf{s}}}$ by dropping the time index $t$, since we focus on the system within a certain fading block.

Among the channel parameters constituting $\mathbf{F}_{k_{\sf s}}$, we classify the channel parameters into three categories: satellite-to-SU geometry-related parameters (i.e., $\theta_{k_{\sf s}}$, $\varphi_{k_{\sf s}}$, $\mathbf{R}(\zeta_{k_{\sf s}})$, and $\phi_{k_{\sf s}}$), long-term statistical channel parameters (i.e., $\beta_{k_{\sf s}}$, $\kappa_{k_{\sf s}}$, ${\sf XPD}_{k_{\sf s},0}^{\sf sat}$, and ${\sf XPD}_{k_{\sf s}}^{\sf sat}$), and small-scale fading coefficients $\tilde{g}^{i,j}_{k_{\sf s}}$, $\forall i,j\in \{{\sf v},{\sf h}\}$. 
The small-scale fading coefficients are governed by the randomly varying local scattering environment surrounding each SU. Consequently, frequent feedback of the instantaneous CSI incurs substantial signaling overhead due to the rapidly varying small-scale coefficients \cite{you2020massive,li2021downlink,you2022beam,10787138}. Moreover, owing to the high velocity of the LEO satellite and the long propagation delay, the CSI available at the LEO satellite can become outdated before data transmission \cite{you2020massive,li2021downlink,you2022beam,10787138}.
In contrast, the geometry-related and long-term statistical channel parameters vary on a much slower time scale and can therefore be assumed to remain approximately constant over a relatively long time interval \cite{you2020massive,li2021downlink,you2022beam,10787138}. Accordingly, following \cite{you2020massive,li2021downlink,you2022beam,10787138}, we consider a system in which only these slowly varying parameters are periodically updated at the satellite, whereas the small-scale fading coefficients are unavailable.

%These parameters are primarily determined by the relative position and motion between the LEO satellite and each SU, whereas the small-scale fading coefficient $\tilde{g}^{i,j}_{k_{\sf s}}(t)$, $\forall i,j \in \{\sf v,\sf h\}$, is governed by randomly varying local scatterers around the SUs. Due to the severe propagation delay and short channel coherence time in LEO SATCOM, acquiring instantaneous CSIT is generally infeasible. Instead, we consider that only the geometry-related parameters and statistical information of the channel, which vary much more slowly than the small-scale fading coefficients, are available at the LEO satellite, as in \cite{you2020massive, li2021downlink, you2022beam, 10787138, Jaehyup:jsac:24}.

% 3. 9번, 10번 식 추가

\begin{figure}[!t]
\centering
 \includegraphics[width=0.6\linewidth]{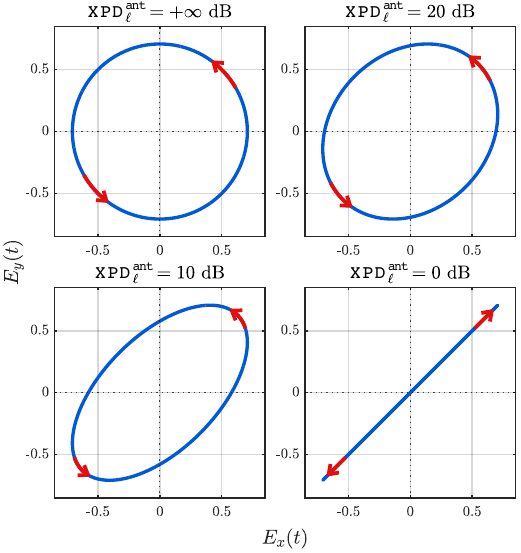}
             %\vspace{-3mm}
 		\caption{Polarization trajectories of $\hat{\boldsymbol{\rho}}_{\ell}^{\sf{r}}$ under different antenna XPD values, where $E_x(t)$ and $E_y(t)$ denote the electric-field components. The red arrows indicate the temporal evolution of the electric-field vector.}
    	\label{RHCP_XPD_ant}%\vspace{-5mm}
\end{figure}

From (\ref{SU_channel}), the effective channel vector between the LEO satellite and the SUs can be constructed using the unit-norm polarization vectors for RHCP and LHCP, respectively, defined as
$\boldsymbol{\rho}_{\sf r} \triangleq \frac{1}{\sqrt{2}}[1,\ -j]^{\sf T}$ and 
$\boldsymbol{\rho}_{\sf l} \triangleq \frac{1}{\sqrt{2}}[1,\ j]^{\sf T}$. These vectors represent the ideal RHCP and LHCP states. In practical antennas, however, polarization purity can be reduced in off-boresight directions \cite{9164895,11393662}. Thus, when the propagation direction deviates from the boresight of the transmit or receive antenna, additional cross-polarized leakage can arise. 
%Although antenna designs employing parasitic elements have been investigated to suppress resulting polarization leakage \cite{yang2018low}, residual leakage can remain owing to the non-ideal characteristics of practical antennas. 
Although advanced antenna design techniques have been investigated to suppress resulting polarization leakage \cite{11393662}, the residual leakage can remain owing to the non-ideal characteristics of practical antennas.
Accordingly, we incorporate the antenna depolarization effect into the RHCP and LHCP polarization vectors using the antenna XPD matrix \cite{castellanos2023linear,lee2026integrated}, as
\begin{align}
\label{eff_rhcp}
       \hat{\boldsymbol{\rho}}_{\ell}^{\sf{r}} =  \mathbf{V}_{\ell} \boldsymbol{\rho}_{\sf{r}} = 
    \begin{bmatrix} 
\sqrt{1-\chi_{\ell}^{\sf ant}}&  \sqrt{\chi_{\ell}^{\sf ant}} \\
\sqrt{\chi_{\ell}^{\sf ant}} &  \sqrt{1-\chi_{\ell}^{\sf ant}}
    \end{bmatrix} \boldsymbol{\rho}_{\sf{r}},
\end{align}
for all ${\ell} \in \mathcal{K}_{{\sf s},{\sf r}} \cup \{ {\sf sat} \}$, and
\begin{align}
\label{eff_lhcp}
       \hat{\boldsymbol{\rho}}_{\ell}^{\sf{l}} =  \mathbf{V}_{\ell} \boldsymbol{\rho}_{\sf{l}} =
    \begin{bmatrix} 
\sqrt{1-\chi_{\ell}^{\sf ant}}&  \sqrt{\chi_{\ell}^{\sf ant}} \\
\sqrt{\chi_{\ell}^{\sf ant}} &  \sqrt{1-\chi_{\ell}^{\sf ant}}
    \end{bmatrix} \boldsymbol{\rho}_{\sf{l}},  
\end{align}
for all ${\ell} \in \mathcal{K}_{{\sf s},{\sf l}} \cup \{ {\sf sat} \}$. The parameter $\chi_{\ell}^{\sf ant}$ models the effect of antenna depolarization, satisfying $0 \le \chi_{\ell}^{\sf ant} \le 1$, where the relationship between the antenna XPD and $\chi_{\ell}^{\sf ant}$ is given by ${\sf{XPD}}_{\ell}^{\sf ant} = \frac{1-{\chi}_{\ell}^{\sf ant}}{{\chi}_{\ell}^{\sf ant}}$.
Fig.~\ref{RHCP_XPD_ant} illustrates the impact of antenna XPD on the polarization purity of
$\hat{\boldsymbol{\rho}}_{\ell}^{\sf r}$ in (\ref{eff_rhcp}), which can be decomposed as
$\hat{\boldsymbol{\rho}}_{\ell}^{\sf r}
=\mathbf{V}_{\ell}\boldsymbol{\rho}_{\sf r}
=\sqrt{1-\chi_{\ell}^{\sf ant}}\boldsymbol{\rho}_{\sf r}
-j\sqrt{\chi_{\ell}^{\sf ant}}\boldsymbol{\rho}_{\sf l}$.
When ${\sf XPD}_{\ell}^{\sf ant}=+\infty$ dB, equivalently $\chi_{\ell}^{\sf ant}=0$, the ideal RHCP state is preserved without cross-polarization leakage. For a finite positive value of the antenna XPD, the resulting vector comprises an RHCP component and a cross-polarized LHCP component with magnitudes $\sqrt{1-\chi_{\ell}^{\sf ant}}$ and $\sqrt{\chi_{\ell}^{\sf ant}}$, respectively, thereby reducing the RHCP purity and producing an elliptical polarization trajectory. As ${\sf XPD}_{\ell}^{\sf ant}$ decreases, the LHCP leakage and associated polarization distortion become more pronounced. At ${\sf XPD}_{\ell}^{\sf ant}=0$ dB, corresponding to
$\chi_{\ell}^{\sf ant}=\frac{1}{2}$, the RHCP and LHCP components have equal magnitudes, yielding linear polarization. Then, the effective channel vector between the LEO satellite and the $k_{{\sf{s}},{\sf{r}}}$-th SU employing RHCP, for all $ k_{{\sf{s}},{\sf{r}}} \in \mathcal{K}_{{\sf s},{\sf r}}$, can be formulated as 
\begin{align}
\label{RP_channel}
    \mathbf{f}_{k_{{\sf{s}},{\sf{r}}}} = {\sf vec}\left(\left((\mathbf{I}_{N_{\sf{s}}} \otimes (\hat{\boldsymbol{\rho}}_{k_{{\sf{s}},{\sf{r}}}}^{\sf{r}})^{\sf H}) \mathbf{F}_{k_{{\sf{s}},{\sf{r}}}} [\hat{\boldsymbol{\rho}}_{{\sf{sat}}}^{\sf{r}}, \hat{\boldsymbol{\rho}}_{{\sf{sat}}}^{\sf{l}}]\right)^{\sf{T}}\right),
\end{align}
whose size is $2N_{\sf{s}}\times1$. The derivation procedure of the equation (\ref{RP_channel}) is provided in Section~\ref{Appendix}. Moreover, the effective channel vector between the LEO satellite and the $k_{{\sf s},{\sf l}}$-th SU employing LHCP, for all $k_{{\sf s},{\sf l}} \in \mathcal{K}_{{\sf s},{\sf l}}$, is expressed as
\begin{align}
\label{LP_channel}
    \mathbf{f}_{k_{{\sf{s}},{\sf{l}}}} = {\sf vec}\left(\left((\mathbf{I}_{N_{\sf{s}}} \otimes (\hat{\boldsymbol{\rho}}_{k_{{\sf{s}},{\sf{l}}}}^{\sf{l}})^{\sf H}) \mathbf{F}_{k_{{\sf{s}},{\sf{l}}}} [\hat{\boldsymbol{\rho}}_{{\sf{sat}}}^{\sf{r}}, \hat{\boldsymbol{\rho}}_{{\sf{sat}}}^{\sf{l}}]\right)^{\sf{T}}\right),
\end{align}
whose size is $2N_{\sf{s}}\times1$.

The interference channel matrix between the LEO satellite equipped with $N_{\sf{s}}$ antenna pairs and the $k_{\sf{t}}$-th CU equipped with a single antenna pair is given by
\begin{align}
\label{CU_intf_channel}
    \mathbf{Z}_{{k_{\sf{t}}}} = \mathbf{a}_{{k_{\sf{t}}}} \otimes \mathbf{G}_{{k_{\sf{t}}}}\in \mathbb{C}^{2N_{\sf{s}}\times2}.
\end{align}
Analogously, for the channel in \eqref{CU_intf_channel}, we consider that only the satellite-to-CU geometry-related parameters and long-term statistical channel parameters are periodically updated at the LEO satellite, whereas the small-scale fading coefficients $\tilde{g}^{i,j}_{k_{\sf t}}$, $\forall i,j\in\{{\sf v},{\sf h}\}$, are unavailable at the LEO satellite. 

Then, from (\ref{CU_intf_channel}), the effective interference channel vector between the LEO satellite and the $k_{{\sf t},{\sf v}}$-th CU, employing VP, for all $k_{{\sf t},{\sf v}} \in \mathcal{K}_{{\sf t},{\sf v}}$, is expressed as
\begin{align}
\label{VP_channel}
    \mathbf{z}_{k_{{\sf{t}},{\sf{v}}}} = {\sf vec}\left(\left((\mathbf{I}_{N_{\sf{s}}} \otimes (\hat{\boldsymbol{\rho}}_{k_{{\sf{t}},{\sf{v}}}}^{\sf{v}})^{\sf H}) \mathbf{Z}_{k_{{\sf{t}},{\sf{v}}}} [\hat{\boldsymbol{\rho}}_{{\sf{sat}}}^{\sf{r}}, \hat{\boldsymbol{\rho}}_{{\sf{sat}}}^{\sf{l}}]\right)^{\sf{T}}\right),
\end{align}
whose size is $2N_{\sf{s}}\times1$.
Moreover, the effective interference channel vector between the LEO satellite and the $k_{{\sf t},{\sf h}}$-th CU, employing HP, for all $k_{{\sf t},{\sf h}} \in \mathcal{K}_{{\sf t},{\sf h}}$, is represented as
\begin{align}
\label{HP_channel}
    \mathbf{z}_{k_{{\sf{t}},{\sf{h}}}} = {\sf vec}\left(\left((\mathbf{I}_{N_{\sf{s}}} \otimes (\hat{\boldsymbol{\rho}}_{k_{{\sf{t}},{\sf{h}}}}^{\sf{h}})^{\sf H}) \mathbf{Z}_{k_{{\sf{t}},{\sf{h}}}} [\hat{\boldsymbol{\rho}}_{{\sf{sat}}}^{\sf{r}}, \hat{\boldsymbol{\rho}}_{{\sf{sat}}}^{\sf{l}}]\right)^{\sf{T}}\right),
\end{align}
whose size is $2N_{\sf{s}}\times1$. In (\ref{VP_channel}), the effective polarization vector of the $k_{{\sf t},{\sf v}}$-th CU, employing VP, is given by $\hat{\boldsymbol{\rho}}_{k_{{\sf t},{\sf v}}}^{\sf v}=\mathbf{V}_{k_{{\sf t},{\sf v}}}\boldsymbol{\rho}_{\sf v}$, for all $k_{{\sf t},{\sf v}}\in\mathcal{K}_{{\sf t},{\sf v}}$, where $\boldsymbol{\rho}_{\sf v}\triangleq[1,0]^{\sf T}$ is the ideal unit-norm polarization vector for VP. Similarly, in (\ref{HP_channel}), the effective polarization vector of the $k_{{\sf t},{\sf h}}$-th CU, employing HP, is given by $\hat{\boldsymbol{\rho}}_{k_{{\sf t},{\sf h}}}^{\sf h}=\mathbf{V}_{k_{{\sf t},{\sf h}}}\boldsymbol{\rho}_{\sf h}$, for all $k_{{\sf t},{\sf h}}\in\mathcal{K}_{{\sf t},{\sf h}}$, where $\boldsymbol{\rho}_{\sf h}\triangleq[0,1]^{\sf T}$ is the ideal unit-norm polarization vector for HP.

%In equation (\ref{VP_channel}), $\hat{\boldsymbol{\rho}}_{k_{{\sf{t}},{\sf{v}}}}^{\sf{v}}$, for all $k_{{\sf t},{\sf v}} \in \mathcal{K}_{\sf v}$, denotes $\hat{\boldsymbol{\rho}}_{k_{{\sf{t}},{\sf{v}}}}^{\sf{v}} = \mathbf{V}_{k_{{\sf{t}},{\sf{v}}}}\boldsymbol{\rho}_{\sf v}$, where $\boldsymbol{\rho}_{\sf v}$ denotes the unit-norm polarization vector for vertical polarization, defined as  $\boldsymbol{\rho}_{\sf v} \triangleq [1, 0]^{\sf T}$. Moreover, in equation (\ref{HP_channel}), $\hat{\boldsymbol{\rho}}_{k_{{\sf{t}},{\sf{h}}}}^{\sf{h}}$, for all $k_{{\sf t},{\sf h}} \in \mathcal{K}_{\sf h}$, denotes $\hat{\boldsymbol{\rho}}_{k_{{\sf{t}},{\sf{h}}}}^{\sf{h}} = \mathbf{V}_{k_{{\sf{t}},{\sf{h}}}}\boldsymbol{\rho}_{\sf h}$, where $\boldsymbol{\rho}_{\sf h}$ denotes the unit-norm polarization vector for horizontal polarization, defined as  $\boldsymbol{\rho}_{\sf h} \triangleq [0, 1]^{\sf T}$.

%Herein, $\boldsymbol{\rho}_{\sf h}$ denotes the unit-norm polarization vector for horizontal polarization, defined as 
%$\boldsymbol{\rho}_{\sf h} \triangleq [0, 1]^{\sf T}$. %In equations (\ref{VP_channel}) and (\ref{HP_channel}), for all ${k_{\sf{t}}} \in \mathcal{K}_{\sf t}$, the complex matrix $\mathbf{Z}_{{k_{\sf{t}}}}$ is given by

\subsubsection{Terrestrial Channel Model}

For the terrestrial networks, we consider a dense urban environment, where direct propagation paths are blocked by surrounding buildings and urban clutter. Since the resulting terrestrial channels are dominated by diffuse multipath components, we model the terrestrial BS-to-user channels according to a Rayleigh fading model, as in \cite{Coldrey:vtc:08}. Therefore, for all ${k_{\sf{t}}} \in \mathcal{K}_{\sf t}$, the complex matrix between the terrestrial BS equipped with $N_{\sf t}$ antenna pairs and the $k_{\sf t}$-th CU equipped with single antenna pair is given by
\begin{align}
\label{CU_channel}
    \mathbf{H}_{{k_{\sf{t}}}} = \left(\mathbf{1}_{N_{\sf t}\times1}\otimes\begin{bmatrix} 
        \sqrt{1-\tilde{\chi}_{k_{\sf{t}}}} \!\!\!\!\!\!&  \sqrt{\tilde{\chi}_{k_{\sf{t}}}} \\
\sqrt{\tilde{\chi}_{k_{\sf{t}}}} \!\!\!\!\!\! &  \sqrt{1-\tilde{\chi}_{k_{\sf{t}}}}
    \end{bmatrix}\right)\odot \tilde{\mathbf{H}}_{{k_{\sf{t}}}},
\end{align}
whose size is $2N_{\sf t}\times2$.
Herein, the matrix $\tilde{\mathbf{H}}_{k_{\sf t}} \in \mathbb{C}^{2N_{\sf t} \times 2}$ 
has i.i.d. complex Gaussian entries following $\mathcal{CN}\!\big(0,\frac{G_{\sf bs}^{\sf Tx} G_{k_{\sf{t}}}^{\sf Rx}}{k_{\sf B} T_{\sf sys} B}
\big(\frac{c}{4\pi f_{\sf c}}\big)^{2} \big(\frac{1}{d_{k_{\sf{t}}}^{\sf bs}}\big)^{\eta}\big)$. Herein, $G_{\sf bs}^{\sf Tx}$, $G_{k_{\sf t}}^{\sf Rx}$, $d_{k_{\sf t}}^{\sf bs}$, and $\eta$ denote the terrestrial BS transmit antenna gain, the $k_{\sf t}$-th CU receive antenna gain, the BS-to-$k_{\sf t}$-th CU distance, and the path-loss exponent. Moreover, $\tilde{\chi}_{k_{\sf t}}$ is related to the terrestrial channel XPD defined as 
${\sf XPD}_{k_{\sf t}}^{\sf bs} = \frac{1 - \tilde{\chi}_{k_{\sf t}}}{\tilde{\chi}_{k_{\sf t}}}$ \cite{Coldrey:vtc:08}. 

Then, from (\ref{CU_channel}), the effective channel vector between the terrestrial BS and the $k_{{\sf{t}},{\sf{v}}}$-th CU, employing VP, for all $k_{{\sf{t}},{\sf{v}}}\in \mathcal{K}_{{\sf t},{\sf v}}$, is formulated as follows:
\begin{align}
\label{VP_TN_channel}
    \mathbf{h}_{k_{{\sf{t}},{\sf{v}}}} = {\sf vec}\left(\left((\mathbf{I}_{N_{\sf{t}}} \otimes (\hat{\boldsymbol{\rho}}_{k_{{\sf{t}},{\sf{v}}}}^{\sf{v}})^{\sf H}) \mathbf{H}_{k_{{\sf{t}},{\sf{v}}}} [\hat{\boldsymbol{\rho}}_{{\sf{bs}}}^{\sf{v}}, \hat{\boldsymbol{\rho}}_{{\sf{bs}}}^{\sf{h}}]\right)^{\sf{T}}\right),
\end{align}
whose size is $2N_{\sf{t}}\times1$.
Further, the effective channel vector between the BS and the $k_{{\sf{t}},{\sf{h}}}$-th CU, employing HP, for all $k_{{\sf{t}},{\sf{h}}}\in \mathcal{K}_{{\sf t},{\sf h}}$, is expressed as
\begin{align}
\label{HP_TN_channel} 
    \mathbf{h}_{k_{{\sf{t}},{\sf{h}}}} = {\sf vec}\left(\left((\mathbf{I}_{N_{\sf{t}}} \otimes (\hat{\boldsymbol{\rho}}_{k_{{\sf{t}},{\sf{h}}}}^{\sf{h}})^{\sf H}) \mathbf{H}_{k_{{\sf{t}},{\sf{h}}}} [\hat{\boldsymbol{\rho}}_{{\sf{bs}}}^{\sf{v}}, \hat{\boldsymbol{\rho}}_{{\sf{bs}}}^{\sf{h}}]\right)^{\sf{T}}\right),
\end{align}
whose size is $2N_{\sf{t}}\times1$.
In (\ref{VP_TN_channel}) and (\ref{HP_TN_channel}), the effective polarization vectors of the terrestrial BS are given by
$\hat{\boldsymbol{\rho}}_{\sf bs}^{\sf v}=\mathbf{V}_{\sf bs}\boldsymbol{\rho}_{\sf v}$ and $\hat{\boldsymbol{\rho}}_{\sf bs}^{\sf h}=\mathbf{V}_{\sf bs}\boldsymbol{\rho}_{\sf h}$.
In the terrestrial networks, we assume that the antenna orientations of the BS and the CUs are perfectly aligned (i.e., no polarization mismatch), and that perfect CSIT for the CUs is available at the BS.
\begin{remark}{\rm\textbf{(Acquisition of Slowly Varying Channel Parameters at LEO Satellite)}:}
\rm{The geometry-related parameters can be obtained from the position, velocity, and attitude information of the LEO satellite and the users. The satellite obtains its position and velocity from ephemeris data and its attitude from an onboard attitude determination system. Each user obtains its position and velocity using a global navigation satellite system \cite{10787138} and measures its attitude using an inertial measurement unit \cite{rezaie2022deep}. The user then periodically reports this information to the satellite. Based on the collected information, the LEO satellite determines, for each satellite-to-user link, the AoDs, the relative antenna orientation, and the propagation delay and Doppler shift of the LOS path. The long-term statistical channel parameters can be estimated from downlink pilot observations accumulated over multiple blocks. Specifically, each user estimates the average channel gain, Rician K-factor, and LOS and NLOS XPDs from the empirical statistics of the observed co- and cross-polarized channel coefficients. These statistical estimates are periodically reported to the satellite.}
\end{remark}
%\vspace{-3mm}
\begin{remark}{\rm\textbf{(Distinct Interference Structure of Mixed Dual-Polarized Networks)}:}
\rm{Existing approaches account for intra-/inter-network interference arising from radio resource sharing without considering polarization-related effects, which attenuate the desired polarization component and cause interference between distinct polarization states within and across networks \cite{Longfei:twc:23,jh:tvt:24,10908461,xu2024distributed,zhang2024interference,zhang2025metars}. Meanwhile, the conventional dual-polarized RSMA precoder designs account for co-/cross-polar interference in networks with a homogeneous pair of orthogonal polarization states \cite{Sena:gc:22,Sena:wcl:22,Sena:twc:23,sun2025dual}. In contrast, the considered mixed dual-polarized network comprises heterogeneous pairs of orthogonal polarization states across its sub-networks. Within each sub-network, power leakage between two polarization states causes cross-polar interference in addition to co-polar interference. Across the sub-networks, neither RHCP nor LHCP is orthogonal to VP or HP; thus, each CP state can interfere with both LP states, and vice versa. Consequently, the MDP-ISTN experiences co-polar interference among signals sharing the same polarization state, cross-polar interference within each orthogonal polarization pair, and interference between polarization states belonging to heterogeneous orthogonal pairs.}
\end{remark}

\subsection{Signal Model for MDP-RSMA in MDP-ISTN}
\begin{figure}[!t]
\centering
 \includegraphics[width=1\linewidth]{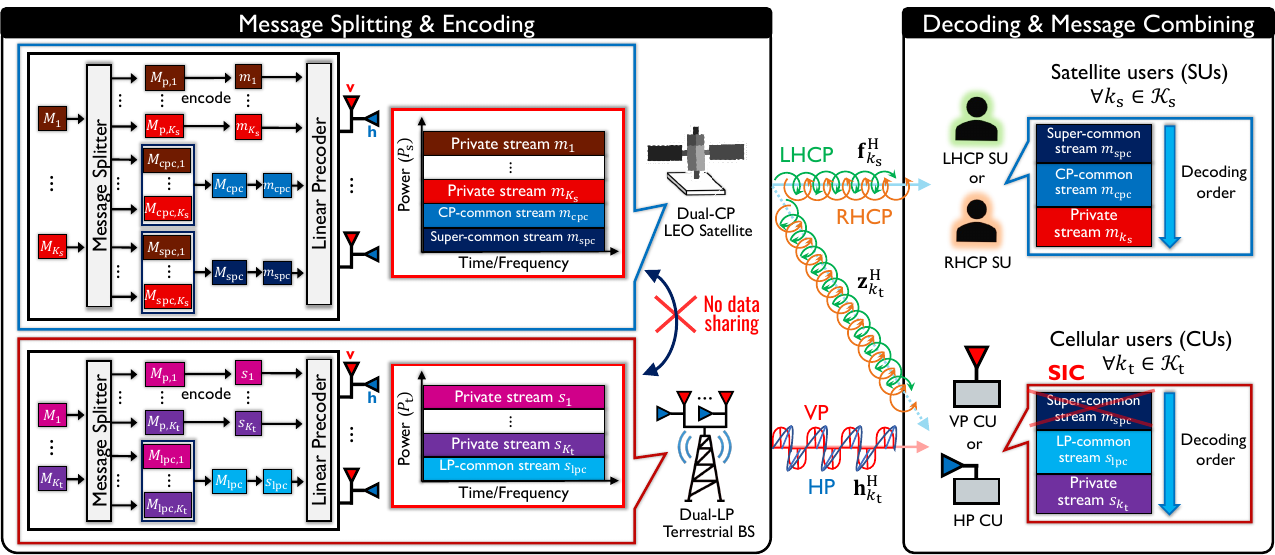}
             %\vspace{-5mm}
 		\caption{Message splitting architecture of the proposed MDP-RSMA scheme.}
    	\label{Fig_2}%\vspace{-5mm}
\end{figure}

%\vspace{-1mm}
To jointly manage the intra-/inter-network interference arising in the MDP-ISTN without requiring inter-network user data sharing, we propose the MDP-RSMA framework illustrated in Fig.~\ref{Fig_2}. The satellite messages intended for SUs, denoted as $\{M_{1}, \cdots, M_{K_{\sf s}}\}$, are split into super-common, CP-common, and private parts as $M_{k_{\sf s}} \rightarrow \{M_{{\sf{spc}}, k_{\sf s}}, M_{{\sf{cpc}}, k_{\sf s}}, M_{{\sf{p}}, k_{\sf s}} \}, \forall k_{\sf s} \in \mathcal{K}_{\sf s}$. All super-common messages $\{M_{{\sf{spc}},1}, \cdots, M_{{\sf{spc}},K_{\sf s}}\}$ are aggregated into a single super-common message $M_{\sf{spc}}$, which is encoded into a super-common stream $m_{\sf{spc}}$ to be decodable by all SUs and CUs. Moreover, all CP-common messages $\{M_{{\sf{cpc}},1}, \cdots, M_{{\sf{cpc}},K_{\sf s}}\}$ are aggregated into a single CP-common message $M_{\sf{cpc}}$, which is encoded into a CP-common stream $m_{\sf{cpc}}$ to be decodable by all SUs. On the other hand, each SU’s private message is encoded into a private stream $m_{k_{\sf s}}$ for all $k_{\sf s} \in \mathcal{K}_{\sf s}$, which is decodable only by the corresponding SU.
The messages of the BS intended for CUs, denoted as $\{M_{1}, \cdots, M_{K_{\sf t}}\}$, are split into LP-common and private parts as $M_{k_{\sf t}} \rightarrow \{M_{{\sf{lpc}}, k_{\sf t}}, M_{{\sf{p}}, k_{\sf t}} \}, \forall k_{\sf t} \in \mathcal{K}_{\sf t}$. All LP-common messages $\{M_{{\sf{lpc}},1}, \cdots, M_{{\sf{lpc}},K_{\sf t}}\}$ are aggregated into a single LP-common message $M_{\sf{lpc}}$, which is encoded into an LP-common stream $s_{\sf{lpc}}$ to be decodable by all CUs. Each CU’s private message is encoded into a private stream $s_{k_{\sf t}}$, for all $k_{\sf t} \in \mathcal{K}_{\sf t}$, to be decodable only by the corresponding CU. 
By doing so, the super-common stream facilitates the suppression of inter-network interference from the dual-CP LEO satellite sub-network to the dual-LP terrestrial sub-network.
Meanwhile, the CP-common and LP-common streams serve to effectively mitigate co-/cross-polar interference within each sub-network, respectively. The super-common, CP-common, LP-common, and private streams are drawn from an i.i.d. complex Gaussian distribution such that $m_{\sf{spc}}, m_{\sf{cpc}}, m_{k_{\sf s}}, s_{\sf{lpc}}, s_{k_{\sf t}} \sim \mathcal{CN}{(0,1)}$.

At the LEO satellite, using the precoding matrix $\mathbf{W} =[\mathbf{w}_{\sf{spc}},\mathbf{w}_{\sf{cpc}},\mathbf{w}_{1},\cdots,\mathbf{w}_{K_{\sf s}}]\in\mathbb{C}^{2N_{\sf s}\times(K_{\sf s}+2)}$, consisting of the super-common precoding vector $\mathbf{w}_{\sf{spc}}\in \mathbb{C}^{{2N_{\sf{s}}}\times{1}}$, CP-common precoding vector $\mathbf{w}_{\sf{cpc}}\in \mathbb{C}^{{2N_{\sf{s}}}\times{1}}$, and private precoding vectors $\mathbf{w}_{k_{\sf s}} \in \mathbb{C}^{2N_{\sf{s}}\times{1}}$, $\forall k_{\sf s} \in \mathcal{K}_{\sf s}$, a stream vector $\mathbf{m}=[m_{\sf{spc}},m_{\sf{cpc}},m_{1},\cdots,m_{K_{\sf s}}]^{\sf T}\in\mathbb{C}^{(K_{\sf s}+2)\times1}$ is linearly combined. Then, the LEO satellite transmits the signal 
\begin{align}
\mathbf{x}_{\sf sat} = \mathbf{W}\mathbf{m} = \mathbf{w}_{\sf spc}m_{\sf spc} + \mathbf{w}_{\sf cpc}m_{\sf cpc} + \sum_{j=1}^{K_{\sf s}}\mathbf{w}_{j}m_{j}.
\end{align}
At the terrestrial BS, using the precoding matrix $\mathbf{P} =[\mathbf{p}_{\sf{lpc}},\mathbf{p}_{1},\cdots,\mathbf{p}_{K_{\sf t}}]\in\mathbb{C}^{2N_{\sf t}\times(K_{\sf t}+1)}$, consisting of the LP-common precoding vector $\mathbf{p}_{\sf{lpc}}\in \mathbb{C}^{{2N_{\sf{t}}}\times{1}}$ and private precoding vectors $\mathbf{p}_{k_{\sf t}} \in \mathbb{C}^{2N_{\sf{t}}\times{1}}$, $\forall k_{\sf t} \in \mathcal{K}_{\sf t}$, a stream vector $\mathbf{s}=[s_{\sf{lpc}},s_{1},\cdots,s_{K_{\sf t}}]^{\sf T}\in\mathbb{C}^{(K_{\sf t}+1)\times1}$ is linearly combined. Subsequently, the terrestrial BS transmits the signal
\begin{align}
\mathbf{x}_{\sf bs} = \mathbf{P}\mathbf{s} = \mathbf{p}_{\sf lpc}s_{\sf lpc} + \sum_{j=1}^{K_{\sf t}} \mathbf{p}_{j}s_{j}.
\end{align}

Therefore, the received signals at the $k_{\sf s}$-th SU and the $k_{\sf t}$-th CU are expressed as follows:  
\begin{align}
 &y_{k_{\sf s}} = \mathbf{f}_{k_{\sf s}}^{\sf H}\mathbf{W}\mathbf{m} + {n}_{k_{\sf s}}, \,\, y_{k_{\sf t}} = \mathbf{h}_{k_{\sf t}}^{\sf H}\mathbf{P}\mathbf{s} + \mathbf{z}_{k_{\sf t}}^{\sf H}\mathbf{Wm} + {n}_{k_{\sf t}},
\end{align}
where ${n}_{k_{\sf s}}$ and ${n}_{k_{\sf t}}$ denote the additive white Gaussian noise (AWGN) at the $k_{\sf s}$-th SU and the $k_{\sf t}$-th CU, respectively. Herein, ${n}_{k_{\sf s}}$ and ${n}_{k_{\sf t}}$ are i.i.d. such that $n_{k_{\sf s}}\sim\mathcal{CN}(0,\sigma_{{\sf{n}},k_{\sf s}}^2)$, $\forall k_{\sf s} \in \mathcal{K}_{\sf s}$, and $n_{k_{\sf t}}\sim\mathcal{CN}(0,\sigma_{{\sf{n}},k_{\sf t}}^2)$, $\forall k_{\sf t} \in \mathcal{K}_{\sf t}$. 
For both SUs and CUs, the super-common stream is first decoded while treating the remaining signals as noise. Accordingly, the instantaneous spectral efficiencies for the super-common stream at the $k_{\sf s}$-th SU and the $k_{\sf t}$-th CU are formulated as
\begin{align}
\label{Rate_spc_SU}
{R}_{{\sf{spc}},k_{\sf s}} \!\!=\! \log_{2}\! \Bigg( \! 1+ \frac{\vert{\mathbf{f}}_{k_{\sf s}}^{\sf H}\mathbf{w}_{\sf{spc}}\vert^{2}}{\vert{\mathbf{f}}_{k_{\sf s}}^{\sf H}\mathbf{w}_{\sf{cpc}}\vert^{2} \!+\! \sum_{j=1}^{K_{\sf s}}\vert{\mathbf{f}}_{k_{\sf s}}^{\sf H} \mathbf{w}_{j}\vert^{2} \!+\! \sigma_{{\sf n},k_{\sf s}}^{2}} \!\Bigg)\!,
%\\[0.6mm]
\end{align}
\begin{align}
\label{Rate_spc_CU}
{R}_{{\sf{spc}},k_{\sf t}} \!\!=\! \log_{2}\! \Bigg( \! 1+\! \frac{\vert{\mathbf{z}}_{k_{\sf t}}^{\sf H}\mathbf{w}_{\sf{spc}}\vert^{2}}{\vert{\mathbf{z}}_{k_{\sf t}}^{\sf H}\mathbf{w}_{\sf{cpc}}\vert^{2} \!+\!\! \sum_{j=1}^{K_{\sf s}}\!\vert{\mathbf{z}}_{k_{\sf t}}^{\sf H} \mathbf{w}_{j}\vert^{2} \!+\!\! I_{{\sf{spc}},k_{\sf t}}} \!\Bigg)\!, 
\end{align}
where $I_{{\sf{spc}},k_{\sf t}} \triangleq \vert{\mathbf{h}}_{k_{\sf t}}^{\sf H}\mathbf{p}_{\sf{lpc}}\vert^{2} + \sum_{j=1}^{K_{\sf t}}\vert{\mathbf{h}}_{k_{\sf t}}^{\sf H} \mathbf{p}_{j}\vert^{2} + \sigma_{{\sf n},k_{\sf{t}}}^{2}$.
To design a robust MDP-RSMA precoder under unknown small-scale fading coefficients at the LEO satellite, we define the ergodic super-common spectral efficiencies at the $k_{\sf s}$-th SU and the $k_{\sf t}$-th CU as
$\mathbb{E}_{\tilde{\mathbf{g}}_{k_{\sf s}}}\!\left[R_{{\sf spc},k_{\sf s}}\right]$
and
$\mathbb{E}_{\tilde{\mathbf{g}}_{k_{\sf t}}}\!\left[R_{{\sf spc},k_{\sf t}}\right]$,
respectively, where
$\tilde{\mathbf{g}}_{k_{\sf s}}
\triangleq
[\tilde{g}^{\sf v,\sf v}_{k_{\sf s}},
\tilde{g}^{\sf h,\sf v}_{k_{\sf s}},
\tilde{g}^{\sf v,\sf h}_{k_{\sf s}},
\tilde{g}^{\sf h,\sf h}_{k_{\sf s}}]^{\sf T}$
and
$\tilde{\mathbf{g}}_{k_{\sf t}}
\triangleq
[\tilde{g}^{\sf v,\sf v}_{k_{\sf t}},
\tilde{g}^{\sf h,\sf v}_{k_{\sf t}},
\tilde{g}^{\sf v,\sf h}_{k_{\sf t}},
\tilde{g}^{\sf h,\sf h}_{k_{\sf t}}]^{\sf T}$.
As obtaining closed-form expressions for the ergodic equations is generally intractable, we employ the sample average approximation (SAA) technique to approximate them with empirical averages. Specifically, the LEO satellite constructs $S$ channel samples using the available geometry-related and long-term statistical channel parameters together with $S$ independently generated realizations of the small-scale fading coefficients. The resulting channel samples are given by $\mathbf{F}_{k_{\sf s}}^{(s)}=\mathbf{a}_{k_{\sf s}}\otimes\mathbf{G}_{k_{\sf s}}^{(s)}$ and $\mathbf{Z}_{k_{\sf t}}^{(s)}=\mathbf{a}_{k_{\sf t}}\otimes\mathbf{G}_{k_{\sf t}}^{(s)}$,  $\forall s\in\mathcal{S}\triangleq\{1,\ldots,S\}$. Herein, $s$ denotes the index of an independently generated realization of the small-scale fading coefficients, with the corresponding coefficients satisfying $\tilde{\mathbf{g}}_{k_{\sf s}}^{(s)}\sim\mathcal{CN}(\mathbf{0},\mathbf{I}_4)$ and $\tilde{\mathbf{g}}_{k_{\sf t}}^{(s)}\sim\mathcal{CN}(\mathbf{0},\mathbf{I}_4)$, for all $s\in\mathcal{S}$. Based on these channel samples, the ergodic super-common spectral efficiencies at the $k_{\sf s}$-th SU and the $k_{\sf t}$-th CU are approximated by their empirical averages as 
\begin{align}
\label{empirical average super common rate SU}
\mathbb{E}_{\tilde{\mathbf{g}}_{k_{\sf s}}}
\left[R_{{\sf spc},k_{\sf s}}\right]
&\approx
\bar{R}_{{\sf spc},k_{\sf s}}
\triangleq
\frac{1}{S}
\sum_{s=1}^{S}
R_{{\sf spc},k_{\sf s}}^{(s)},\\
\label{empirical average super common rate CU}
\mathbb{E}_{\tilde{\mathbf{g}}_{k_{\sf t}}}
\left[R_{{\sf spc},k_{\sf t}}\right]
&\approx
\bar{R}_{{\sf spc},k_{\sf t}}
\triangleq
\frac{1}{S}
\sum_{s=1}^{S}
R_{{\sf spc},k_{\sf t}}^{(s)},
\end{align}
where the approximations become tight as $S\rightarrow\infty$.
Herein,
$R_{{\sf spc},k_{\sf s}}^{(s)}$
and
$R_{{\sf spc},k_{\sf t}}^{(s)}$
denote the instantaneous spectral efficiencies of the super-common stream evaluated for the channel realizations
$\mathbf{F}_{k_{\sf s}}^{(s)}$
and
$\mathbf{Z}_{k_{\sf t}}^{(s)}$,
respectively. To ensure all SUs and CUs can decode the super-common stream $m_{\sf spc}$, the corresponding approximated ergodic spectral efficiency is given by
%To ensure that all SUs and CUs can decode the super-common stream $m_{\sf spc}$, the ergodic spectral efficiency of the super-common stream is given by
\begin{align}
\label{spc-rate}
\min \left( \min_{k_{\sf s}\in \mathcal{K}_{\sf s}}\bar{R}_{{\sf{spc}},k_{\sf s}}, \min_{k_{\sf t}\in \mathcal{K}_{\sf t}}\bar{R}_{{\sf{spc}},k_{\sf t}} \right) = \sum_{j=1}^{K_{\sf s}}C_{{\sf{spc}},j},
\end{align}
where $C_{{\sf{spc}},k_{\sf s}}$ denotes the rate portion allocated to the $k_{\sf s}$-th SU for the super-common stream.

After the successful decoding, the super-common stream is re-encoded and subtracted from the received signal via SIC. After the SIC, the CP-common and LP-common streams are decoded at the SUs and CUs, respectively, while treating the remaining signals as noise. Thus, the instantaneous spectral efficiencies for the CP-common stream at the $k_{\sf s}$-th SU and the LP-common stream at the $k_{\sf t}$-th CU are formulated as 
\begin{align}
\label{Rate_cpc_SU}
R_{{\sf{cpc}}, k_{\sf s}} = \log_{2}\Bigg( 1+ \frac{\vert\mathbf{f}_{k_{\sf s}}^{\sf H} \mathbf{w}_{\sf{cpc}}\vert^{2}} {\sum_{j=1}^{K_{\sf s}}\vert\mathbf{f}_{k_{\sf s}}^{\sf H} \mathbf{w}_{j}\vert^{2} + \sigma_{{\sf n},k_{\sf{s}}}^2}\Bigg),
\end{align}
\begin{align}
\label{Rate_lpc_CU}
& R_{{\sf lpc},{k_{\sf t}}} \!\!=\!  \log_{2} \!\!\Bigg(\! 1 \!+\! \frac{\vert{\mathbf{h}}_{k_{\sf t}}^{\sf H}\mathbf{p}_{k_{\sf{lpc}}}\vert^{2}}{\vert{\mathbf{z}}_{k_{\sf t}}^{\sf H}\mathbf{w}_{\sf{cpc}}\vert^{2} \!+\! \sum_{j=1}^{K_{\sf s}}\vert{\mathbf{z}}_{k_{\sf t}}^{\sf H} \mathbf{w}_{j}\vert^{2} \!+\! I_{{\sf lpc}, k_{\sf t}}} \!\Bigg)\!, 
\end{align}
where $I_{{\sf lpc}, k_{\sf t}} \triangleq \sum_{j=1}^{K_{\sf t}}\vert{\mathbf{h}}_{k_{\sf t}}^{\sf H} \mathbf{p}_{j}\vert^{2} + \sigma_{{\sf n},k_{\sf{t}}}^{2} $.
Following the same procedure as for the super-common stream, we then approximate the ergodic CP-common spectral efficiency at the $k_{\sf s}$-th SU and the ergodic LP-common spectral efficiency at the $k_{\sf t}$-th CU using the SAA technique as 
\begin{align}
\label{empirical average CPC rate SU}
\mathbb{E}_{\tilde{\mathbf{g}}_{k_{\sf s}}}
\left[R_{{\sf cpc},k_{\sf s}}\right]
&\approx
\bar{R}_{{\sf cpc},k_{\sf s}}
\triangleq
\frac{1}{S}
\sum_{s=1}^{S}
R_{{\sf cpc},k_{\sf s}}^{(s)},\\
\label{empirical average LPC rate CU}
\mathbb{E}_{\tilde{\mathbf{g}}_{k_{\sf t}}}
\left[R_{{\sf lpc},k_{\sf t}}\right]
&\approx
\bar{R}_{{\sf lpc},k_{\sf t}}
\triangleq
\frac{1}{S}
\sum_{s=1}^{S}
R_{{\sf lpc},k_{\sf t}}^{(s)}.
\end{align}
Since the CP-common and LP-common streams should be decodable by all SUs and CUs, respectively, the corresponding approximated ergodic spectral efficiencies are given by
\begin{align}
\label{cpc_lpc_rate} 
\min_{k_{\sf s}\in \mathcal{K}_{\sf s}}\bar{R}_{{\sf{cpc}},k_{\sf s}}
= \sum_{j=1}^{K_{\sf s}}C_{{\sf{cpc}},j}, \,\, \min_{k_{\sf t}\in \mathcal{K}_{\sf t}}\bar{R}_{{\sf{lpc}},k_{\sf t}}
= \sum_{j=1}^{K_{\sf t}}C_{{\sf lpc},j}.
\end{align}
Herein, $C_{{\sf{cpc}},{k_{\sf s}}}$ and $C_{{\sf{lpc}},{k_{\sf t}}}$ denote the rate portions allocated to the $k_{\sf s}$-th SU and the $k_{\sf t}$-th CU for the CP-common and LP-common streams, respectively.

Once the CP-common and LP-common streams are decoded and subtracted using the SIC technique at the SUs and CUs, respectively, the corresponding private streams are decoded while treating the remaining signals as noise. Accordingly, the instantaneous
spectral efficiencies for the private streams at the $k_{\sf s}$-th SU and the $k_{\sf t}$-th CU are represented as
\begin{align}
\label{Rate_p_SU}
R_{{\sf{p}}, k_{\sf s}} = \log_{2} \Bigg( 1+  \frac{\vert\mathbf{f}_{k_{\sf s}}^{\sf H} \mathbf{w}_{k_{\sf s}}\vert^{2}} {\sum_{j=1, j \neq k_{\sf s}}^{K_{\sf s}}\vert\mathbf{f}_{k_{\sf s}}^{\sf H} \mathbf{w}_{j}\vert^{2} + \sigma_{{\sf n},k_{\sf{s}}}^2} \Bigg),
\end{align}
\begin{align}
\label{Rate_p_CU}
R_{{\sf p},{k_{\sf t}}} \!=\!  \log_{2} \Bigg( 1 \!+\! \frac{\vert{\mathbf{h}}_{k_{\sf t}}^{\sf H}\mathbf{p}_{k_{\sf{t}}}\vert^{2}}{\vert{\mathbf{z}}_{k_{\sf t}}^{\sf H}\mathbf{w}_{\sf{cpc}}\vert^{2} \!+\! \sum_{j=1}^{K_{\sf s}}\vert{\mathbf{z}}_{k_{\sf t}}^{\sf H} \mathbf{w}_{j}\vert^{2} \!+\! I_{{\sf p}, k_{\sf t}}} \Bigg), 
\end{align}
where $I_{{\sf p}, k_{\sf t}} \triangleq \sum_{j=1,j \neq k_{\sf t}}^{K_{\sf t}}\vert{\mathbf{h}}_{k_{\sf t}}^{\sf H} \mathbf{p}_{j}\vert^{2} + \sigma_{{\sf n},k_{\sf{t}}}^{2}$.
Then, the ergodic private spectral efficiencies at the $k_{\sf s}$-th SU and the $k_{\sf t}$-th CU are approximated as follows:
\begin{align}
\label{empirical average private rate SU}
\mathbb{E}_{\tilde{\mathbf{g}}_{k_{\sf s}}}
\left[R_{{\sf p},k_{\sf s}}\right]
&\approx
\bar{R}_{{\sf p},k_{\sf s}}
\triangleq
\frac{1}{S}
\sum_{s=1}^{S}
R_{{\sf p},k_{\sf s}}^{(s)},\\
\label{empirical average private rate CU}
\mathbb{E}_{\tilde{\mathbf{g}}_{k_{\sf t}}}
\left[R_{{\sf p},k_{\sf t}}\right]
&\approx
\bar{R}_{{\sf p},k_{\sf t}}
\triangleq
\frac{1}{S}
\sum_{s=1}^{S}
R_{{\sf p},k_{\sf t}}^{(s)}.
\end{align}
Consequently, the approximated total ergodic spectral efficiencies for the $k_{\sf s}$-th SU and the $k_{\sf t}$-th CU are given by
\begin{align}
\label{R_tot} 
    \bar{R}_{k_{\sf s}} = C_{{\sf{spc}},k_{\sf s}}+C_{{\sf{cpc}},k_{\sf s}} + \bar{R}_{{\sf{p}},k_{\sf s}}, \, \bar{R}_{k_{\sf t}} = C_{{\sf{lpc}},k_{\sf t}} + \bar{R}_{{\sf{p}}, k_{\sf t}}.
\end{align}

%\vspace{-3mm}
\subsection{Problem Formulation}
We aim to maximize the minimum user spectral efficiency among all SUs and CUs in MDP-ISTN under imperfect CSIT at the LEO satellite. This objective can be formulated as the following optimization problem.
\begin{align}
\nonumber
{\mathscr P_1:} \maximize_{{\mathbf{W}}, {\mathbf{P}}, {\mathbf{c}}_{\sf{spc}}, {\mathbf{c}}_{\sf{cpc}}, {\mathbf{c}}_{\sf{lpc}}} \,\, 
\min \left( \!\min_{k_{\sf s}\in \mathcal{K}_{\sf s}}\bar{R}_{{k_{\sf s}}}, \min_{k_{\sf t}\in \mathcal{K}_{\sf t}}\bar{R}_{{k_{\sf t}}} \!\right)
\nonumber
\end{align}
\setcounter{equation}{38}\vspace{-3mm}
{\allowdisplaybreaks[4]
\begin{subequations}\label{condition1}
\begin{align}
\mathrm{s.t.}\,\,\,\,\,\,
\label{PF1CST1}
&\bar{R}_{{\sf{spc}},k_{\sf{s}}} \geq \sum_{j=1}^{K_{\sf s}}C_{{\sf spc},j}, \,\, \forall k_{\sf{s}} \in \mathcal{K}_{\sf{s}},\\
\label{PF1CST2}
&\bar{R}_{{\sf{spc}},k_{\sf{t}}} \geq \sum_{j=1}^{K_{\sf s}}C_{{\sf spc},j}, \,\, \forall k_{\sf{t}} \in \mathcal{K}_{\sf{t}},\\
\label{PF1CST3}
&\bar{R}_{{\sf{cpc}},k_{\sf{s}}} \geq \sum_{j=1}^{K_{\sf s}}C_{{\sf cpc},j}, \,\, \forall k_{\sf{s}} \in \mathcal{K}_{\sf{s}},\\
\label{PF1CST4}
&\bar{R}_{{\sf{lpc}},k_{\sf{t}}} \geq \sum_{j=1}^{K_{\sf t}}C_{{\sf lpc},j}, \,\, \forall k_{\sf{t}} \in \mathcal{K}_{\sf{t}},\\
\label{PF1CST5}
&C_{{\sf spc}, k_{\sf{s}}} \geq 0, \,\, C_{{\sf cpc}, k_{\sf{s}}} \geq 0, \,\, \forall k_{\sf{s}} \in \mathcal{K}_{\sf{s}}, \\
\label{PF1CST6}
&C_{{\sf lpc}, k_{\sf{t}}} \geq 0, \,\, \forall k_{\sf{t}} \in \mathcal{K}_{\sf{t}}, \\
\label{PF1CST7}
&{\sf{tr}}\left(\mathbf{W} \mathbf{W}^{\sf H}\right) \leq P_{\sf s}, \,\, {\sf{tr}}\left(\mathbf{P} \mathbf{P}^{\sf H}\right) \leq P_{\sf t}.
\end{align}
\end{subequations}
}In the formulated problem $\mathscr P_1$, the vectors $\mathbf{c}_{\sf spc}=[C_{{\sf spc},1},\cdots, C_{{\sf spc},K_{\sf s}}]^{\sf{T}}$, $\mathbf{c}_{\sf cpc}=[C_{{\sf cpc},1},\cdots, C_{{\sf cpc},K_{\sf s}}]^{\sf{T}}$, and $\mathbf{c}_{\sf lpc}=[C_{{\sf lpc},1},\cdots, C_{{\sf lpc},K_{\sf t}}]^{\sf{T}}$ denote the vectors composed of the super-common portions assigned to each SU, the CP-common portions assigned to each SU, and the LP-common portions assigned to each CU, respectively. The constraints (\ref{PF1CST1}) and (\ref{PF1CST2}) guarantee that the super-common stream remains decodable for all SUs and CUs, whereas (\ref{PF1CST3}) and (\ref{PF1CST4}) ensure that the CP-common stream and the LP-common stream remain decodable for all SUs and CUs, respectively. Meanwhile, (\ref{PF1CST5}) and (\ref{PF1CST6}) enforce the non-negativity of the super-common portions, CP-common portions, and LP-common portions. The constraint (\ref{PF1CST7}) imposes the total transmit power constraint for the LEO satellite and the terrestrial BS, where $P_{\sf{s}}$ and $P_{\sf{t}}$ denote the total transmit power budgets of the LEO satellite and the terrestrial BS, respectively.

%\vspace{-3mm}
\section{Proposed MDP-RSMA Precoder Design}
%\vspace{-1mm}

The optimization problem $\mathscr{P}_1$ is a non-convex problem; therefore, directly solving it is challenging. We reformulate it into a more tractable form via introducing auxiliary variables $\alpha_{{\sf{p}},{k_{\sf s}}}$, $\alpha_{{\sf{p}},{k_{\sf t}}}$, and $R_{\sf min}$ corresponding to $\bar{R}_{{\sf{p}},{k_{\sf s}}}$, $\bar{R}_{{\sf{p}},{k_{\sf t}}}$, and the minimum user spectral efficiency, respectively, as
\begin{align}
\nonumber
{\mathscr P_2:} \maximize_{
\mathbf{W}, \mathbf{P}, \mathbf{c}_{\sf spc}, 
\mathbf{c}_{\sf cpc}, \mathbf{c}_{\sf lpc}, 
\boldsymbol{\alpha}_{{\sf p},{\sf sat}}, 
\boldsymbol{\alpha}_{{\sf p},{\sf bs}}, R_{\sf min}} \,\, 
R_{\sf min}
\nonumber
\end{align}
\setcounter{equation}{39}\vspace{-3mm}
\begin{subequations}\label{condition2}
\begin{align}
\mathrm{s.t.}\,\,\,\,\,\,
\label{PF2CST1}
& \alpha_{{\sf{p}}, k_{\sf s}} + C_{{\sf cpc}, k_{\sf s}} + C_{{\sf spc}, k_{\sf s}} \geq R_{\sf min}, \,\, \forall k_{\sf s} \in \mathcal{K}_{\sf s}, \\
\label{PF2CST2}
& \alpha_{{\sf{p}}, k_{\sf t}} + C_{{\sf lpc}, k_{\sf t}}  \geq R_{\sf min}, \,\, \forall k_{\sf t} \in \mathcal{K}_{\sf t}, \\
\label{PF2CST3}
&\bar{R}_{{\sf{p}},k_{\sf s}} \geq \alpha_{{\sf{p}}, k_{\sf s}},  \,\, \forall k_{\sf s} \in \mathcal{K}_{\sf s}, \\
\label{PF2CST4}
&\bar{R}_{{\sf{p}},k_{\sf t}} \geq \alpha_{{\sf{p}}, k_{\sf t}},  \,\, \forall k_{\sf t} \in \mathcal{K}_{\sf t}, \\
\label{PF2CST5}
&\alpha_{{\sf{p}}, k_{\sf s}} \geq 0,  \,\, \forall k_{\sf s} \in \mathcal{K}_{\sf s}, \,\, \alpha_{{\sf{p}}, k_{\sf t}} \geq 0,  \,\, \forall k_{\sf t} \in \mathcal{K}_{\sf t}, \\
&\textrm{(\ref{PF1CST1})}, \,\, \textrm{(\ref{PF1CST2})}, \,\,  \textrm{(\ref{PF1CST3})}, \,\, \textrm{(\ref{PF1CST4})}, \,\, \textrm{(\ref{PF1CST5})}, \,\, \textrm{(\ref{PF1CST6})}, \,\, \textrm{(\ref{PF1CST7})}, \nonumber
\end{align}
\end{subequations}
where $\boldsymbol{\alpha}_{{\sf p}, {\sf sat}}$ and $\boldsymbol{\alpha}_{{\sf p}, {\sf bs}}$ denote $\boldsymbol{\alpha}_{{\sf p}, {\sf sat}}=[\alpha_{{\sf p},1},\cdots, \alpha_{{\sf p},K_{\sf s}}]^{\sf{T}}$ and  $\boldsymbol{\alpha}_{{\sf p}, {\sf bs}}=[\alpha_{{\sf p},1},\cdots, \alpha_{{\sf p}, K_{\sf t}}]^{\sf{T}}$, respectively.
Nonetheless, the problem $\mathscr{P}_2$ remains non-convex due to the constraints (\ref{PF1CST1})--(\ref{PF1CST4}) and (\ref{PF2CST3})--(\ref{PF2CST4}), where the expressions of spectral efficiency are non-convex functions. We address this issue by establishing a problem-specific WMMSE reformulation that reflects the message-splitting, decoding, and interference management structure of the proposed MDP-RSMA, enabling tractable AO-based precoder design.

%We address this issue and develop a WMMSE-based algorithm tailored to the proposed MDP-RSMA framework. 
From equations (\ref{Rate_spc_SU}) and (\ref{Rate_spc_CU}), it is observed that the super-common spectral efficiencies for the $k_{\sf s}$-th SU, $R_{{\sf spc}, k_{\sf s}}$, and the $k_{\sf t}$-th CU, $R_{{\sf spc}, k_{\sf t}}$, can be obtained by decoding the super-common stream $m_{\sf{spc}}$ from the received signals $\mathbf{f}_{k_{\sf s}}^{\sf{H}}\mathbf{w}_{\sf spc}m_{\sf{spc}} + \mathbf{f}_{k_{\sf s}}^{\sf{H}}\mathbf{w}_{\sf cpc}m_{\sf cpc} + \sum_{j=1}^{K_{\sf s}}\mathbf{f}_{k_{\sf s}}^{\sf{H}}\mathbf{w}_{j}m_{j} + n_{k_{\sf s}}$ and $\mathbf{z}_{k_{\sf t}}^{\sf{H}}\mathbf{w}_{\sf spc}m_{\sf spc} + \mathbf{z}_{k_{\sf t}}^{\sf{H}}\mathbf{w}_{\sf cpc}m_{\sf cpc} + \sum_{j=1}^{K_{\sf s}}\mathbf{z}_{k_{\sf t}}^{\sf{H}}\mathbf{w}_{j}m_{j} + \mathbf{h}_{k_{\sf t}}^{\sf H}\mathbf{p}_{\sf lpc}s_{\sf lpc} + \sum_{j=1}^{K_{\sf t}}\mathbf{h}_{k_{\sf t}}^{\sf H}\mathbf{p}_{j}s_{j} + n_{k_{\sf t}}$, respectively. Building upon these observations, we formulate the super-common mean square errors (MSEs) for the $k_{\sf s}$-th SU and the $k_{\sf t}$-th CU, denoted as $\epsilon_{{\sf{spc}},k_{\sf s}}$ and $\epsilon_{{\sf{spc}},k_{\sf t}}$, with corresponding equalizers $q_{{\sf{spc}},k_{\sf s}}$ and $q_{{\sf{spc}},k_{\sf t}}$, respectively, as
\begin{align}
\label{MSE_spc_SU}
   \epsilon_{{\sf{spc}}, k_{\sf s}} & = \mathbb{E}\bigg[ \big\vert q_{{\sf{spc}},k_{\sf s}} \big(\mathbf{f}_{k_{\sf s}}^{\sf{H}}\mathbf{w}_{\sf spc}m_{\sf spc} + \mathbf{f}_{k_{\sf s}}^{\sf{H}}\mathbf{w}_{\sf cpc}m_{\sf cpc} 
   + \sum_{j=1}^{K_{\sf s}}\mathbf{f}_{k_{\sf s}}^{\sf{H}}\mathbf{w}_{j}m_{j} + n_{k_{\sf s}}\big) - m_{{\sf{spc}}} \big\vert^{2}\bigg] \nonumber \\
   & = \vert q_{{\sf{spc}}, k_{\sf s}} \vert^{2} T_{{\sf{spc}}, k_{\sf s}} - 2{\sf{Re}}\{q_{{\sf{spc}}, k_{\sf s}} \mathbf{f}_{k_{\sf s}}^{\sf{H}}\mathbf{w}_{\sf spc}\} + 1 
\end{align}
and
\begin{align}
\label{MSE_spc_CU}
   \epsilon_{{\sf{spc}}, k_{\sf t}} & \!=\! \mathbb{E}\bigg[ \big\vert q_{{\sf{spc}},k_{\sf t}} \big(\mathbf{z}_{k_{\sf t}}^{\sf{H}}\mathbf{w}_{\sf spc}m_{\sf spc} \!+\! \mathbf{z}_{k_{\sf t}}^{\sf{H}}\mathbf{w}_{\sf cpc}m_{\sf cpc} \!+\! \sum_{j=1}^{K_{\sf s}}\mathbf{z}_{k_{\sf t}}^{\sf{H}}\mathbf{w}_{j}m_{j}
   + \mathbf{h}_{k_{\sf t}}^{\sf H}\mathbf{p}_{\sf lpc}s_{\sf lpc} + \sum_{j=1}^{K_{\sf t}}\mathbf{h}_{k_{\sf t}}^{\sf H}\mathbf{p}_{j}s_{j} + n_{k_{\sf t}} \big) - m_{{\sf{spc}}} \big\vert^{2}\bigg] \nonumber \\
   & = \vert q_{{\sf{spc}}, k_{\sf t}} \vert^{2} T_{{\sf{spc}}, k_{\sf t}} - 2{\sf{Re}}\{q_{{\sf{spc}}, k_{\sf t}} \mathbf{z}_{k_{\sf t}}^{\sf{H}}\mathbf{w}_{\sf spc}\} + 1,   
\end{align}
where $T_{{\sf{spc}}, k_{\sf s}} \triangleq \vert{\mathbf{f}}_{k_{\sf s}}^{\sf H}\mathbf{w}_{\sf{spc}}\vert^{2} + \vert{\mathbf{f}}_{k_{\sf s}}^{\sf H}\mathbf{w}_{\sf{cpc}}\vert^{2} + \sum_{j=1}^{K_{\sf s}}\vert{\mathbf{f}}_{k_{\sf s}}^{\sf H} \mathbf{w}_{j}\vert^{2} + \sigma_{{\sf n},k_{\sf s}}^{2}$, and $T_{{\sf{spc}}, k_{\sf t}} \triangleq \vert{\mathbf{z}}_{k_{\sf t}}^{\sf H}\mathbf{w}_{\sf{spc}}\vert^{2} + \vert{\mathbf{z}}_{k_{\sf t}}^{\sf H}\mathbf{w}_{\sf{cpc}}\vert^{2} + \sum_{j=1}^{K_{\sf s}}\vert{\mathbf{z}}_{k_{\sf t}}^{\sf H} \mathbf{w}_{j}\vert^{2} + I_{{\sf{spc}}, k_{\sf t}}$.
In an analogous  manner, equations (\ref{Rate_cpc_SU}) and (\ref{Rate_lpc_CU}) indicate that the CP-common spectral efficiency for the $k_{\sf s}$-th SU, $R_{{\sf cpc}, k_{\sf s}}$, and the LP-common spectral efficiency for the $k_{\sf t}$-th CU, $R_{{\sf lpc}, k_{\sf t}}$, can be obtained by decoding the CP-common stream $m_{\sf cpc}$ from the signal $\mathbf{f}_{k_{\sf s}}^{\sf{H}}\mathbf{w}_{\sf cpc}m_{\sf cpc} + \sum_{j=1}^{K_{\sf s}}\mathbf{f}_{k_{\sf s}}^{\sf{H}}\mathbf{w}_{j}m_{j} + n_{k_{\sf s}}$ and decoding the LP-common stream $s_{\sf lpc}$ from the signal $\mathbf{z}_{k_{\sf t}}^{\sf{H}}\mathbf{w}_{\sf cpc}m_{\sf cpc} + \sum_{j=1}^{K_{\sf s}}\mathbf{z}_{k_{\sf t}}^{\sf{H}}\mathbf{w}_{j}m_{j} + \mathbf{h}_{k_{\sf t}}^{\sf H}\mathbf{p}_{\sf lpc}s_{\sf lpc} + \sum_{j=1}^{K_{\sf t}}\mathbf{h}_{k_{\sf t}}^{\sf H}\mathbf{p}_{j}s_{j} + n_{k_{\sf t}}$, respectively. Therefore, the CP-common MSE for the $k_{\sf s}$-th SU and the LP-common MSE for the $k_{\sf t}$-th CU, denoted as $\epsilon_{{\sf{cpc}},k_{\sf s}}$ and $\epsilon_{{\sf{lpc}},k_{\sf t}}$, with corresponding equalizers $q_{{\sf{cpc}},k_{\sf s}}$ and $q_{{\sf{lpc}},k_{\sf t}}$, are respectively given as
\begin{align}
\label{MSE_cpc_SU}
   \epsilon_{{\sf{cpc}}, k_{\sf s}} & \!=\! \mathbb{E}\bigg[ \big\vert q_{{\sf{cpc}},k_{\sf s}} \big(\mathbf{f}_{k_{\sf s}}^{\sf{H}}\mathbf{w}_{\sf cpc}m_{\sf cpc} \!+\!\! \sum_{j=1}^{K_{\sf s}}\mathbf{f}_{k_{\sf s}}^{\sf{H}}\mathbf{w}_{j}m_{j} \!+\! n_{k_{\sf s}}\big) \!-\! m_{{\sf{cpc}}} \big\vert^{2}\bigg] \nonumber \\
   & = \vert q_{{\sf{cpc}}, k_{\sf s}} \vert^{2} T_{{\sf{cpc}}, k_{\sf s}} - 2{\sf{Re}}\{q_{{\sf{cpc}}, k_{\sf s}} \mathbf{f}_{k_{\sf s}}^{\sf{H}}\mathbf{w}_{\sf cpc}\} + 1
\end{align}  
and
\begin{align}
\label{MSE_lpc_CU}
   \epsilon_{{\sf{lpc}}, k_{\sf t}} & = \mathbb{E}\bigg[ \big\vert q_{{\sf{lpc}},k_{\sf t}} \big(\mathbf{z}_{k_{\sf t}}^{\sf{H}}\mathbf{w}_{\sf cpc}m_{\sf cpc} + \sum_{j=1}^{K_{\sf s}}\mathbf{z}_{k_{\sf t}}^{\sf{H}}\mathbf{w}_{j}m_{j} + \mathbf{h}_{k_{\sf t}}^{\sf H}\mathbf{p}_{\sf lpc}s_{\sf lpc} + \sum_{j=1}^{K_{\sf t}}\mathbf{h}_{k_{\sf t}}^{\sf H}\mathbf{p}_{j}s_{j} + n_{k_{\sf t}} \big) - s_{{\sf{lpc}}} \big\vert^{2}\bigg] \nonumber \\
   & = \vert q_{{\sf{lpc}}, k_{\sf t}} \vert^{2} T_{{\sf{lpc}}, k_{\sf t}} - 2{\sf{Re}}\{q_{{\sf{lpc}}, k_{\sf t}} \mathbf{h}_{k_{\sf t}}^{\sf{H}}\mathbf{p}_{\sf lpc}\} + 1 ,
\end{align}
where $T_{{\sf{cpc}}, k_{\sf s}} \triangleq \vert{\mathbf{f}}_{k_{\sf s}}^{\sf H}\mathbf{w}_{\sf{cpc}}\vert^{2} + \sum_{j=1}^{K_{\sf s}}\vert{\mathbf{f}}_{k_{\sf s}}^{\sf H} \mathbf{w}_{j}\vert^{2} + \sigma_{{\sf n},k_{\sf s}}^{2}$, and $T_{{\sf{lpc}}, k_{\sf t}} \triangleq  \vert{\mathbf{z}}_{k_{\sf t}}^{\sf H}\mathbf{w}_{\sf{cpc}}\vert^{2} + \sum_{j=1}^{K_{\sf s}}\vert{\mathbf{z}}_{k_{\sf t}}^{\sf H} \mathbf{w}_{j}\vert^{2} +  \vert{\mathbf{h}}_{k_{\sf t}}^{\sf H}\mathbf{p}_{\sf{lpc}}\vert^{2} + I_{{\sf lpc}, k_{\sf t}}$.

Furthermore, equations (\ref{Rate_p_SU}) and (\ref{Rate_p_CU}) indicate that the private spectral efficiencies for the $k_{\sf s}$-th SU, $R_{{\sf p}, k_{\sf s}}$, and the $k_{\sf t}$-th CU, $R_{{\sf p}, k_{\sf t}}$, can be obtained by decoding the private stream $m_{k_{\sf s}}$ from the signal $\sum_{j=1}^{K_{\sf s}}\mathbf{f}_{k_{\sf s}}^{\sf{H}}\mathbf{w}_{j}m_{j} + n_{k_{\sf s}}$ and decoding the private stream $s_{k_{\sf t}}$ from the signal $\mathbf{z}_{k_{\sf t}}^{\sf{H}}\mathbf{w}_{\sf cpc}m_{\sf cpc} + \sum_{j=1}^{K_{\sf s}}\mathbf{z}_{k_{\sf t}}^{\sf{H}}\mathbf{w}_{j}m_{j} + \sum_{j=1}^{K_{\sf t}}\mathbf{h}_{k_{\sf t}}^{\sf H}\mathbf{p}_{j}s_{j} + n_{k_{\sf t}}$, respectively. 
Thus, the private MSEs for the $k_{\sf s}$-th SU and the $k_{\sf t}$-th CU, denoted by $\epsilon_{{\sf p},k_{\sf s}}$ and $\epsilon_{{\sf p},k_{\sf t}}$, with corresponding equalizers $q_{{\sf{p}},k_{\sf s}}$ and $q_{{\sf{p}},k_{\sf t}}$, are as
\begin{align}
\label{MSE_p_SU}
   \epsilon_{{\sf{p}}, k_{\sf s}} & = \mathbb{E}\bigg[ \big\vert q_{{\sf{p}},k_{\sf s}} \big(\sum_{j=1}^{K_{\sf s}}\mathbf{f}_{k_{\sf s}}^{\sf{H}}\mathbf{w}_{j}m_{j} + n_{k_{\sf s}}\big) - m_{k_{\sf s}} \big\vert^{2} \bigg] = \vert q_{{\sf{p}}, k_{\sf s}} \vert^{2} T_{{\sf{p}}, k_{\sf s}} - 2{\sf{Re}}\{q_{{\sf{p}}, k_{\sf s}} \mathbf{f}_{k_{\sf s}}^{\sf{H}}\mathbf{w}_{k_{\sf s}}\} + 1
\end{align}  
and
\begin{align}
\label{MSE_p_CU}
   \epsilon_{{\sf{p}}, k_{\sf t}} & = \mathbb{E}\bigg[ \big\vert q_{{\sf{p}},k_{\sf t}} \big(\mathbf{z}_{k_{\sf t}}^{\sf{H}}\mathbf{w}_{\sf cpc}m_{\sf cpc} + \sum_{j=1}^{K_{\sf s}}\mathbf{z}_{k_{\sf t}}^{\sf{H}}\mathbf{w}_{j}m_{j} 
   + \sum_{j=1}^{K_{\sf t}}\mathbf{h}_{k_{\sf t}}^{\sf H}\mathbf{p}_{j}s_{j} + n_{k_{\sf t}} \big) - s_{k_{\sf t}} \big\vert^{2}\bigg] \nonumber \\
   & = \vert q_{{\sf{p}}, k_{\sf t}} \vert^{2} T_{{\sf{p}}, k_{\sf t}} - 2{\sf{Re}}\{q_{{\sf{p}}, k_{\sf t}} \mathbf{h}_{k_{\sf t}}^{\sf{H}}\mathbf{p}_{k_{\sf t}}\} + 1, 
\end{align}
respectively. Herein, $T_{{\sf{p}}, k_{\sf s}} \triangleq \sum_{j=1}^{K_{\sf s}}\vert{\mathbf{f}}_{k_{\sf s}}^{\sf H} \mathbf{w}_{j}\vert^{2} + \sigma_{{\sf n}, k_{\sf s}}^{2}$, and $T_{{\sf{p}}, k_{\sf t}} \triangleq  \vert{\mathbf{z}}_{k_{\sf t}}^{\sf H}\mathbf{w}_{\sf{cpc}}\vert^{2} + \sum_{j=1}^{K_{\sf s}}\vert{\mathbf{z}}_{k_{\sf t}}^{\sf H} \mathbf{w}_{j}\vert^{2} + \vert{\mathbf{h}}_{k_{\sf t}}^{\sf H} \mathbf{p}_{k_{\sf t}}\vert^{2} + I_{{\sf p},k_{\sf t}}$.
    
Applying the first-order optimality conditions to (\ref{MSE_spc_SU})--(\ref{MSE_p_CU}), i.e.,  $\frac{\partial \epsilon_{{\sf{spc}}, k_{\sf s}}}{\partial q_{{\sf{spc}}, k_{\sf s}}} = 0$, $\frac{\partial \epsilon_{{\sf{spc}}, k_{\sf t}}}{\partial q_{{\sf{spc}}, k_{\sf t}}} = 0$, $\frac{\partial \epsilon_{{\sf{cpc}}, k_{\sf s}}}{\partial q_{{\sf{cpc}}, k_{\sf s}}} = 0$, $\frac{\partial \epsilon_{{\sf{lpc}}, k_{\sf t}}}{\partial q_{{\sf{lpc}}, k_{\sf t}}} = 0$, $\frac{\partial \epsilon_{{\sf{p}}, k_{\sf s}}}{\partial q_{{\sf{p}}, k_{\sf s}}} = 0$, and  $\frac{\partial \epsilon_{{\sf{p}}, k_{\sf t}}}{\partial q_{{\sf{p}}, k_{\sf t}}} = 0$, the minimum values of (\ref{MSE_spc_SU})--(\ref{MSE_p_CU}) are obtained when the equalizers are set as
\begin{align}
\label{MMSE_spc}
q_{{\sf{spc}}, k_{\sf s}}^{\sf{MMSE}} = T_{{\sf{spc}}, k_{\sf s}}^{-1} \mathbf{w}_{\sf spc}^{\sf{H}} \mathbf{f}_{k_{\sf s}}, \,\, q_{{\sf{spc}}, k_{\sf t}}^{\sf{MMSE}} = T_{{\sf{spc}}, k_{\sf t}}^{-1} \mathbf{w}_{\sf spc}^{\sf{H}} \mathbf{z}_{k_{\sf t}},
\end{align}
\begin{align}
\label{MMSE_cpc_lpc}
q_{{\sf{cpc}}, k_{\sf s}}^{\sf{MMSE}} = T_{{\sf{cpc}}, k_{\sf s}}^{-1} \mathbf{w}_{\sf cpc}^{\sf{H}} \mathbf{f}_{k_{\sf s}}, \,\, q_{{\sf{lpc}}, k_{\sf t}}^{\sf{MMSE}} = T_{{\sf{lpc}}, k_{\sf t}}^{-1} \mathbf{p}_{\sf lpc}^{\sf{H}} \mathbf{h}_{k_{\sf t}},
\end{align}
\begin{align}
\label{MMSE_p}
q_{{\sf{p}}, k_{\sf s}}^{\sf{MMSE}} = T_{{\sf{p}}, k_{\sf s}}^{-1} \mathbf{w}_{k_{\sf s}}^{\sf{H}} \mathbf{f}_{k_{\sf s}}, \,\, q_{{\sf{p}}, k_{\sf t}}^{\sf{MMSE}} = T_{{\sf{p}}, k_{\sf t}}^{-1} \mathbf{p}_{k_{\sf t}}^{\sf{H}} \mathbf{h}_{k_{\sf t}}.
\end{align}

The substitutions of $q_{{\sf{spc}}, k_{\sf s}}^{\sf{MMSE}}$, $q_{{\sf{spc}}, k_{\sf t}}^{\sf{MMSE}}$, $q_{{\sf{cpc}}, k_{\sf s}}^{\sf{MMSE}}$, $q_{{\sf{lpc}}, k_{\sf t}}^{\sf{MMSE}}$, $q_{{\sf{p}}, k_{\sf s}}^{\sf{MMSE}}$, and $q_{{\sf{p}}, k_{\sf t}}^{\sf{MMSE}}$ are performed in (\ref{MSE_spc_SU}), (\ref{MSE_spc_CU}), (\ref{MSE_cpc_SU}), (\ref{MSE_lpc_CU}), (\ref{MSE_p_SU}), and (\ref{MSE_p_CU}), respectively. Consequently, the corresponding minimum MSE (MMSE) values for the super-common stream at the $k_{\sf s}$-th SU, the super-common stream at the $k_{\sf t}$-th CU, the CP-common stream at the $k_{\sf s}$-th SU, the LP-common stream at the $k_{\sf t}$-th CU, the private stream at the $k_{\sf s}$-th SU, and the private stream at the $k_{\sf t}$-th CU are obtained as
\begin{align}
\label{MMSE_value_spc_SU}
\epsilon_{{\sf{spc}}, k_{\sf s}}^{\sf{MMSE}} = T_{{\sf{spc}}, k_{\sf s}}^{-1}(T_{{\sf{spc}}, k_{\sf s}} - \vert \mathbf{f}_{k_{\sf s}}^{\sf{H}} \mathbf{w}_{\sf spc} \vert^{2}), 
\end{align}
\begin{align}
\label{MMSE_value_spc_CU}
\epsilon_{{\sf{spc}}, k_{\sf t}}^{\sf{MMSE}} = T_{{\sf{spc}}, k_{\sf t}}^{-1}(T_{{\sf{spc}}, k_{\sf t}} - \vert \mathbf{z}_{k_{\sf t}}^{\sf{H}} \mathbf{w}_{\sf spc} \vert^{2}), 
\end{align}
\begin{align}
\label{MMSE_value_cpc_SU}
\epsilon_{{\sf{cpc}}, k_{\sf s}}^{\sf{MMSE}} = T_{{\sf{cpc}}, k_{\sf s}}^{-1}(T_{{\sf{cpc}}, k_{\sf s}} - \vert \mathbf{f}_{k_{\sf s}}^{\sf{H}} \mathbf{w}_{\sf cpc} \vert^{2}), 
\end{align}
\begin{align}
\label{MMSE_value_lpc_CU}
\epsilon_{{\sf{lpc}}, k_{\sf t}}^{\sf{MMSE}} = T_{{\sf{lpc}}, k_{\sf t}}^{-1}(T_{{\sf{lpc}}, k_{\sf t}} - \vert \mathbf{h}_{k_{\sf t}}^{\sf{H}} \mathbf{p}_{\sf lpc} \vert^{2}), 
\end{align}
\begin{align}
\label{MMSE_value_p_SU}
\epsilon_{{\sf{p}}, k_{\sf s}}^{\sf{MMSE}} = T_{{\sf{p}}, k_{\sf s}}^{-1}(T_{{\sf{p}}, k_{\sf s}} - \vert \mathbf{f}_{k_{\sf s}}^{\sf{H}} \mathbf{w}_{k_{\sf s}} \vert^{2}), 
\end{align}
\begin{align}
\label{MMSE_value_p_CU}
\epsilon_{{\sf{p}}, k_{\sf t}}^{\sf{MMSE}} = T_{{\sf{p}}, k_{\sf t}}^{-1}(T_{{\sf{p}}, k_{\sf t}} - \vert \mathbf{h}_{k_{\sf t}}^{\sf{H}} \mathbf{p}_{k_{\sf t}} \vert^{2}). 
\end{align}

Subsequently, by introducing the positive real weights $u_{{\sf{spc}}, k_{\sf s}}$, $u_{{\sf{spc}}, k_{\sf t}}$, $u_{{\sf{cpc}}, k_{\sf s}}$, $u_{{\sf{lpc}}, k_{\sf t}}$, $u_{{\sf{p}}, k_{\sf s}}$, and $u_{{\sf{p}}, k_{\sf t}}$, we formulate the augmented weighted mean square errors (WMSEs) for the super-common stream at the $k_{\sf s}$-th SU, the super-common stream at the $k_{\sf t}$-th CU, the CP-common stream at the $k_{\sf s}$-th SU, the LP-common stream at the $k_{\sf t}$-th CU, the private stream at the $k_{\sf s}$-th SU, and the private stream at the $k_{\sf t}$-th CU. For notational compactness, we define the set of valid stream-user pairs as $\mathcal{D} \triangleq \{ ({\sf spc},k_{\sf s}), ({\sf spc},k_{\sf t}), ({\sf cpc},k_{\sf s}), ({\sf lpc},k_{\sf t}), ({\sf p},k_{\sf s}), ({\sf p},k_{\sf t}) \}$. Accordingly, the augmented WMSE can be expressed as
\begin{align}
\label{WMSE}
\xi_{\ell, k} = u_{\ell, k} \epsilon_{\ell, k} - \log_2{u_{\ell, k}},\,\, \forall (\ell, k) \in \mathcal{D}. 
\end{align}
According to the first-order optimality conditions, the optimum equalizers and weights that minimize augmented WMSE are obtained by solving 
%$\frac{\partial \xi_{\ell, k}}{\partial q_{\ell, k}} = \frac{\partial \xi_{\ell, k}}{\partial u_{\ell, k}} = 0$, $\forall (\ell, k) \in \mathcal{D}$.
\begin{align}
    \frac{\partial \xi_{\ell, k}}{\partial q_{\ell, k}} = \frac{\partial \xi_{\ell, k}}{\partial u_{\ell, k}} = 0, \,\, \forall (\ell, k) \in \mathcal{D}.
\end{align}
Therefore, the optimum equalizers and weights that minimize augmented WMSE are obtained as follows:
\begin{align}
\label{WMSE_value}
q_{\ell, k}^{\star} = q_{\ell, k}^{\sf{MMSE}}, \,\, u_{\ell, k}^{\star} = 1/\epsilon_{\ell, k}^{\sf{MMSE}}, \,\, \forall (\ell, k) \in \mathcal{D}.
\end{align}
Then, by substituting the optimal values in (\ref{WMSE_value}) into the augmented WMSE expression (\ref{WMSE}), $\forall (\ell, k) \in \mathcal{D}$, the relationship between the WMMSE and spectral efficiency is established as
\begin{align}
    \label{WMMSE_SE_relationship}
       \xi_{\ell, k}^{\star} & = \min_{q_{\ell, k}, u_{\ell, k}} {\xi_{\ell, k}} = 1 + \log_2{\epsilon_{\ell, k}^{\sf{MMSE}}} = 1 - R_{\ell, k}, \,\, \forall (\ell, k) \in \mathcal{D}.
\end{align}
Based on the relationship in (\ref{WMMSE_SE_relationship}), we construct $S$ independent $(\xi_{{\sf{spc}}, k_{\sf s}}^{\star})^{\left(s\right)}$, $(\xi_{{\sf{spc}}, k_{\sf t}}^{\star})^{\left(s\right)}$, $(\xi_{{\sf{cpc}}, k_{\sf s}}^{\star})^{\left(s\right)}$, $(\xi_{{\sf{lpc}}, k_{\sf t}}^{\star})^{\left(s\right)}$, $(\xi_{{\sf{p}}, k_{\sf s}}^{\star})^{\left(s\right)}$, and $(\xi_{{\sf{p}}, k_{\sf t}}^{\star})^{\left(s\right)}$, by generating
$\mathbf{F}_{k_{\sf s}}^{(s)}=\mathbf{a}_{k_{\sf s}}\otimes\mathbf{G}_{k_{\sf s}}^{(s)}$ and $\mathbf{Z}_{k_{\sf t}}^{(s)}=\mathbf{a}_{k_{\sf t}}\otimes\mathbf{G}_{k_{\sf t}}^{(s)}$, for all $s\in\mathcal{S}$.
%By doing so, the equations (\ref{Rate_spc_SU}), (\ref{Rate_spc_CU}), (\ref{Rate_cpc_SU}), (\ref{Rate_lpc_CU}), (\ref{Rate_p_SU}), and (\ref{Rate_p_CU}) are rewritten as follows:
By doing so, the empirical average of spectral
efficiency expressions are rewritten as follows:
\begin{align}
   \label{spc_SU_relationship_average}
    \bar{R}_{{\sf{spc}},k_{\sf s}} &= 1 - \bar{\xi}_{{\sf{spc}},k_{\sf s}}^{\star} = 1 - \frac{1}{S}\sum\limits_{s=1}^{S} (\xi_{{\sf{spc}},k_{\sf s}}^{\star})^{\left(s\right)},
%\end{align}
%\begin{align}
\\
   \label{spc_CU_relationship_average}
    \bar{R}_{{\sf{spc}},k_{\sf t}} &= 1 - \bar{\xi}_{{\sf{spc}},k_{\sf t}}^{\star} = 1 - \frac{1}{S}\sum\limits_{s=1}^{S} (\xi_{{\sf{spc}},k_{\sf t}}^{\star})^{\left(s\right)},
%\end{align}
%\begin{align}
\\
    \label{cpc_SU_relationship_average}
    \bar{R}_{{\sf{cpc}},k_{\sf s}} &= 1 - \bar{\xi}_{{\sf{cpc}},k_{\sf s}}^{\star} = 1 - \frac{1}{S}\sum\limits_{s=1}^{S} (\xi_{{\sf{cpc}},k_{\sf s}}^{\star})^{\left(s\right)},
%\end{align}
%\begin{align}
\\
    \label{lpc_CU_relationship_average}
    \bar{R}_{{\sf{lpc}},k_{\sf t}} &= 1 - \bar{\xi}_{{\sf{lpc}},k_{\sf t}}^{\star} = 1 - \frac{1}{S}\sum\limits_{s=1}^{S} (\xi_{{\sf{lpc}},k_{\sf t}}^{\star})^{\left(s\right)},
%\end{align}
%\begin{align}
\\
    \label{p_SU_relationship_average}
    \bar{R}_{{\sf{p}},k_{\sf s}} &= 1 - \bar{\xi}_{{\sf{p}},k_{\sf s}}^{\star} = 1 - \frac{1}{S}\sum\limits_{s=1}^{S} (\xi_{{\sf{p}},k_{\sf s}}^{\star})^{\left(s\right)},
%\end{align}
%\begin{align}
\\
    \label{p_CU_relationship_average}
    \bar{R}_{{\sf{p}},k_{\sf t}} &= 1 - \bar{\xi}_{{\sf{p}},k_{\sf t}}^{\star} = 1 - \frac{1}{S}\sum\limits_{s=1}^{S} (\xi_{{\sf{p}},k_{\sf t}}^{\star})^{\left(s\right)}.
\end{align}

Subsequently, we define the set of sampled equalizers as
$\mathbf{Q}_{\sf sat} \triangleq\{\mathbf{q}_{{\sf{spc}}, k_{\sf s}}, \mathbf{q}_{{\sf{cpc}}, k_{\sf s}},\mathbf{q}_{{\sf{p}}, k_{\sf s}}  \, \vert \, \forall k_{\sf s} \in \mathcal{K}_{\sf s}\}$ and $\mathbf{Q}_{\sf bs} \triangleq\{\mathbf{q}_{{\sf{spc}}, k_{\sf t}}, \mathbf{q}_{{\sf{lpc}}, k_{\sf t}},\mathbf{q}_{{\sf{p}}, k_{\sf t}}  \, \vert \, \forall k_{\sf t} \in \mathcal{K}_{\sf t}\}$ for notational compactness, where $\mathbf{q}_{{\sf{spc}}, k_{\sf s}} \triangleq \{q_{{\sf{spc}},k_{\sf s}}^{(s)} \, \vert \, \forall s \in \mathcal{S}\}$, $\mathbf{q}_{{\sf{spc}}, k_{\sf t}} \triangleq \{q_{{\sf{spc}},k_{\sf t}}^{(s)} \, \vert \, \forall s \in \mathcal{S}\}$, $\mathbf{q}_{{\sf{cpc}}, k_{\sf s}} \triangleq \{q_{{\sf{cpc}},k_{\sf s}}^{(s)} \, \vert \, \forall s \in \mathcal{S}\}$, $\mathbf{q}_{{\sf{lpc}}, k_{\sf t}} \triangleq \{q_{{\sf{lpc}},k_{\sf t}}^{(s)} \, \vert \, \forall s \in \mathcal{S}\}$, $\mathbf{q}_{{\sf{p}}, k_{\sf s}} \triangleq \{q_{{\sf{p}},k_{\sf s}}^{(s)} \, \vert \, \forall s \in \mathcal{S}\}$, and  $\mathbf{q}_{{\sf{p}}, k_{\sf t}} \triangleq \{q_{{\sf{p}},k_{\sf t}}^{(s)} \, \vert \, \forall s \in \mathcal{S}\}$. In an analogous manner, we define the set of sampled weights as $\mathbf{U}_{\sf sat} \triangleq\{\mathbf{u}_{{\sf{spc}}, k_{\sf s}}, \mathbf{u}_{{\sf{cpc}}, k_{\sf s}},\mathbf{u}_{{\sf{p}}, k_{\sf s}}  \, \vert \, \forall k_{\sf s} \in \mathcal{K}_{\sf s}\}$ and $\mathbf{U}_{\sf bs} \triangleq\{\mathbf{u}_{{\sf{spc}}, k_{\sf t}}, \mathbf{u}_{{\sf{lpc}}, k_{\sf t}},\mathbf{u}_{{\sf{p}}, k_{\sf t}}  \, \vert \, \forall k_{\sf t} \in \mathcal{K}_{\sf t}\}$, where $\mathbf{u}_{{\sf{spc}}, k_{\sf s}} \triangleq \{u_{{\sf{spc}},k_{\sf s}}^{(s)} \, \vert \, \forall s \in \mathcal{S}\}$, $\mathbf{u}_{{\sf{spc}}, k_{\sf t}} \triangleq \{u_{{\sf{spc}},k_{\sf t}}^{(s)} \, \vert \, \forall s \in \mathcal{S}\}$, $\mathbf{u}_{{\sf{cpc}}, k_{\sf s}} \triangleq \{u_{{\sf{cpc}},k_{\sf s}}^{(s)} \, \vert \, \forall s \in \mathcal{S}\}$, $\mathbf{u}_{{\sf{lpc}}, k_{\sf t}} \triangleq \{u_{{\sf{lpc}},k_{\sf t}}^{(s)} \, \vert \, \forall s \in \mathcal{S}\}$, $\mathbf{u}_{{\sf{p}}, k_{\sf s}} \triangleq \{u_{{\sf{p}},k_{\sf s}}^{(s)} \, \vert \, \forall s \in \mathcal{S}\}$, and $\mathbf{u}_{{\sf{p}}, k_{\sf t}} \triangleq \{u_{{\sf{p}},k_{\sf t}}^{(s)} \, \vert \, \forall s \in \mathcal{S}\}$. We then reformulate $\mathscr{P}_2$ as $\mathscr{P}_3$ based on the relationship in (\ref{spc_SU_relationship_average})--(\ref{p_CU_relationship_average}).
\begin{align}
\nonumber
{\mathscr P_3:} \quad
\maximize_{\substack{
\mathbf{W}, \mathbf{P}, \mathbf{c}_{\sf spc}, 
\mathbf{c}_{\sf cpc}, \mathbf{c}_{\sf lpc}, \boldsymbol{\alpha}_{{\sf p},{\sf sat}}, 
\boldsymbol{\alpha}_{{\sf p},{\sf bs}}, 
\\ R_{\sf min}, \mathbf{Q}_{\sf sat}, \mathbf{Q}_{\sf bs}, \mathbf{U}_{\sf sat}, \mathbf{U}_{\sf bs}}}
\, R_{\sf min}
\nonumber
\end{align}
\setcounter{equation}{65}\vspace{-3mm}
{\allowdisplaybreaks[4]
\begin{subequations}\label{condition3}
\begin{align}
\mathrm{s.t.}\,\,\,\,\,\,
\label{PF3CST1}
& 1- \bar{\xi}_{{\sf{spc}},k_{\sf{s}}}  \geq \sum_{j=1}^{K_{\sf s}}C_{{\sf spc},j}, \,\, \forall k_{\sf{s}} \in \mathcal{K}_{\sf{s}},\\
\label{PF3CST2}
& 1- \bar{\xi}_{{\sf{spc}},k_{\sf{t}}} \geq \sum_{j=1}^{K_{\sf s}}C_{{\sf spc},j}, \,\, \forall k_{\sf{t}} \in \mathcal{K}_{\sf{t}},\\
\label{PF3CST3}
& 1- \bar{\xi}_{{\sf{cpc}},k_{\sf{s}}} \geq \sum_{j=1}^{K_{\sf s}}C_{{\sf cpc},j}, \,\, \forall k_{\sf{s}} \in \mathcal{K}_{\sf{s}},\\
\label{PF3CST4}
& 1- \bar{\xi}_{{\sf{lpc}},k_{\sf{t}}} \geq \sum_{j=1}^{K_{\sf t}}C_{{\sf lpc},j}, \,\, \forall k_{\sf{t}} \in \mathcal{K}_{\sf{t}},\\
\label{PF3CST5}
& 1- \bar{\xi}_{{\sf{p}},k_{\sf{s}}} \geq \alpha_{{\sf{p}}, k_{\sf s}},  \,\, \forall k_{\sf s} \in \mathcal{K}_{\sf s}, \\
\label{PF3CST6}
& 1- \bar{\xi}_{{\sf{p}},k_{\sf{t}}} \geq \alpha_{{\sf{p}}, k_{\sf t}},  \,\, \forall k_{\sf t} \in \mathcal{K}_{\sf t}, \\
&\textrm{(\ref{PF1CST5})}, \,\, \textrm{(\ref{PF1CST6})}, \,\, \textrm{(\ref{PF1CST7})}, \,\, \textrm{(\ref{PF2CST1})}, \,\, \textrm{(\ref{PF2CST2})}, \,\,   \textrm{(\ref{PF2CST5})}, \nonumber
\end{align}
\end{subequations}
}where the sample-averaged augmented WMSEs are defined as follows: 
$\bar{\xi}_{{\sf spc},k_{\sf s}} \triangleq \frac{1}{S}\sum_{s=1}^{S}\xi_{{\sf spc},k_{\sf s}}^{(s)}$, 
$\bar{\xi}_{{\sf spc},k_{\sf t}} \triangleq \frac{1}{S}\sum_{s=1}^{S}\xi_{{\sf spc},k_{\sf t}}^{(s)}$, 
$\bar{\xi}_{{\sf cpc},k_{\sf s}} \triangleq \frac{1}{S}\sum_{s=1}^{S}\xi_{{\sf cpc},k_{\sf s}}^{(s)}$, 
$\bar{\xi}_{{\sf lpc},k_{\sf t}} \triangleq \frac{1}{S}\sum_{s=1}^{S}\xi_{{\sf lpc},k_{\sf t}}^{(s)}$, 
$\bar{\xi}_{{\sf p},k_{\sf s}} \triangleq \frac{1}{S}\sum_{s=1}^{S}\xi_{{\sf p},k_{\sf s}}^{(s)}$, and 
$\bar{\xi}_{{\sf p},k_{\sf t}} \triangleq \frac{1}{S}\sum_{s=1}^{S}\xi_{{\sf p},k_{\sf t}}^{(s)}$. Although $\mathscr{P}_3$ is non-convex with respect to the joint variable set, it is addressed via an alternating optimization technique that sequentially optimizes a subset of variables while keeping the remaining ones fixed. This procedure is described step-by-step below.

{\textbf{Step I}:} At the $[n]$-th iteration, $\mathbf{Q}_{\sf{sat}}^{[n]}$, $\mathbf{Q}_{\sf{bs}}^{[n]}$, $\mathbf{U}_{\sf{sat}}^{[n]}$, and $\mathbf{U}_{\sf{bs}}^{[n]}$ are first updated based on (\ref{WMSE_value}), using $\mathbf{W}^{[n-1]}$ and $\mathbf{P}^{[n-1]}$ obtained from the $[n-1]$-th iteration. Then, for all $k_{\sf s} \in \mathcal{K}_{\sf s}$, $k_{\sf t} \in \mathcal{K}_{\sf t}$, and $s \in \mathcal{S}$, the following variables are determined by the updated $\mathbf{Q}_{\sf{sat}}^{[n]}$, $\mathbf{Q}_{\sf{bs}}^{[n]}$, $\mathbf{U}_{\sf{sat}}^{[n]}$, and $\mathbf{U}_{\sf{bs}}^{[n]}$.
\begin{align}
 \label{step1_spc_beta}
    \delta^{(s)}_{{\sf{spc}},k_{\sf s}} = u^{(s)}_{{\sf{spc}},k_{\sf s}} \vert q^{(s)}_{{\sf{spc}},k_{\sf s}} \vert^2, \,\, \delta^{(s)}_{{\sf{spc}},k_{\sf t}} = u^{(s)}_{{\sf{spc}},k_{\sf t}} \vert q^{(s)}_{{\sf{spc}},k_{\sf t}} \vert^2, 
\end{align}
\begin{align}
 \label{step1_c_beta}
    \delta^{(s)}_{{\sf{cpc}},k_{\sf s}} = u^{(s)}_{{\sf{cpc}},k_{\sf s}} \vert q^{(s)}_{{\sf{cpc}},k_{\sf s}} \vert^2, \,\, \delta^{(s)}_{{\sf{lpc}},k_{\sf t}} = u^{(s)}_{{\sf{lpc}},k_{\sf t}} \vert q^{(s)}_{{\sf{lpc}},k_{\sf t}} \vert^2, 
\end{align}
\begin{align}
 \label{step1_p_beta}
    \delta^{(s)}_{{\sf{p}},k_{\sf s}} = u^{(s)}_{{\sf{p}},k_{\sf s}} \vert q^{(s)}_{{\sf{p}},k_{\sf s}} \vert^2, \,\, \delta^{(s)}_{{\sf{p}},k_{\sf t}} = u^{(s)}_{{\sf{p}},k_{\sf t}} \vert q^{(s)}_{{\sf{p}},k_{\sf t}} \vert^2, 
\end{align}
\begin{align}
 \label{step2_spc_phi_SU}
\boldsymbol{\Psi}^{(s)}_{{\sf{spc}},k_{\sf s}} = \delta^{(s)}_{{\sf{spc}},k_{\sf s}} \mathbf{f}^{(s)}_{k_{\sf s}} (\mathbf{f}^{(s)}_{k_{\sf s}})^{\sf{H}},
\end{align}
\begin{align}
 \label{step2_spc_phi_CU}
\boldsymbol{\Psi}^{(s)}_{{\sf{spc}},k_{\sf t}} = \delta^{(s)}_{{\sf{spc}},k_{\sf t}} \mathbf{h}_{k_{\sf t}} \mathbf{h}_{k_{\sf t}}^{\sf{H}}, 
(\boldsymbol{\Psi}^{\sf{intf}}_{{\sf{spc}},k_{\sf t}})^{(s)} \!\!=\! \delta^{(s)}_{{\sf{spc}},k_{\sf t}}
\mathbf{z}^{(s)}_{k_{\sf t}} (\mathbf{z}^{(s)}_{k_{\sf t}})^{\sf{H}}, 
\end{align}
\begin{align}
 \label{step2_cpc_phi_SU}
\boldsymbol{\Psi}^{(s)}_{{\sf{cpc}},k_{\sf s}} = \delta^{(s)}_{{\sf{cpc}},k_{\sf s}} \mathbf{f}^{(s)}_{k_{\sf s}} (\mathbf{f}^{(s)}_{k_{\sf s}})^{\sf{H}}, 
\end{align}
\begin{align}
 \label{step2_lpc_phi_CU}
 \boldsymbol{\Psi}^{(s)}_{{\sf{lpc}},k_{\sf t}} = \delta^{(s)}_{{\sf{lpc}},k_{\sf t}} \mathbf{h}_{k_{\sf t}} \mathbf{h}_{k_{\sf t}}^{\sf{H}}, \,\,
(\boldsymbol{\Psi}^{\sf{intf}}_{{\sf{lpc}},k_{\sf t}})^{(s)} \!\!=\! \delta^{(s)}_{{\sf{lpc}},k_{\sf t}}
\mathbf{z}^{(s)}_{k_{\sf t}} (\mathbf{z}^{(s)}_{k_{\sf t}})^{\sf{H}}, 
\end{align}
\begin{align}
 \label{step2_p_phi_SU}
\boldsymbol{\Psi}^{(s)}_{{\sf{p}},k_{\sf s}} = \delta^{(s)}_{{\sf{p}},k_{\sf s}}
\mathbf{f}^{(s)}_{k_{\sf s}} (\mathbf{f}^{(s)}_{k_{\sf s}})^{\sf{H}}, 
\end{align}
\begin{align}
 \label{step2_p_phi_CU}
 \boldsymbol{\Psi}^{(s)}_{{\sf{p}},k_{\sf t}} = \delta^{(s)}_{{\sf{p}},k_{\sf t}}
\mathbf{h}_{k_{\sf t}} \mathbf{h}_{k_{\sf t}}^{\sf{H}}, \,\,
(\boldsymbol{\Psi}^{\sf{intf}}_{{\sf{p}},k_{\sf t}})^{(s)} \!\!=\! \delta^{(s)}_{{\sf{p}},k_{\sf t}}
\mathbf{z}^{(s)}_{k_{\sf t}} (\mathbf{z}^{(s)}_{k_{\sf t}})^{\sf{H}}, 
\end{align}
\begin{align}
\label{step3_spc_omeg_SU}
\boldsymbol{\omega}^{(s)}_{{\sf{spc}},k_{\sf s}} = u^{(s)}_{{\sf{spc}},k_{\sf s}} (q^{(s)}_{{\sf{spc}},k_{\sf s}})^{*} \mathbf{f}^{(s)}_{k_{\sf s}}, %\,\, 
\end{align}
\begin{align}
\label{step3_spc_omeg_CU}
\boldsymbol{\omega}^{(s)}_{{\sf{spc}},k_{\sf t}} = u^{(s)}_{{\sf{spc}},k_{\sf t}} (q^{(s)}_{{\sf{spc}},k_{\sf t}})^{*} \mathbf{z}^{(s)}_{k_{\sf t}},
\end{align}
\begin{align}
 \label{step3_c_omeg_SU}
\boldsymbol{\omega}^{(s)}_{{\sf{cpc}},k_{\sf s}} = u^{(s)}_{{\sf{cpc}},k_{\sf s}} (q^{(s)}_{{\sf{cpc}},k_{\sf s}})^{*} \mathbf{f}^{(s)}_{k_{\sf s}}, %\,\, 
\end{align}
\begin{align}
\label{step3_c_omeg_CU}
\boldsymbol{\omega}^{(s)}_{{\sf{lpc}},k_{\sf t}} = u^{(s)}_{{\sf{lpc}},k_{\sf t}} (q^{(s)}_{{\sf{lpc}},k_{\sf t}})^{*} \mathbf{h}_{k_{\sf t}},
\end{align}
\begin{align}
 \label{step3_p_omeg}
  \boldsymbol{\omega}^{(s)}_{{\sf{p}},k_{\sf s}} = u^{(s)}_{{\sf{p}},k_{\sf s}} (q^{(s)}_{{\sf{p}},k_{\sf s}})^{*} \mathbf{f}^{(s)}_{k_{\sf s}}, \,\, \boldsymbol{\omega}^{(s)}_{{\sf{p}},k_{\sf t}} = u^{(s)}_{{\sf{p}},k_{\sf t}} (q^{(s)}_{{\sf{p}},k_{\sf t}})^{*} \mathbf{h}_{k_{\sf t}},
\end{align}
\begin{align}
 \label{step4_spc_v}
    \nu^{(s)}_{{\sf{spc}},k_{\sf s}} = \log_2 u^{(s)}_{{\sf{spc}},k_{\sf s}}, \,\, \nu^{(s)}_{{\sf{spc}},k_{\sf t}} = \log_2 u^{(s)}_{{\sf{spc}},k_{\sf t}},
\end{align}
\begin{align}
 \label{step4_c_v}
    \nu^{(s)}_{{\sf{cpc}},k_{\sf s}} = \log_2 u^{(s)}_{{\sf{cpc}},k_{\sf s}}, \,\, \nu^{(s)}_{{\sf{lpc}},k_{\sf t}} = \log_2 u^{(s)}_{{\sf{lpc}},k_{\sf t}},
\end{align}
\begin{align}
 \label{step4_p_v}
    \nu^{(s)}_{{\sf{p}},k_{\sf s}} = \log_2 u^{(s)}_{{\sf{p}},k_{\sf s}}, \,\, \nu^{(s)}_{{\sf{p}},k_{\sf t}} = \log_2 u^{(s)}_{{\sf{p}},k_{\sf t}}.
\end{align}

\begin{figure*}[!t]
%\vspace{-3mm}
%\noindent\rule{\textwidth}{.5pt}%\vskip3pt
\small \begin{align}
\nonumber
{\mathscr P_3^{[n]}:} \maximize_{\substack{
\mathbf{W}, \mathbf{P}, \mathbf{c}_{\sf spc}, 
\mathbf{c}_{\sf cpc}, \mathbf{c}_{\sf lpc}, 
 \boldsymbol{\alpha}_{{\sf p},{\sf sat}}, 
\boldsymbol{\alpha}_{{\sf p},{\sf bs}}, R_{\sf min}}} \,\, 
R_{\sf min}
\nonumber
\end{align}
\setcounter{equation}{83}\vspace{-3mm}
\begin{subequations}\label{condition4}
\begin{align}
\mathrm{s.t.}\,\,\,\,\,\,
\label{WMMSECST1_n}
&1 - \mathbf{w}_{\sf spc}^{\sf{H}}\bar{\boldsymbol{\Psi}}_{{\sf{spc}}, k_{\sf s}}\mathbf{w}_{\sf spc} - \mathbf{w}_{\sf cpc}^{\sf{H}}\bar{\boldsymbol{\Psi}}_{{\sf{spc}}, k_{\sf s}}\mathbf{w}_{\sf cpc} - \sum_{j=1}^{K_{\sf s}}\mathbf{w}_{j}^{\sf{H}}\bar{\boldsymbol{\Psi}}_{{\sf{spc}}, k_{\sf s}}\mathbf{w}_{j} \nonumber \\%[-3.3mm] 
& \,\,\, +2{\sf{Re}}\{\bar{\boldsymbol{\omega}}^{\sf{H}}_{{\sf{spc}},k_{\sf s}}\mathbf{w}_{\sf spc}\}  - \bar{\delta}_{{\sf{spc}},k_{\sf s}}\sigma_{{\sf{n}},k_{\sf s}}^{2} - \bar{u}_{{\sf{spc}},k_{\sf s}}  + \bar{\nu}_{{\sf{spc}},k_{\sf s}} \geq \sum_{j=1}^{K_{\sf s}} C_{{\sf spc},j},  \,\, \forall k_{\sf s} \in \mathcal{K}_{\sf s}, \\%[-1.3mm]
\label{WMMSECST2_n}
&1 - \mathbf{w}_{\sf spc}^{\sf{H}}\bar{\boldsymbol{\Psi}}_{{\sf{spc}}, k_{\sf t}}^{\sf intf}\mathbf{w}_{\sf spc} - \mathbf{w}_{\sf cpc}^{\sf{H}}\bar{\boldsymbol{\Psi}}_{{\sf{spc}}, k_{\sf t}}^{\sf intf}\mathbf{w}_{\sf cpc} - \sum_{j=1}^{K_{\sf s}}\mathbf{w}_{j}^{\sf{H}}\bar{\boldsymbol{\Psi}}_{{\sf{spc}}, k_{\sf t}}^{\sf intf}\mathbf{w}_{j} - \mathbf{p}_{\sf lpc}^{\sf{H}}\bar{\boldsymbol{\Psi}}_{{\sf{spc}}, k_{\sf t}}\mathbf{p}_{\sf lpc}   - \sum_{j=1}^{K_{\sf t}}\mathbf{p}_{j}^{\sf{H}}\bar{\boldsymbol{\Psi}}_{{\sf{spc}}, k_{\sf t}}\mathbf{p}_{j} \nonumber \\%[-3.3mm]
& \,\,\, + 2{\sf{Re}}\{\bar{\boldsymbol{\omega}}^{\sf{H}}_{{\sf{spc}},k_{\sf t}}\mathbf{w}_{\sf spc}\}  - \bar{\delta}_{{\sf{spc}},k_{\sf t}}\sigma_{{\sf{n}},k_{\sf t}}^{2} - \bar{u}_{{\sf{spc}},k_{\sf t}} + \bar{\nu}_{{\sf{spc}},k_{\sf t}} \geq \sum_{j=1}^{K_{\sf s}} C_{{\sf spc},j},  \,\, \forall k_{\sf t} \in \mathcal{K}_{\sf t}, \\%[-1.3mm]
\label{WMMSECST3_n}
&1 \!-\! \mathbf{w}_{\sf cpc}^{\sf{H}}\bar{\boldsymbol{\Psi}}_{{\sf{cpc}}, k_{\sf s}}\mathbf{w}_{\sf cpc} \!-\!\! \sum_{j=1}^{K_{\sf s}}\mathbf{w}_{j}^{\sf{H}}\bar{\boldsymbol{\Psi}}_{{\sf{cpc}}, k_{\sf s}}\mathbf{w}_{j} \nonumber \\%[-3.3mm]
& \,\,\, + 2{\sf{Re}}\{\bar{\boldsymbol{\omega}}^{\sf{H}}_{{\sf{cpc}},k_{\sf s}}\mathbf{w}_{\sf cpc}\} \!-\! \bar{\delta}_{{\sf{cpc}},k_{\sf s}}\sigma_{{\sf{n}},k_{\sf s}}^{2}  \!-\! \bar{u}_{{\sf{cpc}},k_{\sf s}} \!+\! \bar{\nu}_{{\sf{cpc}},k_{\sf s}} \!\geq\! \sum_{j=1}^{K_{\sf s}} C_{{\sf cpc},j}, \forall k_{\sf s} \in \mathcal{K}_{\sf s}, \\%[-1.3mm]
\label{WMMSECST4_n}
&1 - \mathbf{w}_{\sf cpc}^{\sf{H}}\bar{\boldsymbol{\Psi}}_{{\sf{lpc}}, k_{\sf t}}^{\sf intf}\mathbf{w}_{\sf cpc} - \sum_{j=1}^{K_{\sf s}}\mathbf{w}_{j}^{\sf{H}}\bar{\boldsymbol{\Psi}}_{{\sf{lpc}}, k_{\sf t}}^{\sf intf}\mathbf{w}_{j} - \mathbf{p}_{\sf{lpc}}^{\sf{H}}\bar{\boldsymbol{\Psi}}_{{\sf{lpc}}, k_{\sf t}}\mathbf{p}_{\sf{lpc}} - \sum_{j=1}^{K_{\sf t}}\mathbf{p}_{j}^{\sf{H}}\bar{\boldsymbol{\Psi}}_{{\sf{lpc}}, k_{\sf t}}\mathbf{p}_{j} \nonumber \\%[-3.3mm]
& \,\,\, +2{\sf{Re}}\{\bar{\boldsymbol{\omega}}^{\sf{H}}_{{\sf{lpc}},k_{\sf t}}\mathbf{p}_{\sf lpc}\}  - \bar{\delta}_{{\sf{lpc}},k_{\sf t}}\sigma_{{\sf{n}},k_{\sf t}}^{2} - \bar{u}_{{\sf{lpc}},k_{\sf t}} + \bar{\nu}_{{\sf{lpc}},k_{\sf t}} \geq \sum_{j=1}^{K_{\sf t}} C_{{\sf lpc},j},  \,\, \forall k_{\sf t} \in \mathcal{K}_{\sf t}, \\%[-1.3mm]
\label{WMMSECST5_n}
&1 - \sum_{j=1}^{K_{\sf s}} \mathbf{w}_{j}^{\sf{H}}\bar{\boldsymbol{\Psi}}_{{\sf{p}}, k_{\sf s}}\mathbf{w}_{j} + 2{\sf{Re}}\{\bar{\boldsymbol{\omega}}^{\sf{H}}_{{\sf{p}},k_{\sf s}}\mathbf{w}_{k_{\sf s}}\} - \bar{\delta}_{{\sf{p}},k_{\sf s}}\sigma_{{\sf{n}},k_{\sf s}}^{2}   - \bar{u}_{{\sf{p}},k_{\sf s}} + \bar{\nu}_{{\sf{p}},k_{\sf s}} \geq \alpha_{{\sf p},k_{\sf s}},  \,\, \forall k_{\sf s} \in \mathcal{K}_{\sf s}, \\%[-1.3mm]
\label{WMMSECST6_n}
&1 \!-\! \mathbf{w}_{\sf cpc}^{\sf{H}}\bar{\boldsymbol{\Psi}}_{{\sf{p}}, k_{\sf t}}^{\sf intf}\mathbf{w}_{\sf cpc} \!-\!\! \sum_{j=1}^{K_{\sf s}}\mathbf{w}_{j}^{\sf{H}}\bar{\boldsymbol{\Psi}}_{{\sf{p}}, k_{\sf t}}^{\sf intf}\mathbf{w}_{j} \!-\!\! \sum_{j=1}^{K_{\sf t}}\mathbf{p}_{j}^{\sf{H}}\bar{\boldsymbol{\Psi}}_{{\sf{p}}, k_{\sf t}}\mathbf{p}_{j} \nonumber \\%[-3.3mm]
& \,\,\, + 2{\sf{Re}}\{\bar{\boldsymbol{\omega}}^{\sf{H}}_{{\sf{p}},k_{\sf t}}\mathbf{p}_{k_{\sf t}}\}  \!-\! \bar{\delta}_{{\sf{p}},k_{\sf t}}\sigma_{{\sf{n}},k_{\sf t}}^{2}  \!-\! \bar{u}_{{\sf{p}},k_{\sf t}} \!+\! \bar{\nu}_{{\sf{p}},k_{\sf t}} \!\geq\! \alpha_{{\sf p},k_{\sf t}},  \forall k_{\sf t} \in \mathcal{K}_{\sf t}, \\
&\textrm{(\ref{PF1CST5})}, \,\, \textrm{(\ref{PF1CST6})}, \,\, \textrm{(\ref{PF1CST7})}, \,\, \textrm{(\ref{PF2CST1})}, \,\, \textrm{(\ref{PF2CST2})}, \,\,   \textrm{(\ref{PF2CST5})}. \nonumber
\end{align}
\end{subequations}
%\par\vspace{-3mm}
\noindent\rule{\textwidth}{.5pt}%\vskip3pt
%\vspace{-4.5mm}
\end{figure*}

Following this, the sample-averaged values, denoted as $\bar{\delta}_{{\sf{spc}},k_{\sf s}}$, $\bar{\delta}_{{\sf{spc}},k_{\sf t}}$, $\bar{\delta}_{{\sf{cpc}},k_{\sf s}}$, $\bar{\delta}_{{\sf{lpc}},k_{\sf t}}$, $\bar{\delta}_{{\sf{p}},k_{\sf s}}$, $\bar{\delta}_{{\sf{p}},k_{\sf t}}$,
$\bar{\boldsymbol{\Psi}}_{{\sf{spc}},k_{\sf s}}$, $\bar{\boldsymbol{\Psi}}_{{\sf{spc}},k_{\sf t}}$, $\bar{\boldsymbol{\Psi}}_{{\sf{spc}},k_{\sf t}}^{\sf intf}$,
$\bar{\boldsymbol{\Psi}}_{{\sf{cpc}},k_{\sf s}}$, $\bar{\boldsymbol{\Psi}}_{{\sf{lpc}},k_{\sf t}}$, $\bar{\boldsymbol{\Psi}}_{{\sf{lpc}},k_{\sf t}}^{\sf intf}$, $\bar{\boldsymbol{\Psi}}_{{\sf{p}},k_{\sf s}}$, $\bar{\boldsymbol{\Psi}}_{{\sf{p}},k_{\sf t}}$, $\bar{\boldsymbol{\Psi}}_{{\sf{p}},k_{\sf t}}^{\sf intf}$, 
$\bar{\boldsymbol{\omega}}_{{\sf{spc}},k_{\sf s}}$, $\bar{\boldsymbol{\omega}}_{{\sf{spc}},k_{\sf t}}$, $\bar{\boldsymbol{\omega}}_{{\sf{cpc}},k_{\sf s}}$, $\bar{\boldsymbol{\omega}}_{{\sf{lpc}},k_{\sf t}}$, $\bar{\boldsymbol{\omega}}_{{\sf{p}},k_{\sf s}}$, $\bar{\boldsymbol{\omega}}_{{\sf{p}},k_{\sf t}}$, 
$\bar{\nu}_{{\sf{spc}},k_{\sf s}}$, $\bar{\nu}_{{\sf{spc}},k_{\sf t}}$, $\bar{\nu}_{{\sf{cpc}},k_{\sf s}}$, $\bar{\nu}_{{\sf{lpc}},k_{\sf t}}$, $\bar{\nu}_{{\sf{p}},k_{\sf s}}$, $\bar{\nu}_{{\sf{p}},k_{\sf t}}$
$\bar{u}_{{\sf{spc}},k_{\sf s}}$, $\bar{u}_{{\sf{spc}},k_{\sf t}}$, $\bar{u}_{{\sf{cpc}},k_{\sf s}}$, $\bar{u}_{{\sf{lpc}},k_{\sf t}}$, $\bar{u}_{{\sf{p}},k_{\sf s}}$, and $\bar{u}_{{\sf{p}},k_{\sf t}}$ are obtained by averaging over $S$ realizations.

{\textbf{Step II}:} In this step, we update $\mathbf{W}^{[n]}$, $\mathbf{P}^{[n]}$, $\mathbf{c}_{\sf{spc}}^{[n]}$, $\mathbf{c}_{\sf{cpc}}^{[n]}$, $\mathbf{c}_{\sf{lpc}}^{[n]}$, $\boldsymbol{\alpha}_{{\sf{p}},{\sf sat}}^{[n]}$, $\boldsymbol{\alpha}_{{\sf{p}},{\sf bs}}^{[n]}$, and $R_{\sf min}^{[n]}$ based on the updated values of $\mathbf{Q}_{\sf{sat}}^{[n]}$, $\mathbf{Q}_{\sf{bs}}^{[n]}$, $\mathbf{U}_{\sf{sat}}^{[n]}$, and $\mathbf{U}_{\sf{bs}}^{[n]}$ obtained in \textbf{Step I}. Specifically, $\bar{\xi}_{{\sf{spc}}, k_{\sf s}}$, $\bar{\xi}_{{\sf{spc}}, k_{\sf t}}$, $\bar{\xi}_{{\sf{cpc}}, k_{\sf s}}$, $\bar{\xi}_{{\sf{lpc}}, k_{\sf t}}$, $\bar{\xi}_{{\sf{p}}, k_{\sf s}}$, and $\bar{\xi}_{{\sf{p}}, k_{\sf t}}$ in $\mathscr{P}_{3}$ are replaced with the corresponding sample-averaged values computed from \textbf{Step I}. Subsequently, $\mathbf{W}^{[n]}$, $\mathbf{P}^{[n]}$, $\mathbf{c}_{\sf{spc}}^{[n]}$, $\mathbf{c}_{\sf{cpc}}^{[n]}$, $\mathbf{c}_{\sf{lpc}}^{[n]}$, $\boldsymbol{\alpha}_{{\sf{p}},{\sf sat}}^{[n]}$, $\boldsymbol{\alpha}_{{\sf{p}},{\sf bs}}^{[n]}$, and $R_{\sf min}^{[n]}$ are obtained by solving the convex optimization problem $\mathscr{P}_{3}^{[n]}$ at the top of this page. %Since $\mathscr{P}_3^{[n]}$ is a convex problem, it can be solved using the interior-point method.

These steps are iteratively repeated until the absolute difference between two consecutive objective function values satisfies $\vert R_{\sf min}^{[n]} - R_{\sf min}^{[n-1]}\vert \le \epsilon$. Using the optimal solutions $\mathbf{W}^{[\star]}$, $\mathbf{P}^{[\star]}$, $\mathbf{c}_{\sf spc}^{[\star]}$, $\mathbf{c}_{\sf cpc}^{[\star]}$, and $\mathbf{c}_{\sf lpc}^{[\star]}$, we calculate the instantaneous spectral efficiency of each user according to \eqref{Rate_spc_SU}, \eqref{Rate_spc_CU}, \eqref{Rate_cpc_SU}, \eqref{Rate_lpc_CU}, \eqref{Rate_p_SU}, and \eqref{Rate_p_CU}. Notably, the objective function value of $\mathscr{P}_{3}^{[n]}$ forms a monotonically non-decreasing sequence over the iterations and is upper-bounded due to the given total transmit power constraint. Thanks to these features, the convergence of the proposed WMMSE-SAA-based MDP-RSMA precoder design algorithm is guaranteed. The primary computational complexity of the proposed algorithm arises from solving the convex optimization problem in \textbf{Step II}. The reformulated optimization problem $\mathscr{P}_{3}^{[n]}$ can be efficiently solved based on the interior-point method; the overall computational complexity of the proposed algorithm is characterized in big-O notation as $\mathcal{O}\big([2N_{\sf{s}}K_{\sf{s}} + 2N_{\sf{t}}K_{\sf{t}}]^{3.5} \log(\epsilon^{-1})\big)$.

%\begin{figure*}[!b]
%\vspace{-6mm}
%\centering
%\subfigure[]{\raisebox{-2.9mm}{\includegraphics[width=.672\columnwidth]{Fig_2_rev.pdf}}\label{Fig2}}
%\hfil
%\subfigure[]{\raisebox{-2.9mm}{\includegraphics[width=.662\columnwidth]{Fig_3_rev.pdf}}\label{Fig3}}
%\hfil
%\subfigure[]{\includegraphics[width=.673\columnwidth]{Fig_4_bar.pdf}\label{Fig4}}
%\subfigure[]{\raisebox{0.3mm}{\includegraphics[width=.686\columnwidth]{Fig_4_rev.pdf}}\label{Fig4}} 
%\hfil
%\vspace{-5mm}
%\caption{\tcb{Minimum spectral efficiency comparisons under various system settings: (a) As a function of 
%$\sigma_{\sf{e}}$ with $N_{\sf{t}}$ = 2 and $K$ = 20; %(b) As a function of %the number of users 
%$K$ with $N_{\sf{t}}$ = 2 and $\sigma_{\sf{e}}$ = 2; (c) As a function of %the number of transmit antenna feeds 
%$N_{\sf{t}}$ with $\sigma_{\sf{e}}$ = 1, for $K$ = 12 (upper subplot) and $K$ = 24 (lower subplot).}}
%\vspace{-6mm}
%\end{figure*}

% . Thanks to these features, the convergence of the proposed algorithm is guaranteed.
% WMMSE-RATE relationship에서 inequality를 equality로 만드는게 이전 step에서의 solution

%\vspace{-2mm}
 \section{Performance Evaluation}
%\vspace{-1mm}

%\vspace{-1mm}
\subsection{Simulation Setup}
%\vspace{-1mm}
We evaluate the performance of the proposed framework (denoted as {\sf MDP-RSMA-ISTN} in Figs.~\ref{result_1}--\ref{result_9} in comparison with several baseline schemes in various MDP-ISTN scenarios.
Unless explicitly stated otherwise, the simulation parameters are configured as follows.
The LEO satellite, located at an altitude of $\num{530}$ km, covers a service area with a radius of $\num{50}$ km. Within this coverage footprint, the satellite, equipped with $\num{16}$ transmit antenna pairs, i.e., $N_{\sf s}^{\sf x}=N_{\sf s}^{\sf y}=\num{4}$, serves $K_{\sf s}=\num{8}$ SUs over the S-band. The dual-LP terrestrial network consists of a terrestrial BS with a height
of $\num{30}$ m and $N_{\sf t}=\num{6}$ transmit antenna pairs. The BS
serves $K_{\sf t}=\num{4}$ CUs within its $\num{1}$ km coverage radius while
sharing the same bandwidth with the dual-CP satellite network. The SUs lie
outside the $\num{1}$ km terrestrial BS coverage area while remaining within
the $\num{50}$ km satellite coverage footprint.\footnote{Under the considered simulation setup, the terrestrial BS-to-SU channel gain rapidly decreases beyond the 1 km terrestrial coverage boundary due to the larger terrestrial path-loss exponent, supporting the assumed operating scenario.}
The transmit antenna gain for both the satellite and the BS is set to $G^{\sf Tx}_{\sf sat} = G^{\sf Tx}_{\sf bs} = 6$ dBi.
The carrier frequency, bandwidth, and path-loss exponent are set to $f_{\sf c}=2$ GHz, $B=5$ MHz, and $\eta=4$, respectively. 
For all SUs and CUs, the channel XPD factors for the LOS and NLOS paths of the satellite channel and for the terrestrial channel are set to ${\sf XPD}_{0}^{\sf sat}=15$ dB, ${\sf XPD}^{\sf sat}=5$ dB, and ${\sf XPD}^{\sf bs}=5$ dB, respectively \cite{Coldrey:vtc:08}.
%\refstepcounter{footnote}
%\edef\BSfootnotenumber{\number\value{footnote}}
%\footnotetext[\BSfootnotenumber]{%
%\color{blue}
%Under the considered simulation setup, the average powers of the satellite-to-SU and BS-to-SU channel gains scale with the inverse second and fourth powers of their respective link distances. Since the satellite-to-SU distance remains approximately $\num{530}$ km, their relative scaling factor is approximately $\num{9.5}$ at the terrestrial coverage boundary of $\num{1}$ km and rapidly increases to approximately $\num{152}$ and $\num{769}$ at $\num{2}$ km and $\num{3}$ km, respectively, demonstrating the consistency of the simulation settings with the intended operating scenario.}
Moreover, for all users, the Rician K-factor, receive antenna gain, system noise temperature, and variance of noise are set to $\kappa=15$ dB, $G^{\sf Rx}=0$ dBi, $T_{\sf sys}=290$ K, and $\sigma_{\sf n}^2=1$, respectively. The antenna orientation difference $\zeta$ between the satellite and the users is uniformly distributed over $[0,2\pi]$. The antenna XPD factors for the LEO satellite, terrestrial BS, and users are set to ${\sf XPD}^{\sf ant}=10$ dB \cite{castellanos2023linear,lee2026integrated}.
The number of SAA samples and the tolerance value are set to
$S=\num{1000}$ and $\epsilon=10^{-6}$, respectively. Numerical results are averaged over $\num{1000}$ channel realizations
with randomly generated user locations in each realization. The following benchmark schemes are considered for performance evaluation. The schemes in \cite{de2021irs,Sena:wcl:22,Sena:twc:23}, originally developed for intra-network interference management in dual-LP terrestrial networks, are adapted to the considered MDP-ISTN with dual-CP LEO satellite and dual-LP terrestrial sub-networks. The co-/cross-polar interference within each sub-network and the inter-network interference are jointly accounted for in their precoder optimization.
%\vspace{-1.2mm}
\begin{itemize}
\item {\textbf{RSMA-PD-ISTN} \cite{Sena:twc:23}}: This approach builds upon a conventional RSMA-based scheme that employs polarization diversity (PD) \cite{Sena:twc:23}. The satellite and terrestrial networks utilize the same radio resources. Based on the shared CSI, the common and private precoders of the satellite and terrestrial networks are jointly optimized to manage both intra- and inter-network interference. The problem can be solved by setting $\Vert\mathbf{w}_{\sf spc}\Vert^{2}=0$ in the proposed scheme.
\item { \textbf{RSMA-Dual PM-ISTN} \cite{Sena:wcl:22}}: 
This method builds upon a conventional RSMA-based scheme that employs polarization multiplexing (PM) \cite{Sena:wcl:22}. 
In each network, common and private streams are transmitted over separate polarization dimensions, eliminating the need for SIC at the receiver. For instance, the common stream is transmitted using RHCP, while the private streams are transmitted using LHCP at the satellite. An analogous transmit strategy is adopted at the BS using VP and HP. Although SIC is not required, SUs and CUs need to support both RHCP/LHCP and VP/HP, respectively. Based on the shared CSI, the common and private precoders of satellite and terrestrial networks are jointly optimized to manage both intra- and inter-network interference.
\item {\textbf{NOMA-ISTN} \cite{de2021irs}}: This approach builds upon a NOMA-based scheme that groups users according to their polarization states \cite{de2021irs}. The decoding order is independently determined within each polarization group based on the ascending order of the users’ channel strengths. For the dual-CP LEO satellite network, $\Vert\mathbf f_{k_{{\sf s},{\sf r}}}\Vert$ and $\Vert\mathbf f_{k_{{\sf s},{\sf l}}}\Vert$ are used within $\mathcal K_{{\sf s},{\sf r}}$ and $\mathcal K_{{\sf s},{\sf l}}$, respectively. Under imperfect CSIT, their empirical averages over the channel realizations used for SAA are employed. Similarly, $\Vert\mathbf h_{k_{{\sf t},{\sf v}}}\Vert$ and $\Vert\mathbf h_{k_{{\sf t},{\sf h}}}\Vert$ are used within $\mathcal K_{{\sf t},{\sf v}}$ and $\mathcal K_{{\sf t},{\sf h}}$, respectively, for the dual-LP terrestrial network. Based on the shared CSI, the satellite and terrestrial precoders are jointly optimized to manage both intra- and inter-network interference.
\item { \textbf{SDMA-ISTN}}: This method employs SDMA, where the satellite and terrestrial precoders are jointly optimized using the shared CSI between networks to manage both intra- and inter-network interference.
\item { \textbf{RSMA-OMA}}: The available radio resources are equally divided between the satellite and the terrestrial networks. Within each network, the same RSMA transmission strategy as in RSMA-PD-ISTN is employed to manage intra-network interference. Since the two networks operate over orthogonal radio resources, inter-network interference is eliminated, and the satellite and terrestrial precoders are optimized independently within their respective networks.
\item {\textbf{NOMA-OMA}}: The same polarization-based user grouping and decoding order as in NOMA-ISTN are employed within each network to manage intra-network interference. Since orthogonal radio resources are allocated to the satellite and terrestrial networks, inter-network interference is eliminated, and the corresponding precoders are optimized independently within the respective networks.
\item { \textbf{SDMA-OMA}}:
Within each network, the same SDMA transmission strategy as in SDMA-ISTN is employed to manage intra-network interference. Since the satellite and terrestrial networks use orthogonal radio resources, inter-network interference is eliminated, and their precoders are optimized independently within the respective networks.
\end{itemize}

%\vspace{-2mm}
\subsection{Simulation Results}
%\vspace{-1mm}

%\begin{figure}[!t]
%    \centering
%    %\hspace{-2mm}
%  {\includegraphics[width=.765\columnwidth]{Fig_rev/Perfect_dist_range.pdf}}
%    \vspace{-2mm}
%    \caption{\tcb{Minimum spectral efficiency versus the terrestrial BS-to-SU distance range under perfect CSIT for LEO satellite transmit powers of $10$ dBW (upper subplot) and $20$ dBW (lower subplot), where the transmit power of the terrestrial BS is set to $13$ dBW.}}
%    \label{result_0}
%    \vspace{-2mm}
%\end{figure}

\begin{figure}[!t]
\centering
    %\vspace{-1mm}
    \subfigure[Perfect CSIT]{\includegraphics[width=.48\columnwidth]{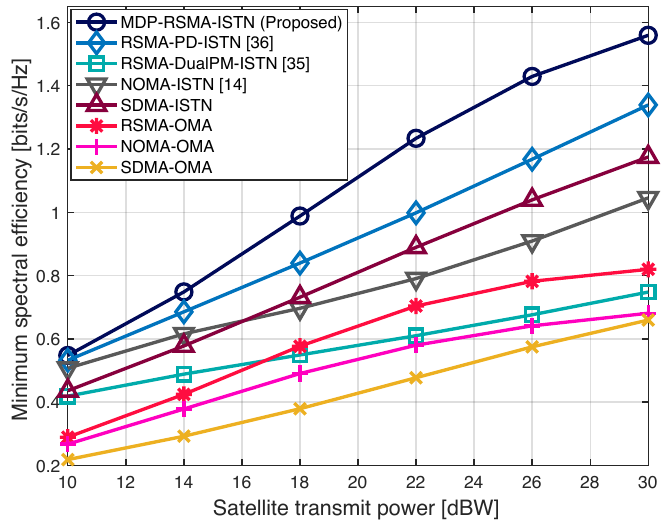}\label{result_1}}
    %\vspace{-2mm}
    \hfill
    \subfigure[Imperfect CSIT]{\includegraphics[width=.48\columnwidth]{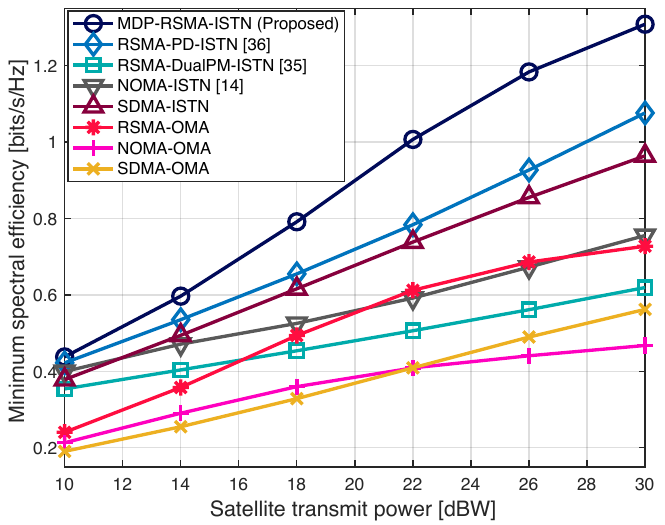}\label{result_2}}
%\vspace{-2mm}
\caption{Minimum spectral efficiency versus the LEO satellite transmit power, where the transmit power of the terrestrial BS is set to $13$ dBW.}
%\vspace{-2mm}
\end{figure}

\begin{figure}[!t]
    \centering
    %\hspace{-2mm}
    {\includegraphics[width=.6\columnwidth]{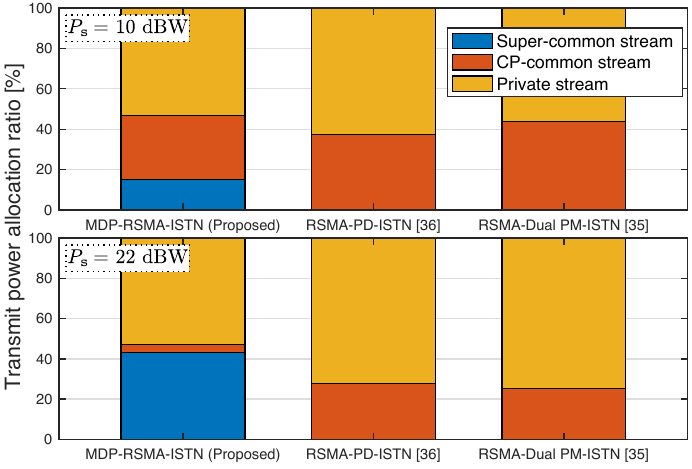}}
    %\vspace{-2mm}
    \caption{Comparison of transmit power allocation ratio for super-common, CP-common, and private streams in RSMA-based ISTN schemes under perfect CSIT, with the BS transmit power set to $13$ dBW. Upper and lower subplots correspond to satellite transmit powers of $10$ dBW and $22$ dBW, respectively.}
    \label{result_2.5}
    %\vspace{-3mm}
\end{figure}

We first compare the minimum spectral efficiency across all SUs and CUs achieved by the proposed {\sf MDP-RSMA-ISTN} framework with the benchmarks as a function of the LEO satellite transmit power under perfect and imperfect CSIT in Figs.~\ref{result_1} and \ref{result_2}, respectively.
The proposed framework consistently outperforms the conventional RSMA-based ISTN schemes as well as other MA techniques, yielding significant improvements in minimum spectral efficiency under both CSI conditions. 
For instance, when the LEO satellite transmit power is set to $\num{22}$ dBW, the proposed scheme with the super-common stream achieves $\num{23}\%$ and $\num{28}\%$ improvements in minimum spectral efficiency over {\sf RSMA-PD-ISTN} under perfect and imperfect CSIT, respectively. Meanwhile, {\sf RSMA-Dual PM-ISTN} performs worse than other RSMA-based ISTN schemes, since the orthogonality of the polarization domain among users cannot be effectively exploited in the precoder design. In {\sf RSMA-Dual PM-ISTN}, the polarization domain is primarily used to separate the common and private streams, thereby limiting the exploitation of user-level polarization orthogonality. Moreover, antenna and channel depolarizations induce residual interference between the common and private streams. Notably, the advantage of employing a super-common stream becomes increasingly significant as the satellite transmit power budget grows, as shown in Fig.~\ref{result_2.5}, because higher satellite transmit power exacerbates inter-network interference. 

\begin{figure}[!t]
\centering
    %\vspace{-1mm}
    \subfigure[Perfect CSIT]
{\includegraphics[width=.48\columnwidth]{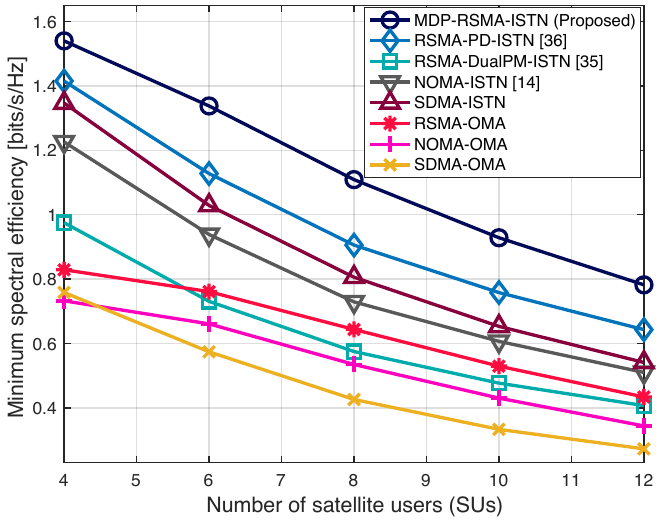}\label{result_3}}
    %\vspace{-2mm}
    \hfill
    \subfigure[Imperfect CSIT]{\includegraphics[width=.48\columnwidth]{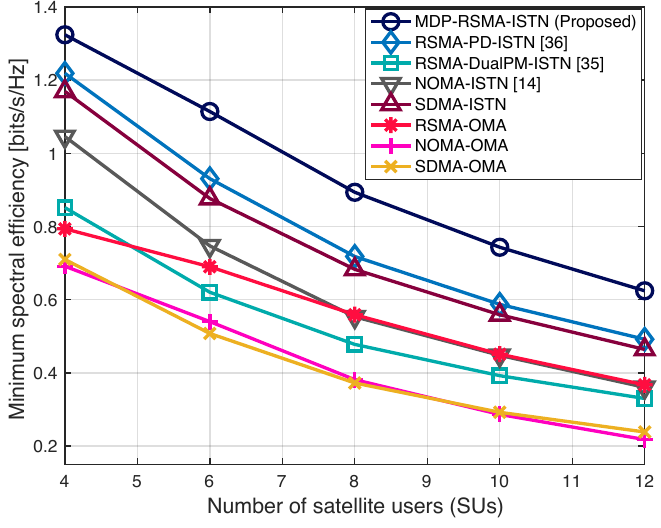}\label{result_4}}
%\vspace{-2mm}
\caption{Minimum spectral efficiency versus the number of SUs, where the transmit powers of the LEO satellite and the terrestrial BS are set to $20$ dBW and $13$ dBW, respectively.}
%\vspace{-2mm}
\end{figure}

%\begin{figure}[!t]
%\centering
%    \hspace{0.5mm}
%    \subfigure[Perfect CSIT]
%    {\includegraphics[width=.785\columnwidth]{Fig_rev/Perfect_XPD_ant.pdf}\label{result_ant_perfect}}
%    \vspace{-2mm}
%    \\
%    \hspace{0.5mm}
%    \subfigure[Imperfect CSIT]{\includegraphics[width=.785\columnwidth]{Fig/imperfect_XPD.pdf}\label{result_ant_imperfect}}
%\vspace{-2mm}
%\caption{Minimum spectral efficiency versus the antenna XPD, where the transmit power of the LEO satellite and terrestrial BS is set to $20$ dBW and $13$ dBW, respectively.}
%\vspace{-4.5mm}
%\end{figure}

Figs.~\ref{result_3} and \ref{result_4} show the minimum spectral efficiency as a function of the number of SUs under perfect and imperfect CSIT, respectively. 
The proposed scheme consistently outperforms the benchmarks regardless of the user density and the CSI accuracy. 
These results indicate that the proposed {\sf MDP-RSMA-ISTN} is well-suited to providing reliable communication services in dense user deployments of heterogeneously polarized satellite-terrestrial integrated networks.

\begin{figure}[!t]
    \centering
{\includegraphics[width=.48\columnwidth]{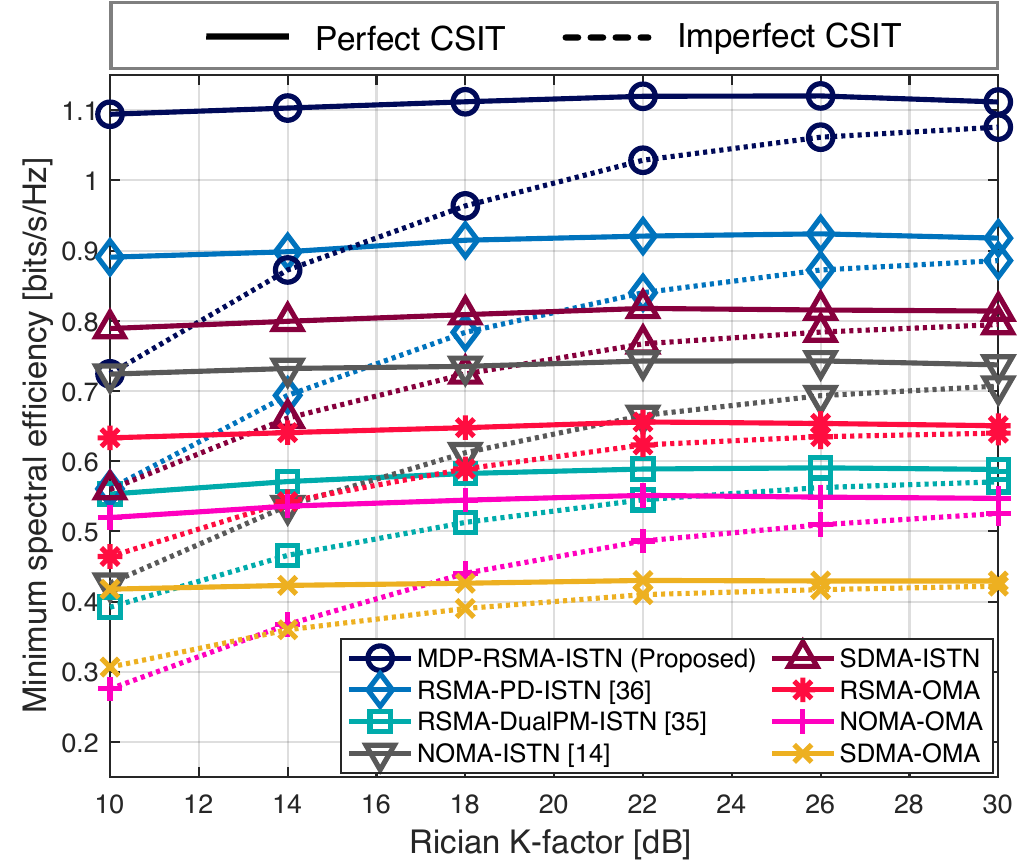}}
    %\vspace{-2mm}
    \caption{Minimum spectral efficiency versus the Rician K-factor of the satellite channel, where the transmit powers of the LEO satellite and the terrestrial BS are set to $20$ dBW and $13$ dBW, respectively.}
    \label{result_5}
    %\vspace{-3mm}
\end{figure}

\begin{figure}[!t]
    \centering  
    %\vspace{-0.5mm}
{\includegraphics[width=.48\columnwidth]{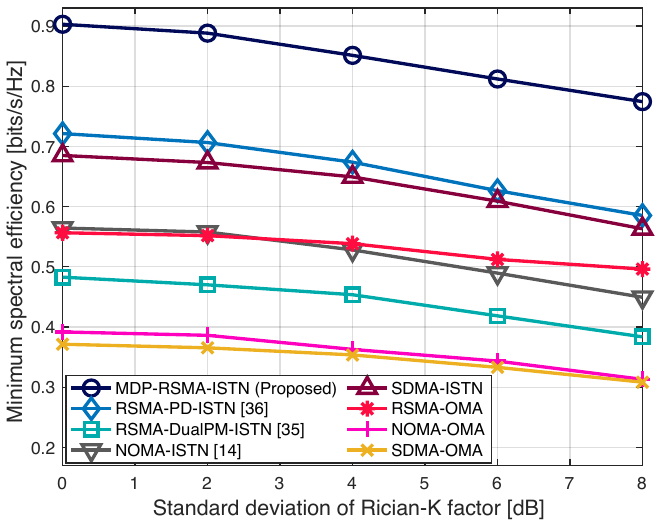}}
    %\vspace{-2mm}
    \caption{Minimum spectral efficiency versus standard deviation of the Rician-K factor under imperfect CSIT, where the transmit power of the LEO satellite and terrestrial BS is set to $20$ dBW and $13$ dBW, respectively.}
    \label{result_6}
    %\vspace{-3mm}
\end{figure}

We then compare the minimum spectral efficiency as a function of the Rician K-factor of the satellite channel through Fig.~\ref{result_5}. 
The solid line represents the performance under perfect CSIT, while the dotted line corresponds to imperfect CSIT.
As demonstrated in the figure, the proposed {\sf MDP-RSMA-ISTN} consistently outperforms the benchmark schemes regardless of the Rician K-factor and the CSI accuracy. 
In particular, when the Rician K-factor exceeds $\num{18}$ dB, the proposed scheme with imperfect CSIT achieves a higher minimum spectral efficiency than {\sf RSMA-PD-ISTN} with perfect CSIT.
The results indicate that the proposed scheme can provide reliable performance for MDP-ISTN despite variations in the Rician K-factor, while remaining robust against CSI uncertainty.
Moreover, the performance gap between the perfect and imperfect CSIT cases for all schemes diminishes as the Rician K-factor increases, since the satellite channel becomes increasingly dominated by the LOS component and thus, more dependent on the geometric relationship between the LEO satellite and users.

\begin{figure}[!t]
\centering
    %\hspace{0.5mm}
    \subfigure[Perfect CSIT]
{\includegraphics[width=.48\columnwidth]{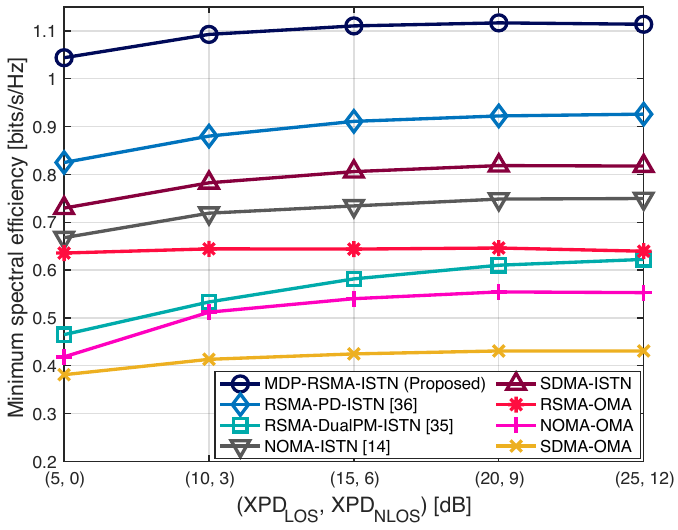}\label{result_7}}
    %\vspace{-2mm}
    \hfill
    %\hspace{0.5mm}
    \subfigure[Imperfect CSIT]{\includegraphics[width=.48\columnwidth]{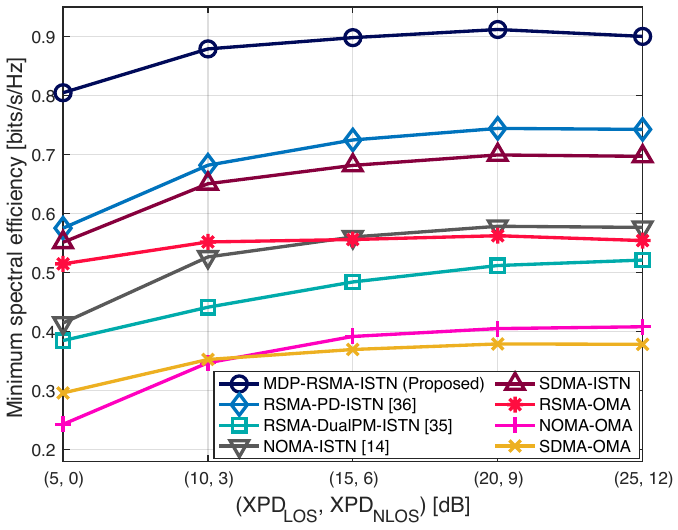}\label{result_8}}
%\vspace{-2mm}
\caption{Minimum spectral efficiency versus the channel XPD, where the transmit powers of the LEO satellite and the terrestrial BS are set to $20$ dBW and $13$ dBW, respectively. In the x-axis label, $\sf XPD_{\sf LOS}$ corresponds to $\sf XPD_0^{\sf sat}$, while $\sf XPD_{\sf NLOS}$ corresponds to $\sf XPD^{\sf sat}$ and $\sf XPD^{\sf bs}$.}
%\vspace{-2mm}
\end{figure}

\begin{figure}[!t]
    \centering
  {\includegraphics[width=.48\columnwidth]{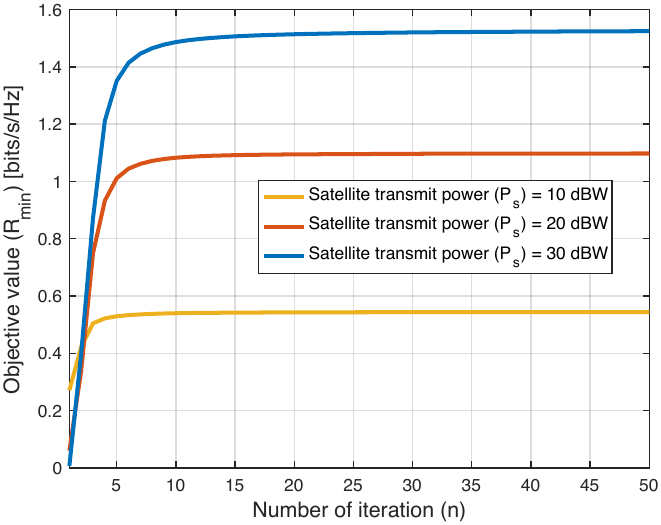}}
    %\vspace{-2mm}
    \caption{Convergence behaviors of the proposed {\sf MDP-RSMA-ISTN} under imperfect
CSIT, with the transmit power of the terrestrial BS $13$ dBW.}
    \label{result_9}
    %\vspace{-3mm}
\end{figure}

Fig.~\ref{result_6} demonstrates the minimum spectral efficiency as a function of the standard deviation of the Rician K-factor under imperfect CSIT. Specifically, the Rician K-factor of each user is independently generated from $\kappa \sim \mathcal{N}\left(\mu_{\kappa},\sigma_{\kappa}^{2}\right)$, where $\mu_{\kappa}=\num{15}$ dB and $\sigma_{\kappa}$ varies from $\num{0}$ to $\num{8}$ dB. As $\sigma_{\kappa}$ increases, the ratios of the LOS component power to the NLOS component power become more heterogeneous across users, reflecting increasingly heterogeneous scattering conditions and levels of CSIT uncertainty. The proposed framework consistently outperforms the benchmark schemes regardless of the value of $\sigma_{\kappa}$. Particularly, the performance gain of the proposed framework over {\sf RSMA-PD-ISTN} increases from $\num{25} \%$ at $\sigma_{\kappa}=\num{0}$ dB to $\num{32} \%$ at $\sigma_{\kappa}=\num{8}$ dB, demonstrating its robustness to heterogeneous CSIT uncertainty across users.

Figs.~\ref{result_7} and~\ref{result_8} show the minimum spectral efficiency as a function of the channel XPD under perfect and imperfect CSIT, respectively. 
As the channel XPD increases, the performance of all schemes improves since the higher channel XPD causes less channel depolarization, i.e., enhancing orthogonality in the polarization domain.
Notably, the proposed scheme consistently outperforms the benchmarks regardless of the channel XPD and the CSIT accuracy. 
%Besides, the performance of {\sf RSMA-Dual PM-ISTN} is more sensitive to the channel XPD than other schemes, as the common and private stream transmissions are confined to certain polarization domains, hindering flexibility for the variations of channel XPD.

Fig.~\ref{result_9} shows the convergence behavior of the proposed WMMSE-SAA-based MDP-RSMA precoder design algorithm under imperfect CSIT. For all considered LEO satellite transmit powers, the objective value increases monotonically and converges, consistent with the theoretically established non-decreasing and upper-bounded objective sequence.

%For example, when the BS transmit power is $16$ dBW, the proposed scheme with the super-common message achieves a $28\%$ and $32\%$ increase in minimum spectral efficiency over {\sf RSMA-PD-ISTN} under perfect and imperfect CSIT, respectively. 

%\vspace{-3mm}
\section{Conclusion}
%\vspace{-1mm}
This paper has investigated the MDP-RSMA framework for heterogeneously polarized satellite-terrestrial integrated networks that combine inter-network RS with a super-common message and intra-network RS with common/private messages.
%We have modeled the MDP-ISTN, accounting for polarization mismatch caused by time-varying antenna orientation differences between the LEO satellite and users, as well as channel depolarization.
We have modeled the practical MDP-ISTN, accounting for Faraday rotation, polarization mismatch, channel depolarization, and antenna depolarization.
Thereafter, the joint optimization of beamforming vector and power allocation has been addressed to maximize the minimum spectral efficiency among all SUs and CUs.
Numerical results across diverse MDP-ISTN scenarios have demonstrated the effectiveness of the proposed framework in jointly managing co- and cross-polar interference as well as inter-network interference.
The results validate that the proposed scheme is a promising solution for next-generation wireless networks, enabling reliable communication service and efficient utilization of dual-polarization.

%Given its demonstrated superiority, the proposed framework shows significant promise for ISTN, providing flexible interference management and enhancing spectral efficiency through dual-polarization utilization.
%Based on this superiority, our proposed framework is promising for ISTN to enhance spectral efficiency with flexible interference management. 
%Therefore, our proposed framework in the polarized ISTN is envisioned to fulfill ubiquitous global connectivity for next-generation wireless networks.
%\vspace{-3mm}
%\vspace{-2.5mm}
\appendix[Derivation of (\ref{RP_channel})]
\label{Appendix}
%\vspace{-1mm}
Throughout the appendix, for notational simplicity, we write $a^{[n_{\sf s}^{\sf x},n_{\sf s}^{\sf y}]}(\theta_{k_{\sf s}},\varphi_{k_{\sf s}})$ as $a_{k_{\sf s}}^{[n_{\sf s}^{\sf x},n_{\sf s}^{\sf y}]}$, for all $n_{\sf s}^{\sf x}\in \{1,\cdots,N_{\sf s}^{\sf x}\}$ and $n_{\sf s}^{\sf y}\in \{1,\cdots,N_{\sf s}^{\sf y}\}$. Then, the effective channel vector between the LEO satellite and the $k_{{\sf s},{\sf r}}$-th SU is given by
\begin{align}
\label{eff_ch_vec}
    & \mathbf{f}_{k_{{\sf{s}},{\sf{r}}}} =  \begin{bmatrix}
(\hat{\boldsymbol{\rho}}_{k_{{\sf{s}},{\sf{r}}}}^{\sf{r}})^{\sf H} \big( \mathbf{G}_{k_{\sf s}} \cdot a_{k_{\sf{s}}}^{[1,1]} \big) \hat{\boldsymbol{\rho}}_{\sf sat}^{\sf{r}}
\\
(\hat{\boldsymbol{\rho}}_{k_{{\sf{s}},{\sf{r}}}}^{\sf{r}})^{\sf H} \big( \mathbf{G}_{k_{\sf s}} \cdot a_{k_{\sf{s}}}^{[1,1]}  \big) \hat{\boldsymbol{\rho}}_{\sf sat}^{\sf{l}}
\\
\vdots
\\
(\hat{\boldsymbol{\rho}}_{k_{{\sf{s}},{\sf{r}}}}^{\sf{r}})^{\sf H} \big( \mathbf{G}_{k_{\sf s}} \cdot a_{k_{\sf{s}}}^{[N_{\sf s}^{\sf x},N_{\sf s}^{\sf y}]}\big) \hat{\boldsymbol{\rho}}_{\sf sat}^{\sf{r}}
\\
(\hat{\boldsymbol{\rho}}_{k_{{\sf{s}},{\sf{r}}}}^{\sf{r}})^{\sf H} \big( \mathbf{G}_{k_{\sf s}} \cdot aa_{k_{\sf{s}}}^{[N_{\sf s}^{\sf x},N_{\sf s}^{\sf y}]} \big) \hat{\boldsymbol{\rho}}_{\sf sat}^{\sf{l}}
\end{bmatrix} \in \mathbb{C}^{2N_{\sf{s}}\times1}.
\end{align}
The effective channel vector (\ref{eff_ch_vec}) can be equivalently obtained by vectorizing the following matrix in a column-wise manner.
\begin{align}
\label{vec_mat}
\begin{bmatrix}
(\hat{\boldsymbol{\rho}}_{k_{{\sf{s}},{\sf{r}}}}^{\sf{r}})^{\sf H} \big( \mathbf{G}_{k_{\sf s}} \!\cdot\! a_{k_{\sf s}}^{[1,1]}  \big) \hat{\boldsymbol{\rho}}_{\sf sat}^{\sf{r}}
& \!\!\!\cdots\!\!\! &
(\hat{\boldsymbol{\rho}}_{k_{{\sf{s}},{\sf{r}}}}^{\sf{r}})^{\sf H} \big( \mathbf{G}_{k_{\sf s}} \!\cdot\! a_{k_{\sf s}}^{[N_{\sf s}^{\sf x},N_{\sf s}^{\sf y}]}\big) \hat{\boldsymbol{\rho}}_{\sf sat}^{\sf{r}}
\\
(\hat{\boldsymbol{\rho}}_{k_{{\sf{s}},{\sf{r}}}}^{\sf{r}})^{\sf H} \big( \mathbf{G}_{k_{\sf s}} \!\cdot\! a_{k_{\sf s}}^{[1,1]} \big) \hat{\boldsymbol{\rho}}_{\sf sat}^{\sf{l}}
& \!\!\!\cdots\!\!\! &
(\hat{\boldsymbol{\rho}}_{k_{{\sf{s}},{\sf{r}}}}^{\sf{r}})^{\sf H} \big( \mathbf{G}_{k_{\sf s}} \!\cdot\! a_{k_{\sf s}}^{[N_{\sf s}^{\sf x},N_{\sf s}^{\sf y}]}\big) \hat{\boldsymbol{\rho}}_{\sf sat}^{\sf{l}}
\end{bmatrix}, \nonumber 
\end{align}
where the above matrix can be rewritten as
\begin{align}
& \begin{bmatrix}
(\hat{\boldsymbol{\rho}}_{k_{{\sf{s}},{\sf{r}}}}^{\sf{r}})^{\sf H} \big( \mathbf{G}_{k_{\sf s}} \!\cdot\! a_{k_{\sf s}}^{[1,1]}  \big) \hat{\boldsymbol{\rho}}_{\sf sat}^{\sf{r}}
& \!\!\!\cdots\!\!\! &
(\hat{\boldsymbol{\rho}}_{k_{{\sf{s}},{\sf{r}}}}^{\sf{r}})^{\sf H} \big( \mathbf{G}_{k_{\sf s}} \!\cdot\! a_{k_{\sf s}}^{[N_{\sf s}^{\sf x},N_{\sf s}^{\sf y}]}\big) \hat{\boldsymbol{\rho}}_{\sf sat}^{\sf{r}}
\\
(\hat{\boldsymbol{\rho}}_{k_{{\sf{s}},{\sf{r}}}}^{\sf{r}})^{\sf H} \big( \mathbf{G}_{k_{\sf s}} \!\cdot\! a_{k_{\sf s}}^{[1,1]} \big) \hat{\boldsymbol{\rho}}_{\sf sat}^{\sf{l}}
& \!\!\!\cdots\!\!\! &
(\hat{\boldsymbol{\rho}}_{k_{{\sf{s}},{\sf{r}}}}^{\sf{r}})^{\sf H} \big( \mathbf{G}_{k_{\sf s}} \!\cdot\! a_{k_{\sf s}}^{[N_{\sf s}^{\sf x},N_{\sf s}^{\sf y}]}\big) \hat{\boldsymbol{\rho}}_{\sf sat}^{\sf{l}}
\end{bmatrix}
\nonumber \\
& = \Bigg(\!\! 
\begin{bmatrix}
(\hat{\boldsymbol{\rho}}_{k_{{\sf{s}},{\sf{r}}}}^{\sf{r}})^{\sf H} \mathbf{G}_{k_{\sf s}}  \!\cdot\! a_{k_{\sf s}}^{[1,1]}
\\
\vdots
\\
(\hat{\boldsymbol{\rho}}_{k_{{\sf{s}},{\sf{r}}}}^{\sf{r}})^{\sf H} \mathbf{G}_{k_{\sf s}} \!\cdot\! a_{k_{\sf s}}^{[N_{\sf s}^{\sf x},N_{\sf s}^{\sf y}]}
\end{bmatrix}
\!\!
[\hat{\boldsymbol{\rho}}_{\sf sat}^{\sf{r}}, \hat{\boldsymbol{\rho}}_{\sf sat}^{\sf{l}}] \!\! \Bigg)^{\!\! {\sf T}} \nonumber \\
& = \Bigg(\!\!
\begin{bmatrix}
(\hat{\boldsymbol{\rho}}_{k_{{\sf{s}},{\sf{r}}}}^{\sf{r}})^{\sf H}  \!\!\!&\!\!\! \mathbf{0}_{1\times2} \!\!\!&\!\!\! \cdots \!\!\!&\!\!\! \mathbf{0}_{1\times2}
\\
\mathbf{0}_{1\times2} \!\!\!&\!\!\! (\hat{\boldsymbol{\rho}}_{k_{{\sf{s}},{\sf{r}}}}^{\sf{r}})^{\sf H} \!\!\!&\!\!\! \cdots \!\!\!&\!\!\! \mathbf{0}_{1\times2}
\\
\mathbf{0}_{1\times2} \!\!\!&\!\!\! \cdots \!\!\!&\!\!\! \ddots \!\!\!&\!\!\! \mathbf{0}_{1\times2}
\\
\mathbf{0}_{1\times2} \!\!\!&\!\!\! \cdots \!\!\!&\!\!\! \mathbf{0}_{1\times2} \!\!&\!\! (\hat{\boldsymbol{\rho}}_{k_{{\sf{s}},{\sf{r}}}}^{\sf{r}})^{\sf H} 
\end{bmatrix}
\!\!\!
\begin{bmatrix}
\mathbf{G}_{k_{\sf s}}  \!\cdot\! a_{k_{\sf s}}^{[1,1]}
\\
\vdots
\\
\mathbf{G}_{k_{\sf s}}  \!\cdot\! a_{k_{\sf s}}^{[N_{\sf s}^{\sf x},N_{\sf s}^{\sf y}]}
\end{bmatrix}
\!\!
[\hat{\boldsymbol{\rho}}_{\sf sat}^{\sf{r}}, \hat{\boldsymbol{\rho}}_{\sf sat}^{\sf{l}}] \!\! \Bigg)^{\!\! {\sf T}} \nonumber \\
& = \left((\mathbf{I}_{N_{\sf{s}}} \otimes (\hat{\boldsymbol{\rho}}_{k_{{\sf{s}},{\sf{r}}}}^{\sf{r}})^{\sf H}) (\mathbf{a}_{k_{\sf{s}}} \otimes \mathbf{G}_{k_{\sf{s}}}) 
[\hat{\boldsymbol{\rho}}_{\sf sat}^{\sf{r}}, \hat{\boldsymbol{\rho}}_{\sf sat}^{\sf{l}}]\right)^{\sf T} \nonumber \\
& = \left((\mathbf{I}_{N_{\sf{s}}} \otimes (\hat{\boldsymbol{\rho}}_{k_{{\sf{s}},{\sf{r}}}}^{\sf{r}})^{\sf H}) \mathbf{F}_{k_{{\sf{s}},{\sf{r}}}} [\hat{\boldsymbol{\rho}}_{\sf sat}^{\sf{r}}, \hat{\boldsymbol{\rho}}_{\sf sat}^{\sf{l}}]\right)^{\sf{T}} \in \mathbb{C}^{2 \times N_{\sf{s}}}.
\end{align}
Thus, the effective channel vector (\ref{eff_ch_vec}) is reformulated as 
\begin{align}
    \mathbf{f}_{k_{{\sf{s}},{\sf{r}}}} = {\sf vec}\left(\left((\mathbf{I}_{N_{\sf{s}}} \otimes (\hat{\boldsymbol{\rho}}_{k_{{\sf{s}},{\sf{r}}}}^{\sf{r}})^{\sf H}) \mathbf{F}_{k_{{\sf{s}},{\sf{r}}}} [\hat{\boldsymbol{\rho}}_{{\sf{sat}}}^{\sf{r}}, \hat{\boldsymbol{\rho}}_{{\sf{sat}}}^{\sf{l}}]\right)^{\sf{T}}\right),
\end{align}
and this completes the derivation. An analogous procedure is used to derive equations (\ref{LP_channel}), (\ref{VP_channel}), (\ref{HP_channel}), (\ref{VP_TN_channel}), and (\ref{HP_TN_channel}).

%\vspace{-3mm}
\bibliographystyle{IEEEtran}

\bibliography{Ref_polar}

@IEEEtranBSTCTL{IEEEexample:BSTcontrol,
  CTLdash_repeated_names = "no"
}

@string{vtc="Proc. IEEE Veh. Technol. Conf. (VTC)"}

@string{globecom="Proc. IEEE Global Commun. Conf. (GLOBECOM)"}

@string{globecomwrk="Proc. IEEE Global Commun. Conf. Workshops (GLOBECOM Workshops) "}

@string{milcom="Proc. IEEE Mil. Commun. Conf. (MILCOM)"}

@string{icc="Proc. IEEE Int. Conf. Commun. (ICC)"}

@string{jsac="IEEE J. Sel. Areas Commun."}

@string{tvt="IEEE Trans. Veh. Technol."}

@string{tsp="IEEE Trans. Signal Process."}

@string{tcom="IEEE Trans. Commun."}

@string{twc="IEEE Trans. Wireless Commun."}

@string{eurawc="EURASIP J. Wireless Commun. Netw."}

@string{tccn="IEEE Trans. Cogn. Commun. Netw."}

@string{access="IEEE Access"}

@string{surtut="IEEE Commun. Surv. Tutor."}

@string{commmag="IEEE Commun. Mag."}

@string{wcl = "IEEE Wireless Commun. Lett."}

@ARTICLE{9164895,
  author={Crocker, Dylan A. and Scott, Waymond R.},
  journal={IEEE Open J. Antennas Propag.}, 
  title={An Unbalanced Sinuous Antenna for Near-Surface Polarimetric Ground-Penetrating Radar}, 
  year={2020},
  volume={1},
  number={},
  pages={435-447}}

@ARTICLE{11393662,
  author={Yang, Yu-Hang and others},
  journal={IEEE Trans. Antennas Propag.}, 
  title={A Dual-Circularly Polarized Phased Array Antenna With Enhanced Polarization Purity for Multiazimuth Plane Scanning}, 
  year={2026},
  volume={74},
  number={5},
  pages={4067-4080}}

@article{sun2025dual,
  title={Dual-polarized stacked metasurface transceiver design with rate splitting for next-generation wireless networks},
  author={Sun, Yifu and others},
  journal=jsac,
  volume={43},
  number={3},
  pages={811--833},
  year={2025},
  publisher={IEEE}
}

@ARTICLE{10908461,
  author={Kim, Doseon and others},
  journal=twc, 
  title={Distributed Precoding for Satellite-Terrestrial Integrated Networks Without Sharing {CSIT}: {A} Rate-Splitting Approach}, 
  year={2025},
  volume={24},
  number={6},
  pages={4750-4766}}

@article{de2021irs,
  title={{IRS}-assisted massive {MIMO}-{NOMA} networks: Exploiting wave polarization},
  author={De Sena, Arthur Sousa and others},
  journal=twc,
  volume={20},
  number={11},
  pages={7166--7183},
  year={2021},
  publisher={IEEE}
}

@article{naeem2025polarization,
  title={Polarization-Aware Antenna Selection for Joint Radar and Communication in {XL-MIMO} Systems},
  author={Naeem, Ahmed and Afeef, Liza and Arslan, H{\"u}seyin},
  journal={IEEE Trans. Aerosp. Electron. Syst.},
  pages={1--13},
  year={2025},
  publisher={IEEE}
}

@article{castellanos2023linear,
  title={Linear polarization optimization for wideband {MIMO} systems with reconfigurable arrays},
  author={Castellanos, Miguel R and Heath, Robert W},
  journal=twc,
  volume={23},
  number={3},
  pages={2282--2295},
  year={2023},
  publisher={IEEE}
}

@article{lee2026integrated,
  title={Integrated polarimetric sensing and communication with polarization-reconfigurable arrays},
  author={Lee, Byunghyun and others},
  journal=twc,
  volume={25},
  pages={10618--10634},
  year={2026},
  publisher={IEEE}
}

@article{rezaie2022deep,
  title={A deep learning approach to location- and orientation-aided {3D} beam selection for mm{W}ave communications},
  author={Rezaie, Sajad and De Carvalho, Elisabeth and Manch{\'o}n, Carles Navarro},
  journal=twc,
  volume={21},
  number={12},
  pages={11110--11124},
  year={2022},
  publisher={IEEE}
}

@article{zhang2025metars,
  title={Meta{RS}: {A} Self-Intelligent Rate-Splitting Approach for Co-Existing Space-Air-Ground Integrated Networks},
  author={Zhang, Shengyu and others},
  journal=twc,
  year={2025},
  publisher={IEEE}
}

@article{zhang2024interference,
  title={Interference management in space-air-ground integrated networks with fully distributed rate-splitting multiple access},
  author={Zhang, Shengyu and others},
  journal=twc,
  volume={24},
  number={1},
  pages={149--164},
  year={2024},
  publisher={IEEE}
}

@article{xu2024distributed,
  title={Distributed rate-splitting multiple access for multilayer satellite communications},
  author={Xu, Yunnuo and others},
  journal=tcom,
  volume={72},
  number={10},
  pages={6131--6144},
  year={2024},
  publisher={IEEE}
}

@INPROCEEDINGS{11432517,
  author={Lee, Juhwan and Lee, Jungwoo and Shin, Wonjae},
  booktitle=globecom, 
  title={Rate-Splitting for Integrated Satellite-Terrestrial Networks with Mixed Dual-Polarization}, 
  year={2025},
  volume={},
  number={},
  pages={5850-5855},
  doi={10.1109/GLOBECOM59602.2025.11432517}}

@ARTICLE{10559954,
  author={Toka, Mesut and others},
  journal={IEEE Commun. Mag.}, 
  title={{RIS}-Empowered {LEO} Satellite Networks for {6G}: Promising Usage Scenarios and Future Directions}, 
  year={2024},
  volume={62},
  number={11},
  pages={128-135},
  doi={10.1109/MCOM.002.2300554}}

@article{jamshed2025tutorial,
  title={A Tutorial on Non-Terrestrial Networks: Towards Global and Ubiquitous {6G} Connectivity},
  author={Jamshed, Muhammad Ali and others},
  journal={Foundations and Trends{\textregistered} in Networking},
  volume={14},
  number={3},
  pages={160--253},
  year={2025},
  publisher={Now Publishers, Inc.}
}

@article{3GPP_ntn_channel,
	title={{S}tudy on {NR} to support non-terrestrial networks ({NTN}) ({R}elease 15)},
    journal={3GPP TR 38.811},
	author={3GPP},
	year={v15.4.0, 2020},
}

@ARTICLE{6829945,
  author={Kawamoto, Yuichi and others},
  journal=commmag, 
  title={Prospects and challenges of context-aware multimedia content delivery in cooperative satellite and terrestrial networks}, 
  year={2014},
  volume={52},
  number={6},
  pages={55-61},
  doi={10.1109/MCOM.2014.6829945}}

@ARTICLE{5962379,
  author={Kwon, Seok-Chul and Stuber, Gordon L.},
  journal=tvt, 
  title={Geometrical Theory of Channel Depolarization}, 
  year={2011},
  volume={60},
  number={8},
  pages={3542-3556},
  doi={10.1109/TVT.2011.2163094}}

@ARTICLE{1033685,
  author={Nabar, R.U. and others},
  journal=tsp, 
  title={Performance of multiantenna signaling techniques in the presence of polarization diversity}, 
  year={2002},
  volume={50},
  number={10},
  pages={2553-2562},
  doi={10.1109/TSP.2002.803322}}

@ARTICLE{1603707,
  author={Shafi, M. and others},
  journal=jsac, 
  title={Polarized {MIMO} channels in {3-D}: Models, measurements and mutual information}, 
  year={2006},
  volume={24},
  number={3},
  pages={514-527},
  doi={10.1109/JSAC.2005.862398}}

@ARTICLE{10787138,
  author={Kim, Seungnyun and others},
  journal=jsac, 
  title={Cell-Free Massive Non-Terrestrial Networks}, 
  year={2025},
  volume={43},
  number={1},
  pages={201-217}}

@article{3gpp_channel,
  title={Study on channel model for frequencies from 0.5 to 100 {GH}z ({R}elease 19)},
  journal={3GPP TR 38.901}, 
  author={3GPP},
  year={V19.1.0, 2025},
}

@ARTICLE{4138008,
  author={Calcev, George and others},
  journal=tvt, 
  title={A Wideband Spatial Channel Model for System-Wide Simulations}, 
  year={2007},
  volume={56},
  number={2},
  pages={389-403},
  doi={10.1109/TVT.2007.891463}}

@article{you2022beam,
  title={Beam squint-aware integrated sensing and communications for hybrid massive {MIMO} {LEO} satellite systems},
  author={You, Li and others},
  journal={IEEE J. Sel. Areas Commun.},
  volume={40},
  number={10},
  pages={2994--3009},
  year={2022},
  publisher={IEEE}
}

@article{li2021downlink,
  title={Downlink transmit design for massive {MIMO} {LEO} satellite communications},
  author={Li, Ke-Xin and others},
  journal={IEEE Trans. Commun.},
  volume={70},
  number={2},
  pages={1014--1028},
  year={2021},
  publisher={IEEE}
}

@article{you2020massive,
  title={Massive {MIMO} transmission for {LEO} satellite communications},
  author={You, Li and others},
  journal={IEEE J. Sel. Areas Commun.},
  volume={38},
  number={8},
  pages={1851--1865},
  year={2020},
  publisher={IEEE}
}

@ARTICLE{DongHyun:tcom:22,
  author={Jung, Dong-Hyun and others},
  journal=tcom, 
  title={Performance Analysis of Satellite Communication System Under the Shadowed-Rician Fading: A Stochastic Geometry Approach}, 
  year={2022},
  volume={70},
  number={4},
  pages={2707-2721}}

@ARTICLE{Sella:tvt:06,
  author={Sellathurai, M. and Guinand, P. and Lodge, J.},
  journal=tvt, 
  title={Space-time coding in mobile Satellite communications using dual-polarized channels}, 
  year={2006},
  volume={55},
  number={1},
  pages={188-199},
  doi={10.1109/TVT.2005.861195}}

@INPROCEEDINGS{Coldrey:vtc:08,
  author={Coldrey, Mikael},
  booktitle=vtc, 
  title={Modeling and Capacity of Polarized {MIMO} Channels}, 
  year={2008},
  volume={},
  number={},
  pages={440-444}}

@ARTICLE{Liolis:tcom:10,
  author={Liolis, Konstantinos P. and others},
  journal=tcom, 
  title={Statistical Modeling of Dual-Polarized {MIMO} Land Mobile Satellite Channels}, 
  year={2010},
  volume={58},
  number={11},
  pages={3077-3083},
  doi={10.1109/TCOMM.2010.091710.090507}}

@ARTICLE{MIMOSat:surv:11,
  author={Arapoglou, Pantelis-Daniel and others},
  journal=surtut, 
  title={{MIMO} over Satellite: A Review}, 
  year={2011},
  volume={13},
  number={1},
  pages={27-51},
  doi={10.1109/SURV.2011.033110.00072}}

@book{ElecWaveAntenna:02,
  title={Electromagnetic waves and antennas},
  author={Orfanidis, Sophocles J},
  year={2002},
  publisher={Rutgers University},
  address = {New Brunswick, NJ, USA},
}

@book{Antenna5G:20,
  title={Advanced antenna systems for 5{G} network deployments: Bridging the gap between theory and practice},
  author={Asplund, Henrik and others},
  year={2020},
  publisher={Academic Press},
  address = {New York, NY, USA}
}

@ARTICLE{EduCP:TEdu:03,
  author={Toh, B.Y. and Cahill, R. and Fusco, V.F.},
  journal={IEEE Trans. Educ.}, 
  title={Understanding and measuring circular polarization}, 
  year={2003},
  volume={46},
  number={3},
  pages={313-318},
  doi={10.1109/TE.2003.813519}}

@manual{ITU:23,
  title={Ionospheric propagation data and prediction methods required for the design of satellite networks and systems},
  author={Series, P},
  year={2023},
  address={Rec. ITU-R P.531-15, Int. Telecommun. Union, Geneva, Switzerland}
}

@ARTICLE{Park:twc:15,
  author={Park, Jaehyun and Clerckx, Bruno},
  journal=twc, 
  title={Multi-User Linear Precoding for Multi-Polarized Massive {MIMO} System Under Imperfect {CSIT}}, 
  year={2015},
  volume={14},
  number={5},
  pages={2532-2547},
  doi={10.1109/TWC.2014.2388207}}

@ARTICLE{Zhang:TVT:15,
  author={Zhang, Yang and others},
  journal=tvt, 
  title={Propagation Characteristics of Circularly and Linearly Polarized Electromagnetic Waves in Urban Macrocell Scenario}, 
  year={2015},
  volume={64},
  number={1},
  pages={209-222},
  doi={10.1109/TVT.2014.2318839}}

@INPROCEEDINGS{Milcom:15,
  author={Ramamurthy, Balachander and others},
  booktitle=milcom, 
  title={On {MIMO} {SATCOM} capacity analysis: Utilising polarization and spatial multiplexing}, 
  year={2015},
  volume={},
  number={},
  pages={163-168},
  doi={10.1109/MILCOM.2015.7357436}}

@INPROCEEDINGS{Linda:icc:15,
  author={Davis, Linda M. and Haley, David},
  booktitle=icc, 
  title={Geometric polarization and Faraday effects for {VHF} satellite communication links}, 
  year={2015},
  volume={},
  number={},
  pages={910-915},
  doi={10.1109/ICC.2015.7248438}}

@ARTICLE{Sena:twc:19,
  author={De Sena, Arthur Sousa and others},
  journal=twc, 
  title={Massive {MIMO}–{NOMA} Networks With Multi-Polarized Antennas}, 
  year={2019},
  volume={18},
  number={12},
  pages={5630-5642}}

@INPROCEEDINGS{Zhu:GC:18,
  author={Zhu, Jing and others},
  booktitle=globecomwrk, 
  title={Dual Polarized Spatial Modulation for Land Mobile Satellite Communications}, 
  year={2018},
  volume={},
  number={},
  pages={1-6},
  doi={10.1109/GLOCOMW.2018.8644297}}

@ARTICLE{Qian:tccn:21,
  author={Qian, Liangxin and others},
  journal=tccn, 
  title={Multi-Dimensional Polarized Modulation for Land Mobile Satellite Communications}, 
  year={2021},
  volume={7},
  number={2},
  pages={383-397},
  doi={10.1109/TCCN.2021.3072593}}

@article{clerckx2016rate,
  title={Rate splitting for {MIMO} wireless networks: {A} promising {PHY}-layer strategy for {LTE} evolution},
  author={Clerckx, Bruno and others},
  journal={IEEE Commun. Mag.},
  volume={54},
  number={5},
  pages={98--105},
  year={2016},
  publisher={IEEE}
}

@article{Mao:eurawc:18,
  title={Rate-splitting multiple access for downlink communication systems: Bridging, generalizing, and outperforming {SDMA} and {NOMA}},
  author={Mao, Yijie and Clerckx, Bruno and Li, Victor OK},
  journal=eurawc,
  volume={2018},
  number={1},
  pages={133},
  year={2018},
  publisher={Springer}
}

@ARTICLE{Shin:Net:24,
  author={Park, Jeonghun and others},
  journal={IEEE Netw.}, 
  title={Rate-Splitting Multiple Access for 6{G} Networks: Ten Promising Scenarios and Applications}, 
  year={2024},
  volume={38},
  number={3},
  pages={128-136},
  doi={10.1109/MNET.2023.3321518}}

@ARTICLE{Jaehyup:jsac:24,
  author={Seong, Jaehyup and others},
  journal=jsac, 
  title={Rate-Splitting for Joint Unicast and Multicast Transmission in {LEO} Satellite Networks With Non-Uniform Traffic Demand}, 
  year={2025},
  volume={43},
  number={1},
  pages={122-138}}

@ARTICLE{jh:tvt:24,
  author={Lee, Juhwan and others},
  journal=tvt, 
  title={Coordinated Rate-Splitting Multiple Access for Integrated Satellite-Terrestrial Networks With Super-Common Message}, 
  year={2024},
  volume={73},
  number={2},
  pages={2989-2994}}

@ARTICLE{Longfei:twc:23,
  author={Yin, Longfei and Clerckx, Bruno},
  journal=twc, 
  title={Rate-Splitting Multiple Access for Satellite-Terrestrial Integrated Networks: Benefits of Coordination and Cooperation}, 
  year={2023},
  volume={22},
  number={1},
  pages={317-332}}

@INPROCEEDINGS{Sena:gc:22,
  author={de Sena, Arthur S. and others},
  booktitle=globecom, 
  title={{RSMA} for Dual-Polarized Massive {MIMO} Networks: A {SIC}-Free Approach}, 
  year={2022},
  volume={},
  number={},
  pages={1643-1648}}

@ARTICLE{Sena:wcl:22,
  author={de Sena, Arthur Sousa and others},
  journal=wcl, 
  title={Dual-Polarized {RSMA} for Massive {MIMO} Systems}, 
  year={2022},
  volume={11},
  number={9},
  pages={2000-2004}}

@ARTICLE{Sena:twc:23,
  author={de Sena, Arthur Sousa and others},
  journal=twc, 
  title={Dual-Polarized Massive {MIMO}-{RSMA} Networks: Tackling Imperfect {SIC}}, 
  year={2023},
  volume={22},
  number={5},
  pages={3194-3215}}

\end{document}